\documentclass[12pt]{article} 
\usepackage[sectionbib]{natbib}
\usepackage{array,epsfig,fancyheadings,rotating}
\usepackage[]{hyperref}  
\usepackage{xcolor}

\usepackage[ruled,vlined]{algorithm2e}
\usepackage{threeparttable} 
\usepackage{adjustbox}      
\usepackage{bm, bbm, booktabs} 
\usepackage{epsfig, epsf, enumerate}
\usepackage{epstopdf}
\usepackage{graphicx}
\usepackage{longtable}
\usepackage{mathrsfs, multirow}
\usepackage{verbatim} 
\usepackage{xr}       
\usepackage{sectsty, secdot}
\sectionfont{\fontsize{12}{14pt plus.8pt minus .6pt}\selectfont}
\renewcommand{\theequation}{\thesection\arabic{equation}}
\subsectionfont{\fontsize{12}{14pt plus.8pt minus .6pt}\selectfont}

\usepackage{amsmath}
\usepackage{amssymb}
\usepackage{amsfonts}
\usepackage{multirow}
\usepackage{amsthm}

\theoremstyle{definition}

\newcommand{\mybeta}{\boldsymbol{\beta}}

\newcommand{\mytau}{\boldsymbol{\tau}}
\newcommand{\mytheta}{\boldsymbol{\theta}}

\newcommand{\mySigma}{\boldsymbol{\Sigma}}
\newcommand{\mysigma}{\boldsymbol{\sigma}}

\newcommand{\myxi}{\boldsymbol{\xi}}

\newcommand{\mya}{\mathbf{a}}
\newcommand{\myb}{\mathbf{b}}

\newcommand{\mye}{\mathbf{e}}
\newcommand{\myf}{\mathbf{f}}

\newcommand{\myq}{\mathbf{q}}

\newcommand{\mys}{\mathbf{s}}

\newcommand{\myu}{\mathbf{u}}

\newcommand{\myx}{\mathbf{x}}
\newcommand{\myy}{\mathbf{y}}
\newcommand{\myz}{\mathbf{z}}

\newcommand{\myA}{\mathbf{A}}
\newcommand{\myB}{\mathbf{B}}
\newcommand{\myD}{\mathbf{D}}

\newcommand{\myH}{\mathbf{H}}
\newcommand{\myI}{\mathbf{I}}
\newcommand{\myL}{\mathbf{L}}

\newcommand{\myM}{\mathbf{M}}
\newcommand{\myW}{\mathbf{W}}
\newcommand{\myX}{\mathbf{X}}

\newcommand{\myP}{\mathbf{P}}

\newcommand{\myQ}{\mathbf{Q}}

\newcommand{\myZ}{\mathbf{Z}}

\newcommand{\mS}{\mathcal{S}}

\theoremstyle{definition}
\newtheorem{mytheorem}{Theorem}[section]
\newtheorem{mylemma}[mytheorem]{Lemma}

\newtheorem{myprop}[mytheorem]{Proposition}

\newtheorem{mycorollary}{Corollary}[section]

\newtheorem{mycondition}{Condition}[section]

\SetKwComment{Comment}{$\triangleright$\ }{}
\SetCommentSty{itshape}
\SetKwBlock{functionalEstimation}{Functional estimation stage:}{end}
\SetKwBlock{PDEidentification}{Model identification stage:}{end}
\makeatletter
\newcommand{\algorithmfootnote}[2][\footnotesize]{%
	\let\old@algocf@finish\@algocf@finish
	\def\@algocf@finish{\old@algocf@finish
		\leavevmode\rlap{\begin{minipage}{\linewidth}
				#1#2
		\end{minipage}}%
	}
}
\makeatother

\graphicspath{{figures/}}

\begin{document}


\renewcommand{\baselinestretch}{2}

\markright{ \hbox{\footnotesize\rm Statistica Sinica
}\hfill\\[-13pt]
\hbox{\footnotesize\rm
}\hfill }

\markboth{\hfill{\footnotesize\rm Yujie Zhao, Xiaoming Huo, Yajun Mei } \hfill}
{\hfill {\footnotesize\rm PDE identification from noisy data} \hfill}

\renewcommand{\thefootnote}{}
$\ $\par


\fontsize{12}{14pt plus.8pt minus .6pt}\selectfont \vspace{0.8pc}
\centerline{\large\bf 	Identification of Partial-Differential-Equations-Based Models  }
\vspace{2pt}
\centerline{\large\bf from Noisy Data via Splines}
\vspace{.4cm}
\centerline{Yujie Zhao\textsuperscript{1}, Xiaoming Huo\textsuperscript{2}, Yajun Mei\textsuperscript{2} }
\vspace{.4cm}
\centerline{\it \textsuperscript{1} Biostatistics and Research Decision Sciences Department, Merck \& Co., Inc}
\centerline{\it \textsuperscript{2} H. Milton Stewart School of Industrial and Systems Engineering, Georgia Tech}
 \vspace{.55cm} \fontsize{9}{11.5pt plus.8pt minus.6pt}\selectfont


\begin{quotation}
\noindent {\it Abstract:}
We propose a two-stage method called \textit{Spline-Assisted Partial Differential Equations-based Model Identification} that can be used to identify models based on partial differential equations (PDEs) from noisy data.
In the first stage, we employ cubic splines to estimate unobservable derivatives.
The underlying PDE is based on a subset of these derivatives.
This stage is computationally efficient. Its computational complexity is the product of a constant and the sample size, which is the lowest possible order of computational complexity.
In the second stage, we apply the least absolute shrinkage and selection operator to identify the underlying PDE-based model.
Statistical properties are developed, including the model identification accuracy.
We validate our theory using numerical examples and a real-data case study based on an National Aeronautics and Space Administration data set.

\vspace{9pt}
\par
\end{quotation}\par

\def\thefigure{\arabic{figure}}
\def\thetable{\arabic{table}}

\renewcommand{\theequation}{\thesection.\arabic{equation}}

\fontsize{12}{14pt plus.8pt minus .6pt}\selectfont

\section{Introduction}
\label{sec: PDE paper -- introduction}
Partial differential equations (PDEs) are widely used to model physical processes in fields such as engineering \cite[][]{wang2019spatiotemporal}, physics \cite[][]{xun2013parameter}, and biology \cite[][]{lagergren2020learning}.
In these applications, there are two classes of technical issues: the \textit{forward problem} and the \textit{inverse problem}.
The forward problem studies the properties of functions that PDEs determine.
It has been extensively studied by mathematicians \cite[]{olver2014introduction, wang2014inhomogeneous}.
Different from forward problems, inverse problems try to identify PDE-based models from the observed noisy data.
Research on the inverse problem is relatively sparse, and the corresponding statistical property is notably less known.
In this paper, we propose a method for solving the inverse problem, which we refer to as a \textit{PDE identification problem}.

With the rise of big data, the PDE identification problem has become indispensable.
A good PDE identification approach offers at least the following two benefits.
First, we can predict future trends using the identified PDE model, conditional that such a model reflects the underlying processes.
Second, interpretable PDE models enable scientists to validate/reexamine the underlying physical/biological laws governing the process.

We propose a new method for the PDE identification problem, called \textit{Spline Assisted Partial Differential Equation based Model Identification (SAPDEMI)}.
SAPDEMI can efficiently identify the underlying PDE model from noisy data $\mathcal D$:
\begin{equation}
	\label{equ: noisy data}
	\begin{array}{ccl}
		\mathcal D
		& =
		\{
		(x_i,t_n, u_i^n):
		&
		x_i \in(0, X_{\max}) \subseteq \mathbb R,\; \forall \; i=0,\ldots,M-1, \;\\
		&  &
		t_n \in(0, T_{\max}) \subseteq \mathbb R, \;\forall \; n=0,\ldots, N-1
		\}
		\in \Omega,
	\end{array}
\end{equation}
where $x_i \in \mathbb{R}$ is a spatial variable, with $x_i \in (0, X_{\max})$, for $i = 0,1,\ldots, M-1$, and we call $M$ the \textit{spatial resolution}.
The variable $t_n \in \mathbb{R}$ is a temporal variable, with $t_n \in (0, T_{\text{max}})$, for $n = 0,1,\ldots, N-1$, and we call $N$ the \textit{temporal resolution}.
We use $T_{\max}$ and $X_{\max}$ to denote the upper bound of the temporal variable and the spatial variable, respectively.
The variable $u_i^n$ is a representation of the ground truth $u(x_i, t_n)$, contaminated by the noise that follows a normal distribution with mean zero and stand deviation $\sigma$:
\begin{equation}
	\label{equ: u = u + noise}
	u_i^n = u(x_i, t_n) + \epsilon_i^n, \text{\; \; \;  \;} \epsilon_i^n \stackrel{i.i.d.}{\sim}  N(0, \sigma^2).
\end{equation}
Here, $u(x, t)$ is the ground truth function, which is determined by an underlying PDE model, and is assumed to satisfy the following equation:
\begin{equation}
	\label{equ: temporal evolutionary PDE}
	\begin{array}{ccl}
		\frac{\partial}{\partial t} u(x,t)
		& = &
		\beta_{00}^*
		+
		\sum\limits_{k=0}^{q_{\max}}
		\sum\limits_{i=1}^{p_{\max}}
		\beta_{k^i}^* \left[ \frac{\partial^k}{\partial^k x} u(x,t) \right]^i
		+ \\
		&   &
		\sum\limits_{\substack{i + j \leq p_{\max} \\ i,j > 0 }} \;
		\sum\limits_{\substack{0 < k < l  \\ l\leq q_{\max} }}
		\beta_{k^i,l^j}^*
		\left[ \frac{\partial^k}{\partial^k x} u(x,t) \right]^i
		\left[ \frac{\partial^l}{\partial^l x} u(x,t) \right]^j.
	\end{array}
\end{equation}
The left-hand side of the above equation is the first-order partial derivative of the underlying function with respect to the temporal variable $t$, and the right-hand side is the $p_{\max}$th-order polynomial of the derivatives with respect to the spatial variable $x$ up to the $q_{\max}$th total order.
For notational simplicity, we denote the ground truth coefficient vector,
$
\mybeta^* =
(\beta_{00}^*, \beta_{0^1}^*, \beta_{1^1}^*, \ldots, \beta_{q_{\max}^{p_{\max}}}^*),
$
as
$
\mybeta^* = (\beta_{1}^*, \beta_{2}^*, \beta_{3}^*, \ldots, \beta_{K}^*)^\top,
$
where
$
K = 1 + (p_{\max}+1)q_{\max} + \frac{1}{2} q_{\max}(q_{\max} + 1) (p_{\max} - 1)!
$
is the total number of coefficients on the right-hand side.
Noted that, in practice, the majority of the entries in $\mybeta^*$ are zero.
For instance, in the transport equation
$
\frac{\partial}{\partial t} u(x,t)
=
a \frac{\partial}{\partial x} u(x,t),
$
with any $a \neq 0$, we have only $\beta^*_3 \neq 0$ and $\beta^*_i = 0$, for any $i \neq 3$ \cite[see][Section 2.2]{olver2014introduction}.
Therefore, it is reasonable to assume that the coefficient $\mybeta^*$ in \eqref{equ: temporal evolutionary PDE} is sparse.

To identify the above model, we need to overcome two technical challenges.
First, the derivatives in \eqref{equ: temporal evolutionary PDE} are unobservable, and need to be estimated from the noisy observations. 
Second, we need to identify the underlying model, which is presumably simple (i.e., sparse) .

We design our proposed SAPDEMI method as a two-stage method to identify the underlying PDE models from the noisy data $\mathcal D$.
The first stage is called the \textit{functional estimation stage}, where we estimate all the derivatives from the noisy data $\mathcal D$, including
$
\frac{\partial}{\partial t} u(x, t),
\frac{\partial}{\partial x} u(x, t),
$
and so on.
In this stage, we first use cubic splines \cite[][]{shridhar1974generalized} to fit the noisy data $\mathcal D$, and then we approximate the derivatives of the underlying function from the derivatives of the estimated cubic splines.
The second stage is called the \textit{model identification stage}, where we apply the least absolute shrinkage and selection operator (Lasso) \cite[][]{tibshirani1996regression} to identify the derivatives (or their combinations) that should be included in the PDE-based models.
To ensure accuracy, we develop sufficient conditions for correct identification and the asymptotic properties of the identified models.
The main tool used in our theoretical analysis is the primal-dual witness (PDW) method \cite[see][Chapter 11]{hastie2015statistical}.

The rest of this section is organized as follows.
In Section \ref{sec: PDE paper -- literature review}, we review existing methods related to the PDE identification problem.
In Section \ref{sec: PDE paper -- contributions}, we summarize our contributions.

\subsection{Literature Review}
\label{sec: PDE paper -- literature review}

A pioneering work in identifying underlying dynamic models from noisy data is that of \cite{liang2008parameter}.
Their method is also a two-stage method. In the functional estimation stage, they use a local polynomial regression to estimate the value of the function and its derivatives.
Subsequently, in the model identification stage, they use the least squares method.
Following this work, various extensions have been proposed.

The first class of extensions modifies the functional estimation stage of \cite{liang2008parameter}, and can be divided into three categories.
{\bf (F1)}. 
In the numerical differentiation category \cite[][]{wu2012numerical},  the derivative $\frac{\partial}{\partial x} u(x,t)$ is simply approximated as
$
\frac{\partial}{\partial x} u(x,t)
\approx
\frac{u(x + \Delta x, t) - u(x - \Delta x, t)}{2 \Delta x},
$
where $(x + \Delta x, t)$ and $(x - \Delta x, t)$ are the two closest points to $(x, t)$ in the $x$-domain.
The essence of numerical differentiation is to approximate the first-order derivative as the slope of a nearby secant line.
Although the implementation is easy, the approximation results can be highly biased, because its accuracy  depends greatly on $\Delta x$:
a small value of $\Delta x$ yields large rounding errors in the subtraction \cite[][]{ueberhuber2012numerical}, and a large value of $\Delta x$ leads to poor performance when estimating the tangent slope using secants. Thus, this naive numerical differentiation is not preferred owing to its bias. 
{\bf (F2)}. 
In the basis expansion category, researchers first approximate the unknown functions using basis expansion methods, and then approximate the derivatives of the underlying function as those of the approximated functions.
There are multiple options for the choice of bases.
The most popular basis is the local polynomial basis \cite[see][]{liang2008parameter, bar1999fitting, schaeffer2017learning, rudy2017data, parlitz2000prediction}.
Another popular choice is the spline basis \cite[see][]{wu2012numerical, xun2013parameter, wang2019spatiotemporal}.
Our proposed method belongs here. 
In this category, the major limitation of existing approaches is the potentially high computational complexity.
For instance, the local polynomial basis requires computational complexity of order $\max\{O(M^2 N), O(M N^2) \}$ in the functional estimation stage.
However, we show that our proposed SAPDEMI method  requires only $O(MN)$.
The sample size of the dataset $\mathcal D$ is $MN$, so it takes at least $MN$ numerical operations to read $\mathcal D$.
Consequently, the lowest possible bound, in theory, is $O(MN)$, As achieved by our proposed SAPDEMI method.
{\bf (F3)}. 
In the machine or deep learning category, researchers first fit unknown functions using machine/deep learning methods, and then approximate the derivatives of the underlying functions as those of the approximated functions.
A popular machine/deep learning method is the neural network (NN) approach.
For instance, \cite{srivastava2020learning} use an artificial neural network (ANN). 
These methods are limited by potential overfitting and the selection of the hyper-parameters. 
	

The second class of extensions modifies the model identification stage of \cite{liang2008parameter}.
Here, existing methods fall within the framework of the (penalized) least squares method, and we can again divide them into three categories.
{\bf (M1)}. 
In the least squares category, researchers study  ordinary differential equation (ODE) identification \cite[][]{miao2009differential} and PDE identification \cite[][]{bar1999fitting, wu2012numerical} , althrough they too have problems with overfitting.
{\bf (M2)}. 
In the $\ell_2$-penalized least squares category, \cite{xun2013parameter} and \cite{wang2019spatiotemporal} penalize the smoothness of the unknown function, which is assumed to be in a prescribed reproducing kernel Hilbert space (RKHS).
Essentially, this method falls within the framework of the $\ell_2$-penalized least squares method.
Although this method helps to avoid overfitting by introducing the $\ell_2$-penalty, it has limited power in terms of ``model selection''.
{\bf (M3)}. 
In the $\ell_1$-penalized least squares method category, \cite{schaeffer2017learning} identifies unknown dynamic models (i.e., functions) using the $\ell_1$-penalized least squares method. The author provides an efficient algorithm, based on the proximal mapping method, but does not discuss the statistical proprieties of the identified model.
Recently, \cite{kang2019ident} use a similar method to that of \cite{schaeffer2017learning}, and demonstrate empirical successes.
However, the derivation of the statistical theory is still missing. Our study addresses this gap in the literature.

In addition to the $\ell_2$- or $\ell_1$-penalized least squares methods, other methods have been proposed for the model selection stage, but are not as widely used. 
Here, examples include the Akaike information criterion (AIC) in \cite{mangan2017model}, smoothly clipped absolute deviation (SCAD) in \cite{lu2011high}, and hard-thresholding in \cite{rudy2017data}.
The first two approaches may lead to NP-hard problems in numerical implementation.
The last one is ad-hoc, and may be difficult to analyze.
Thus, we do not address these alternative approaches.

Although our proposed SAPDEMI method applies to the PDE model, other nonparametric models are possible.
Here, we take PDEs as an initial research project mainly because they are deterministic.
Thus, we can compare our identified model with the true model, and show the model notification accuracy. 
As our initial research project, we prefer the PDE to machine learning (ML) models  (e.g., neural network, random forest), because a PDE offers insight into the physical law.
However, the ML models are usually black-box methods  \citep{loyola2019black}. 
We also prefer the PDE to the time series models, because it behaves like a ``continuous version'' of a time series model \citep{perona2000trajectory, chen2018functional} at a high level.
Furthermore, we prefer the PDE to the Gaussian process (GP) model, because the GP model restricts its response variables to follow a Gaussian distribution \citep{liu2020gaussian, wei2018u}. 
Again, although we take the PDE as our initial research project, we are open to using the aforementioned nonparametric models in future work.

\subsection{Our Contributions}
\label{sec: PDE paper -- contributions}

Here, we summarize the contributions of our proposed method.
{\bf (1)}
In the functional estimation stage, our proposed SAPDEMI method is computationally efficient.
Specifically, we require computational complexity of order $O(MN)$, which is the lowest possible order in this stage.
In comparison, the aforementioned local polynomial regression requires computational complexity of order $\max\{O(M^2 N), O(M N^2)\}$, which is higher.
{\bf (2)}
For our proposed SAPDEMI method, we establish a theoretical guarantee of the model identification accuracy, which, to the best of our knowledge, is a novel result.
{\bf (3)}
We extend our method to PDE-based model identification, and compare it with ODE-based model identification.
The latter has more related work, whereas the former is not yet well understood.

The rest of the paper is organized as follows.
In Section \ref{sec: PDE paper -- model}, we describe the technical details of our proposed SAPDEMI method.
In Section \ref{sec: PDE paper -- theory}, we present our main theory, including the sufficient conditions for correct identification, and the statistical properties of the identified models.
In Section \ref{sec: PDE paper -- simulation}, we conduct numerical experiments to validate the theory from Section \ref{sec: PDE paper -- theory}.
In Section \ref{sec: case study}, we apply SAPDEMI to a real-world case study using data downloaded from the National Aeronautics and Space Administration (NASA).
In Section \ref{sec: conclusion}, we conclude the paper and discuss some future research.

\section{Proposed Method: SAPDEMI}
\label{sec: PDE paper -- model}

The proposed SAPDEMI method is a two-stage method for identifying the underlying PDE model from noisy data $\mathcal{D}$.
The first stage is called the \textit{ functional estimation stage}. Here, we estimate the function and its derivatives from the noisy data $\mathcal D$ in \eqref{equ: noisy data}, and use these as input in the second stage.
The second stage is called the \textit{model identification stage}, where we identify the underlying PDE-based model.

In our notation, scalars are denoted by lowercase letters (e.g., $\beta$).
Vectors are denoted by lowercase bold face letters (e.g., $\mybeta$), and its $i$th entry is denoted as $\beta_i$.
Matrices are denoted by uppercase boldface letter (e.g., $\myB$), and its $(i,j)$th entry is denoted as $B_{ij}$.
For the vector $\mybeta \in \mathbb R^p$, its $k$th norm is defined as
$
\| \mybeta \|_k:= \left( \sum_{i=1}^{p} |\beta_i|^k \right)^{1/k}.
$
For the matrix $\myB \in \mathbb R^{m \times n}$, its Frobenius norm is defined as
$
\left\| \myB \right\|_F
=
\sqrt{ \sum_{i =1}^{m} \sum_{j =1}^{n} |B_{ij}|^2}.
$
We write $f(n) = O(g(n))$ if there exists a $G  \in \mathbb R^{+}$ and an $n_0$ such that
$
|f(n)|
\leq
G g(x),
$
for all $n > n_0$.

This section is organized as follows. 
In Section \ref{sec: solve derivatives by cubic spline}, we introduce the functional estimation stage, and in Section \ref{sec: use LASSO to identify the PDE model}, we describe the model identification stage.

\subsection{Functional Estimation Stage}
\label{sec: solve derivatives by cubic spline}
In this section, we describe the functional estimation stage of our proposed SAPDEMI method.
In this stage, we estimate the functional values and their derivatives from the noisy data $\mathcal D$ in \eqref{equ: noisy data}.
These derivatives include the derivatives with respect to the spatial/temporal variable $x/t$.
We take derivatives with respect to the spatial variable $x$ as an example; the derivatives with respect to the temporal variable $t$ can be derived similarly.

The main tool that we use is the cubic spline.
Suppose there is a cubic spline $s(x)$ over the knots $\left\{(x_i, u_i^n) \right\}_{i=0,1,\ldots, M-1}$ satisfying the  properties in \cite{mckinley1998cubic}:
{\bf (1)} 
$s(x) \in C^2[x_0, x_{M-1}]$, where $C^2[x_0, x_{M-1}]$ denotes the sets of function whose $0$th, first, and second derivatives are continuous in $[x_0, x_{M-1}]$;
{\bf (2)} 
For any $i = 1,\ldots ,M-1$, $s(x)$ is a polynomial of degree three in $[x_{i-1},x_{i}]$;
{\bf (3)}.
For the two end-points, $x_0$ and $x_{M-1}$, we have $s''(x_0) = s''(x_{M-1}) = 0$, where $s''(x)$ is the second derivative of $s(x)$.

By fitting the data $\left\{ (x_i, u_i^n) \right\}_{i=0,1,\ldots,M-1}$ (with a general fixed $n \in \{0,1,\ldots,N-1\}$) into the above cubic spline $s(x)$, one can solve $s(x)$ as the minimizer of the following optimization problem:
\begin{equation}
	\label{equ: smoothing cubic spline -- objective fucntion}
	J_{\alpha} (s)
	=
	\alpha \sum_{i=0}^{M-1} w_i [u_i^n - s(x_i)]^2
	+
	(1-\alpha) \int_{x_0}^{x_{M-1}} s''(x)^2 dx,
\end{equation}
where the first term $\alpha \sum_{i=0}^{M-1} w_i [u_i^n - s(x_i)]^2$ is the weighted sum of squares for the residuals, and we take the weight $w_0 = w_1 = \ldots = w_{M-1} = 1$.
In the second term, $(1-\alpha) \int_{x_0}^{x_{M-1}} s''(x)^2 dx$, the function $s''(x)$ is the second derivative of $s(x)$, and this term is the penalty of the smoothness.
In the above optimization problem, the parameter $\alpha \in (0,1]$ controls the trade-off between the goodness of fit and the smoothness of the cubic spline.
By minimizing the above optimization problem, we obtain an estimate of $s(x)$, together with its first derivative $s'(x)$ and its second derivative $s''(x)$.
If the cubic spline approximates the underlying PDE curves well,  we can declare that the derivatives of the underlying dynamic system can be approximated by the derivatives of the cubic spline $s(x)$, that is, we have
$
\widehat{ u(x,t_n) } \approx \widehat{s(x)},
\widehat{ \frac{\partial}{\partial x}u(x,t_n) }\approx \widehat{ s'(x)},
\widehat{ \frac{\partial^2}{\partial x^2}u(x,t_n) }\approx \widehat{ s''(x) }
$
\cite[][]{ahlberg1967theory, rubin1975cubic, rashidinia2008non}.
Following a similar procedure to obtain the derivatives with respect to the spatial variable $x$, we can get the derivatives with respect to the temporal variable $t$, that is, $\widehat{\frac{\partial}{\partial t} u(x_i,t_n)}$, for any $i = 0, \ldots, M-1$ and $n = 0, \ldots, N-1$. 

A nice property of the cubic spline is that there is a closed-form solution for \eqref{equ: smoothing cubic spline -- objective fucntion}.
First, the value of the cubic spline $s(x)$ at the point $\{ x_0, x_1, \ldots, x_{M-1} \}$, that is,
$
\widehat{ \mys}
=
\left(
\widehat{ s(x_0) },
\widehat{ s(x_1) },
\ldots,
\widehat{ s(x_{M-1}) }
\right)^\top,
$
can be solved as
\begin{equation}
	\label{equ: cubic spline -- zero order derivative estimation}
	\widehat{ \mys }
	=
	[\alpha \myW +(1-\alpha) \myA^\top \myM \myA]^{-1} \alpha \myW \myu_:^n.
\end{equation}
The above closed-form estimation can be used to approximate the function that corresponds to the underlying PDE model, that is,
$
\widehat{ \mys }
\approx
\widehat\myf
$\\
$
=
\left(
\widehat{ u(x_0,t_n) },
\widehat{ u(x_1,t_n) },
\ldots,
\widehat{ u(x_{M-1},t_n) }
\right)^\top.
$
Here, 
$
\myW = \text{diag}(w_0, \ldots, w_{M-1}) \in \mathbb R^{M \times M},
$ 
vector
$
\myu_:^{n} = \left(u_0^n, \ldots,  u_{M-1}^n \right)^\top \in \mathbb R^{M},
$
and the matrices 
$
\myA \in \mathbb R^{(M-2) \times M}
$
and 
$
\myM \in \mathbb R^{(M-2) \times (M-2)}
$
are
\begin{equation}
 	\label{equ: proof -- cubic spline -- derivative estimation -- matrix A}
 	\myA=
 	\left(
 	\begin{array}{ccccccccc}
 		\frac{1}{h_0} & \frac{-1}{h_0} - \frac{1}{h_1} & \frac{1}{h_1}                            & \ldots & 0 & 0& 0 \\
 		0             & \frac{1}{h_1}                  & \frac{-1}{h_1} - \frac{1}{h_2} &  \ldots & 0 & 0& 0  \\
 		\vdots  & \vdots & \vdots & \ddots & \vdots & \vdots  & \vdots \\
 		0             & 0                              & 0                               & \ldots & \frac{1}{h_{M-3}} & \frac{-1}{h_{M-3}} - \frac{1}{h_{M-2}} & \frac{1}{h_{M-2}}\\
 	\end{array}
 	\right),
\end{equation}
 \begin{equation}
 	\label{equ: proof -- cubic spline -- derivative estimation -- matrix M}
 	\myM
 	=
 	\left(
 	\begin{array}{cccccc}
 		\frac{h_0+h_1}{3} & \frac{h_1}{6}     & 0                 & \ldots & 0 & 0 \\
 		\frac{h_1}{6}     & \frac{h_1+h_2}{3} & \frac{h_2}{6}     & \ldots & 0 & 0 \\
 		\vdots            & \vdots            & \vdots            & \ddots & \vdots & \vdots \\
 		0                 & 0                 & 0                 & \ldots & \frac{h_{M-3}}{6} & \frac{h_{M-3} + h_{M-2}}{3}
 	\end{array}
 	\right),
 \end{equation}
respectively, with $h_i = x_{i+1} - x_{i}$, for $i = 0, 1,\ldots, M-2$.

For the mathematical derivation of \eqref{equ: cubic spline -- zero order derivative estimation} from \eqref{equ: smoothing cubic spline -- objective fucntion}, and the derivation of first- and second-order derivatives, please refer to the  Supplementary Material \ref{proof: derivative of the smoothing cubic Spline}.

The advantage of the cubic spline is that its computational complexity is only a linear polynomial of the sample size $MN$.

\begin{myprop}
	\label{theo: computation complexity of cubic spline}
	Given the data $\mathcal D$ in \eqref{equ: noisy data}, if we use the cubic spline in \eqref{equ: smoothing cubic spline -- objective fucntion} in the functional estimation stage, the computation complexity is of order
	$$
	\max\{ O(p_{\max} MN), O(K^3) \},
	$$
	where $p_{\max}$ is the highest polynomial order in \eqref{equ: temporal evolutionary PDE}, $M/N$ is the spatial/temporal resolution, and $K$ is the number of covariates in \eqref{equ: temporal evolutionary PDE}.
\end{myprop}

The proof can be found in the online Supplementary Material \ref{proof: computation complexity in cubic spline (without interaction)}.

As suggested by Proposition \ref{theo: computation complexity of cubic spline}, when $p_{\max}, K \ll M, N$ (which is often the case in practice), it only requires $O(MN)$ numerical operations in the functional estimation stage.
This is the lowest possible order of complexity in this stage, because $MN$ is exactly the sample size of $\mathcal D$, and reading the data is a task of order $O(MN)$.
Therefore, it is very efficient to use a cubic spline, because its computational complexity achieves the lowest possible order of complexity.

By way of comparisons, we discuss the computational complexity of the local polynomial regression, which is widely used in the literature \cite[][]{liang2008parameter, bar1999fitting, schaeffer2017learning, rudy2017data, parlitz2000prediction}. 
This computational complexity is \\
$
  \max\{ O(M^2 N), O(M N^2) , O(p_{\max} MN), O(K^3) \},
$
which is much higher than ours for a generalized polynomial order $p_{\max}$.
Specifically, if one restricts the local polynomial regression method to the same order as that of the cubic spline, its computational complexity is 
$$
  \max\{O(M^2N), O(MN^2), O(K^3)\},
$$
which is still higher than that of the cubic spline method in Proposition \ref{theo: computation complexity of cubic spline}.
The related proposition and proof are available in Supplementary Materials \ref{sec: proof}.
We  summarize the pros and cons of the cubic spline and the local polynomial regression in Table \ref{table: Pros and cons of cubic spline and local polynomial}.

\begin{table}[htbp]
	\caption{Pros and cons of the cubic spline and the local polynomial regression in the functional estimation stage (assume that $p_{\max}, q_{\max}, K \ll M,N$)
	\tiny
		\label{table: Pros and cons of cubic spline and local polynomial}}
		\begin{tabular}{p{0.1\textwidth}|p{0.5\textwidth}|p{0.4\textwidth}}
		\hline
		Method & Cubic spline & Local polynomial regression\\
		\hline
		Pros
		&
		Computational complexity is
		$
		O(MN)
		$
		&
		Derivatives up to any order
		\\
		\hline
		Cons  &
		If higher-than-$2$ order is required, need extensions beyond cubic splines.
		&
		Computational complexity is
		$
		\max\{ (M^2 N), O(M N^2) \}
		$\\
		\hline
	\end{tabular}
\end{table}

\subsection{Model Identification Stage}
\label{sec: use LASSO to identify the PDE model}
In this section, we discuss the model identification stage of our proposed SAPDEMI method.
In this stage, we identify the PDE model in \eqref{equ: temporal evolutionary PDE}.

Note that the model in \eqref{equ: temporal evolutionary PDE} can be regarded as a linear regression model with a response variable that is the first-order derivative with respect to the temporal variable $t$, that is,
$
\frac{\partial u(x, t)}{\partial t},
$
and the covariates are the derivative(s) with respect to the spatial variable $x$, including
$
\frac{\partial}{\partial x} u(x_i, t_n),
\frac{\partial^2}{\partial x^2} u(x_i, t_n),
\ldots,
$
$
\left( \frac{\partial^2}{\partial x^2} u(x_i, t_n) \right)^{p_{\max}}.
$
Because we have $MN$ observations in the data set $\mathcal D$ in \eqref{equ: noisy data}, the response vector is of length $MN$:
\begin{equation}
	\label{equ: estimated response vector}
	\begin{array}{ccccccccc}
		\nabla_t \myu
		& = &
		\left( \right.
		\widehat{ \frac{\partial u(x_0, t_0)}{\partial t} }, &
		\ldots, &
		\widehat{ \frac{\partial u(x_{M-1}, t_0)}{\partial t} }, &
		\ldots, &
		\widehat{ \frac{\partial u(x_{M-1}, t_{N-1})}{\partial t} }
		\left. \right)^\top,
	\end{array}
\end{equation}
and the design matrix is of dimension $MN \times K$:
\begin{equation}
	\label{equ: estimated design matrix}
	\begin{array}{ccccccccc}
		\myX
		& = &
		\left( \right.
		\widehat{\myx_0^0}, &
		\widehat{\myx_1^0}, &
		\ldots, &
		\widehat{\myx_{M-1}^0} ,&
		\widehat{\myx_1^0}, &
		\ldots, &
		\widehat{\myx_{M-1}^{N-1}}
		\left. \right)^\top
		\in \mathbb R^{MN \times K}.
	\end{array}
\end{equation}
For the above design matrix $\myX$, its $(nN+i+1)$st row is
$\widehat{\myx_i^n }  =  \left(	1, 	\widehat{ u(x_i,t_n)  } ,
\right.$
$
\left.
\widehat{ \frac{\partial}{\partial x} u(x_i, t_n) },
\widehat{ \frac{\partial^2}{\partial x^2} u(x_i, t_n)},
\left( \widehat{u(x_i,t_n)} \right)^2 ,
\ldots,
\left( \widehat{ \frac{\partial^2}{\partial x^2} u(x_i, t_n)} \right)^{p_{\max}}
\right)^\top.
$
The $K$ components of $\widehat{\myx}_i^n$ are candidate terms in the PDE model.
Note that all of the derivatives listed in \eqref{equ: estimated response vector} and \eqref{equ: estimated design matrix} are estimated from the functional estimation stage described in Section \ref{sec: solve derivatives by cubic spline}.

Next, we use the Lasso to identify the nonzero coefficients in \eqref{equ: temporal evolutionary PDE}:
\begin{equation}
	\label{equ: LASSO model - matrix algebra}
	\widehat{\mybeta}
	=
	\arg\min_{\mybeta}
	\frac{1}{2MN}
	\left\| \nabla_t \myu - \myX \mybeta \right\|_2^2,
	+
	\lambda \| \mybeta \|_1,
\end{equation}
where $\lambda>0$ is a turning parameter that controls the trade-off between the sparsity of $\mybeta$ and the goodness of fit.
Given the $\ell_1$ penalty in \eqref{equ: LASSO model - matrix algebra}, $\widehat\mybeta$ is sparse, that is, only a few of its entries are likely to be nonzero.
Accordingly, we can identify the underlying PDE model as
\begin{equation}\label{eqy: LASSO model -- identification result}
	\frac{\partial}{\partial t} u(x,t)
	=
	\myx^\top \widehat{\mybeta},
\end{equation}
where $\myx	= \left(	1,  u(x,t),  	\frac{\partial}{\partial x} u(x, t), 	\frac{\partial^2}{\partial x^2} u(x, t),	\left( u(x,t) \right)^2,	\ldots, \left(\frac{\partial^2}{\partial x^2} u(x, t) \right)^{p_{\max}} \right)^\top  \in \mathbb R^{K}$.
To solve equation \eqref{equ: LASSO model - matrix algebra}, one can use the coordinate descent method  \cite[][]{beck2013convergence, tseng2001convergence}; see the online Supplementary Material \ref{appendix: Coordinate Gradient Descent}.
\section{Theory on Statistical Properties}
\label{sec: PDE paper -- theory}

The theoretical evaluation is performed from two aspects. 
\textbf{(S1).}
First, we check whether our identified PDE model contains derivatives that are included in the ``true'' underlying PDE model.
This is called the \textit{support set recovery} property.
Mathematically, we check whether
$
\text{supp}(\widehat{\mybeta})
\subseteq
\text{supp}(\mybeta^*),
$
where $\widehat{\mybeta}$ is the minimizer of \eqref{equ: LASSO model - matrix algebra}, $\mybeta^*$ is the ground truth, and
$\text{supp}(\mybeta) = \{ i: \beta_i \neq 0, \; \forall \; i,  1 \leq i \leq K \}$, for a general vector $\mybeta \in \mathbb R^K$.
However, the support recovery depends on the choice of the penalty parameter $\lambda$:
a large value of $\lambda$ leads to $\text{supp}(\widehat\mybeta) =  \emptyset$ (empty set), whereas a small value of $\lambda$ results in a nonsparse $\widehat \mybeta$.
A proper selection of $\lambda$ hopefully leads to the correct recovery of the support set recovery, that is, we have
$
\text{supp}(\widehat{\mybeta})
\subseteq
\text{supp}(\mybeta^*).
$
We discuss the selection of $\lambda$ to achieve the above goal in Theorem \ref{theo: support set recovary}.
\textbf{(S2).}
Second, we are interested in an upper bound of the estimation error of our estimator.
Specifically, we consider 
$
\left\|
\widehat \mybeta_{\mS}
-
\mybeta^*_{\mS}
\right\|_{\infty},
$
where $\mS = \text{supp}(\mybeta^*)$, and the vectors $\widehat \mybeta_{\mS}$ and  $\mybeta^*_{\mS}$ are subvectors of $\widehat\mybeta$ and $\mybeta^*$, respectively, and contain only elements with indices that are in $\mS$.
An upper bound of the above estimation error is discussed in Theorem \ref{theo: beta estimation error bound}.

This section is organized as follows. 
In Section \ref{sec: Model Assumptions}, we present the conditions for the theorems.
In Section \ref{sec: PDE paper -- main theory}, we state two theorems.


\subsection{Conditions for the Theorems}
\label{sec: Model Assumptions}
In this section, we introduce the conditions we use for our theorems.
We begin with three frequently used conditions in $\ell_1$-regularized regression models.
These conditions provide sufficient conditions for exact sparse recovery \cite[see][Chapter 11]{hastie2015statistical}.
Subsequently, we introduce three conditions that are widely used in cubic spline-based functional estimation 
\cite[see][(2.5)-(2.8)]{silverman1984spline}.


\begin{mycondition}[Mutual Incoherence Condition]
	\label{assumption -- PDW -- mutual incoherence condition}
	For some \textit{incoherence parameter} $\mu\in (0,1]$ and $P_\mu\in [0,1]$, we have
	$
	\mathbb{P}
	\left(
	\left\|
	\myX_{\mathcal S^c}^\top \myX_\mathcal S( \myX_{\mathcal S}^\top \myX_\mathcal S)^{-1}
	\right\|_\infty
	\leq
	1-\mu
	\right)
	\geq
	P_\mu\;,
	$
	where the 
	matrix $\myX_{\mathcal S^c}$ is the complement of $\myX_{\mathcal S}$.
\end{mycondition}
\begin{mycondition}[Minimal Eigenvalue Condition]
	\label{assumption -- PDW -- minimal eigenvalue condition}
	There exists some constant $C_{\min}>0$ such that
	$
	\Lambda_{\min}
	\left(
	\frac{1}{NM} \myX_\mathcal S^\top \myX_\mathcal S
	\right)
	\geq C_{\min},
	\;
	$
	almost surely.
	Here, $\Lambda_{\min}(\mathbf{A})$ denotes the minimal eigenvalue of a square matrix $\mathbf{A}\in\mathbb{R}^{n\times n}$.
	This condition can be considered a stronger version of the invertibility condition \cite[see][Chapter 11]{hastie2015statistical}.
\end{mycondition}

\begin{mycondition}[Knots c.d.f. Convergence Condition]
	\label{assumption -- spline -- convergence cdf}
	
	Suppose the sequence of the empirical distribution function over the design points
	$
	a = x_0 < \ldots < x_{M-1} = b,
	$
	with different sample size $M$, is denoted as $F_M(x)$, that is, we have 
	$
	F_M(x) = \frac{1}{M} \sum_{i=0}^{M-1} \mathbbm 1\{x_i \leq x\}.
	$
	Then, there exists an absolutely continuous distribution function $F$ on $[a, b]$ such that
	$
	F_M \to F
	$
	uniformly as $M \to +\infty$.
	Here, $\mathbbm 1\{A \}$ is the indicator of event $A$.
	A similar condition holds for the temporal variable: suppose the sequence of the empirical distribution function over the design points
	$
	\bar a = t_0 < \ldots < t_{N-1} = \bar b,
	$
	with different sample size $N$, is denoted as $G_N(x)$.
    Then, there exists an absolutely continuous distribution function $G$ on $[\bar a, \bar b]$ such that
	$
	G_N \to G
	$
	uniformly as $N \to +\infty$.
\end{mycondition}

\begin{mycondition}[Knots p.d.f. Convergence Condition]
	\label{assumption -- spline -- bounded pdf}
	Suppose the first derivatives of the functions $F$ and $G$ (defined in Condition \ref{assumption -- spline -- convergence cdf}) are denoted as $f$ and $g$, respectively.
    Then we have
	$$
	0 < \inf_{[x_0, x_{M-1}]} f \leq \sup_{[x_0, x_{M-1}]} f < + \infty
	\text{\;\;and\;\;}
	0 < \inf_{[t_0, t_{N-1}]} g \leq \sup_{[t_0, t_{N-1}]} g < + \infty,
	$$
	and $f$ and $g$ also have bounded first derivatives on $[x_0, x_{M-1}]$, $[t_0, t_{N-1}]$.
\end{mycondition}

\begin{mycondition}[Gentle Decrease of Smoothing Parameter Condition]
	\label{assumption -- spline -- decreasing penalty parameter}
	Suppose that
	$
	\zeta(M)     = \sup_{[x_0, x_{M-1}]} |F_M - F|.
	$
	The smoothing parameter $\alpha$ 
	in \eqref{equ: smoothing cubic spline -- objective fucntion}
	depends on $M$ 
	in such a way that
	$
	\alpha \to 0
	$
	and
	$
	\alpha^{-1/4} \zeta(M)  \to 0
	$
	as $M \to +\infty$.
	A similar condition also holds for the temporal variable.
\end{mycondition}

\subsection{Main Theory}
\label{sec: PDE paper -- main theory}

In the first theorem, we develop the lower bound of $\lambda$ to realize the correct recovery of the support set, that is,
$
\mS(\widehat\mybeta) \subseteq \mS(\mybeta^*).
$

\begin{mytheorem}
	\label{theo: support set recovary}
	Given the data in \eqref{equ: noisy data}, suppose the conditions in Lemma \ref{lemma: u_x and u_x hat difference bound}
	and Corollary \ref{lemma: u_t and u_t hat difference bound}
	(see the online Supplementary Material \ref{sec: Some Preliminaries}) hold, as do Condition
	\ref{assumption -- PDW -- mutual incoherence condition}
	-
	\ref{assumption -- spline -- decreasing penalty parameter}. 
    If we take $M = O(N)$, then there exists a constant
	$
	\mathscr C_{ (\sigma, \|u\|_{L^\infty(\Omega)})} > 0
	$
	that is independent of the spatial resolution $M$ and the temporal resolution $N$.
    Thus, if we set the cubic spline smoothing parameter with the spatial variable $x$ in \eqref{equ: smoothing cubic spline -- objective fucntion} as
	$
	\alpha  = O\left( \left( 1 + M^{-4/7} \right)^{-1} \right),
	$
	set the cubic spline smoothing parameter with temporal variable $t$ as
	$
	\bar\alpha = O\left( \left( 1 + N^{-4/7} \right)^{-1} \right),
	$
	and set the turning parameter
	\begin{equation}
		\label{equ: lower bound of lambda}
		\lambda
		\geq
		\mathscr C(\sigma, \|u\|_{L^\infty(\Omega)})
		\frac{\sqrt K \log(N)}{\mu N^{3/7 - r}}
	\end{equation}
	to identify the PDE model in \eqref{equ: LASSO model - matrix algebra}, for some $r \in \left(0, \frac{3}{7} \right)$, with sufficient large $N$,
	then with  probability greater than
	$
	P_{\mu}
	-
	\underbrace{
		O
		\left(
		N e^{-N^r}
		\right)
	}_{P'},
	$
	we have
	$
	\mS(\widehat{\mybeta}) \subseteq \mS(\mybeta^*).
	$
	Here, $K$ is the number of columns of the design matrix $\myX$ in \eqref{equ: LASSO model - matrix algebra}, and $\mu$ and $P_{\mu}$ are defined in Condition \ref{assumption -- PDW -- mutual incoherence condition}.
\end{mytheorem}

The proof of the above theorem can be found in the Supplementary Material \ref{sec: proof}, along with several lemmas, the  conditions of which are standardized in cubic splines. 
The above theorem provides the lower bound of $\lambda$ to realize the correct recovery of the support set.
As indicated by \eqref{equ: lower bound of lambda}, the lower bound is affected by several factors.
First, it is affected by the temporal resolution $N$: as $N$ increases, there is greater flexibility in tuning the penalty parameter $\lambda$.
Second, the lower bound in \eqref{equ: lower bound of lambda} is affected by the incoherence parameter $\mu$: if $\mu$ is small, then the lower bound increases.
This is because a small $\mu$ means that the feature variable candidates are similar to each other.
This phenomenon is called \textit{multicollinearity}. 
In this case, we have a very limited choice in terms of selecting $\lambda$.
However, we cannot increase the value of $\mu$, because this is decided by the data set $\mathcal D$ (see Condition \ref{assumption -- PDW -- mutual incoherence condition}).
Third, the lower bound in \eqref{equ: lower bound of lambda} is affected by the number of columns of the matrix $\myX$.
If its number of columns is very large, then it requires a larger $\lambda$ to identify the significant feature variables from among potential feature variables.

Note too that the probability
$
P_{\mu} - P'
$
converges to $P_{\mu}$ as $N \to + \infty$.
This limiting probability $P_{\mu}$ is determined by the data $\mathcal D$ (see Condition \eqref{assumption -- PDW -- mutual incoherence condition}).
Thus, when $N$ is very large, our proposed SAPDEMI method can realize
$
\mS(\widehat{\mybeta}) \subseteq \mS(\mybeta^*)
$
with probability close to $P_{\mu}$.


In the second theorem, we develop an upper bound for the estimating error.
\begin{mytheorem}
	\label{theo: beta estimation error bound}
	Suppose the conditions in Theorem \ref{theo: support set recovary} hold. Then with probability greater than
	$
	1 - O(N e^{-N^r}) \to 1,
	$
	there exists an $\dot N>0$, such that when $N > \dot N$, we have
	\begin{eqnarray*}
		\label{equ: main theory -- estimation error bound}
		\left\| \widehat{\mybeta}_\mS - \mybeta^*_\mS \right\|_\infty
		\leq
		\sqrt{K}C_{\min}
		\left(
		\sqrt{K}
		\mathscr C_{ (\sigma, \|u\|_{L^\infty(\Omega)}) }
		\frac{\log(N)}{N^{3/7 - r}}
		+
		\lambda
		\right)  ,
	\end{eqnarray*}
	where $K$ is the number of columns of the matrix $\myX$,
	$
	\mS :=\{i : \beta^*_i \neq 0, \; \forall i = 1,\ldots, K \}
	$
    and the vectors $\widehat\mybeta_{\mS}$, and $\mybeta^*_{\mS}$ are subvectors of $\widehat\mybeta$ and $\mybeta^*$, respectively, that contain only those elements with indices that are in $\mS$.
	The theorem shows that when $N \to +\infty$, the error bound convergences to $0$.
\end{mytheorem}
The proof can be found in the Supplementary Material \ref{sec: proof}.
The previous theorem shows that the estimation error bound for the $\ell_{\infty}$-norm of the coefficient error in \eqref{equ: main theory -- estimation error bound} consists of two components.
The first component is affected by the temporal resolution $N$ and the number of feature variable candidates $K$.
As $N \to +\infty$, this first component converges to zero without an explicit dependence on the feature variable selected from \eqref{equ: LASSO model - matrix algebra}.
The second component is $\sqrt{K} C_{\min} \lambda$.
When $N$ increases to $+\infty$, this second component also converges to zero.
This is because, as stated in Theorem \ref{theo: support set recovary}, when $N \to +\infty$, the lower bound of $\lambda$, which realizes the correct support recovery, converges to zero.
Thus, the accuracy of the coefficient estimation improves as we increase $N$.

By combining  Theorems \ref{theo: support set recovary} and  \ref{theo: beta estimation error bound}, we find that when the minimum absolute value of the nonzero entries of $\mybeta^*$ is sufficiently large,  with an adequate choice of $\lambda$, we can guarantee the exact recovery.
Mathematically, when
$
\min_{i \in \mS} | (\mybeta^*_{\mS})_i |
>
\sqrt{K}C_{\min}
\left(
\sqrt{K}
\mathscr C_{ (\sigma, \|u\|_{L^\infty(\Omega)}) }
\frac{\log(N)}{N^{3/7 - r}}
+
\lambda
\right),
$
where $(\mybeta^*_{\mS})_i$ refers to the $i$th element in the vector $\mybeta^*_{\mS}$, we have a correct signed support of $\widehat\mybeta$.
This helps when selecting the penalty parameters $\lambda$.
In addition, the plot of the solution paths helps with the selection of the penalty parameters $\lambda$; see Section \ref{sec: PDE paper -- simulation}.

\section{Numerical Examples}
\label{sec: PDE paper -- simulation}
We conduct numerical experiments to verify the computational efficiency and the statistical accuracy of our proposed SAPDEMI  method.

Our examples are based on 
(1) the transport equation,
(2) the inviscid Burgers' equation, and
(3) the viscous Burgers' equation.
We select these three PDE models as representatives, because they all play fundamental roles in modeling physical phenomena and demonstrate characteristic behaviors of a more complex system, such as dissipation and shock formation \cite[][]{haberman1983elementary}.
In addition to wide applications, they cover a wide range of categories, including the first-order PDE, second-order PDE, linear PDE, and nonlinear PDE, which cover most of the PDEs frequently seen in practice.
Furthermore, the difficultly of identifying the above PDE models increases from the first example---the transport equation---to the last example---the viscous Burgers' equation. 
We set $p_{\max} = 2$ and $q_{\max} = 2$ in \eqref{equ: temporal evolutionary PDE} for the three numerical examples (see the full formula of the full model in the Supplementary Material \ref{appendix: simulation full model}), that is, we identify the PDE model from the full model.

In terms of computational efficiency, the results of these three examples are the  same, so we present only the result for the first example. We also verify Conditions \ref{assumption -- PDW -- mutual incoherence condition} - \ref{assumption -- spline -- decreasing penalty parameter} for the above three examples.
The details of the verification are provided in the  Supplementary Material \ref{sec: model diagnosis}.

\subsection{Example 1: Transport Equation}
\label{sec: sim -- example 1}

The PDE problem studied in this section is the transport equation.
It is a linear first-order PDE model.
Given its simplicity and straightforward physical meaning, it is widely used to model the concentration of a substance flowing in a fluid at a constant rate, 
For example, it can model a pollutant in a uniform fluid flow that is moving with velocity $a$ \citep[Section 2.2]{olver2014introduction}:
\begin{equation}
	\label{equ: sim -- transport equ}
	\left\{
	\begin{array}{rcll}
		\frac{\partial}{\partial t} u(x,t)
		& = &
		a \frac{\partial}{\partial x} u(x,t), 
		&  \forall \; 0 \leq x \leq X_{\max}, \;0 \leq t \leq T_{\max};\\
		u(x,0)
		& = &
		f(x).
		&
	\end{array}
	\right.
\end{equation}
Here, $a \in \mathbb R$ is a fixed nonzero constant, known as the \textit{wave speed}.
In this section, we set $a = -2, f(x) = 2\sin(4x), X_{\max} =1$ and $T_{\max} = 0.1$.
Given these settings, there is a closed-form solution, 
$ u(x,t) = 2\sin(4x-8t).$

\begin{figure}[htbp]
	\centering
	\begin{tabular}{ccc}
		\includegraphics[width = 0.23\textwidth]{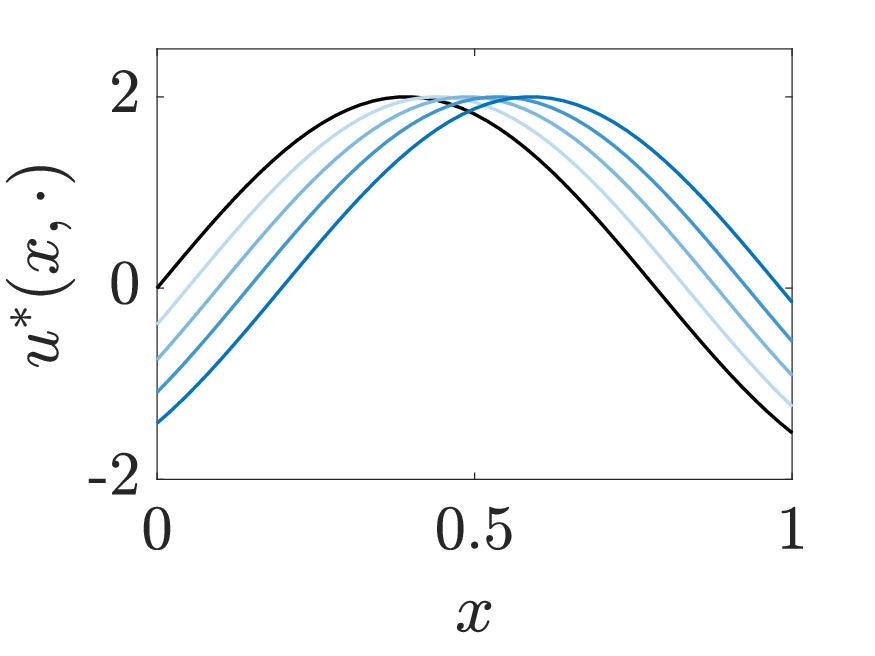} &
		\includegraphics[width = 0.23\textwidth]{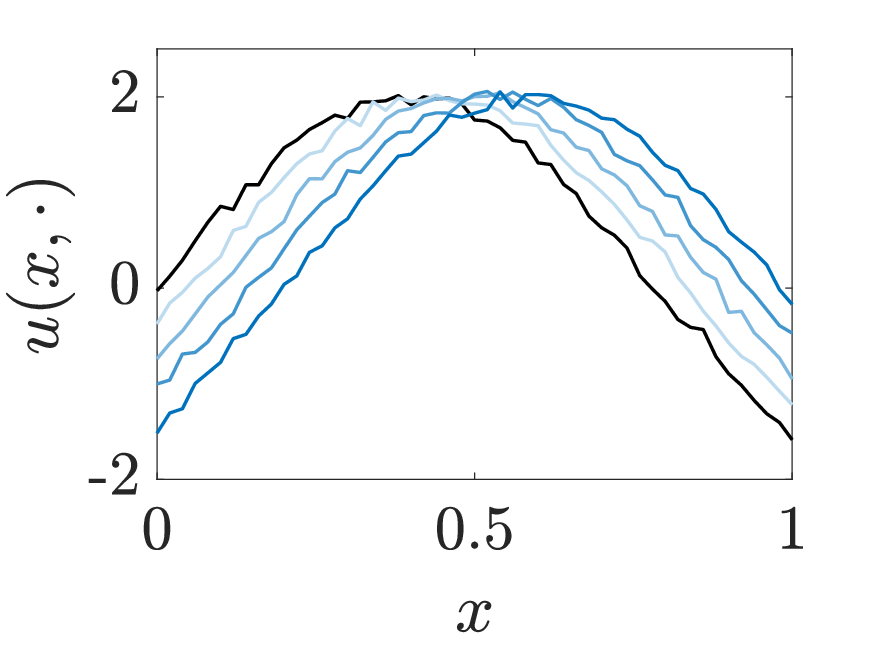} &
		\includegraphics[width = 0.23\textwidth]{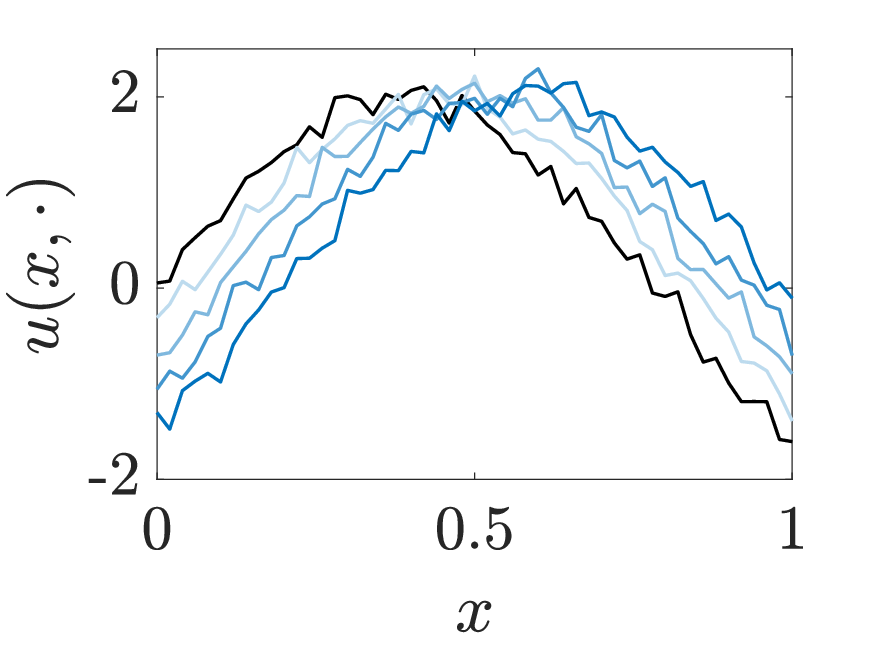} \\
		(a) truth &  (b) $\sigma = 0.05$ & (c) $\sigma = 0.1$ \\
	\end{tabular}
	\caption{Noisy/True curves from \eqref{equ: sim -- transport equ} ($M=N=100$).}
	\label{fig: sim -- transport equation -- true noised noised} 
\end{figure}

The dynamic pattern of the above transport equation is visualized in Fig. \ref{fig: sim -- transport equation -- true noised noised}, 
where the subfigures  (a), (b), and (c) show the ground truth and noisy observations under $\sigma = 0.05$ and $\sigma = 0.1$, respectively.
The figure shows that a larger noise results in the shape of the transport equation being less smooth, potentially leading to additional difficulties in the PDE model identification.

\begin{figure}[htbp]
	\centering
	\begin{tabular}{ccc}
		\includegraphics[width=0.25\textwidth]{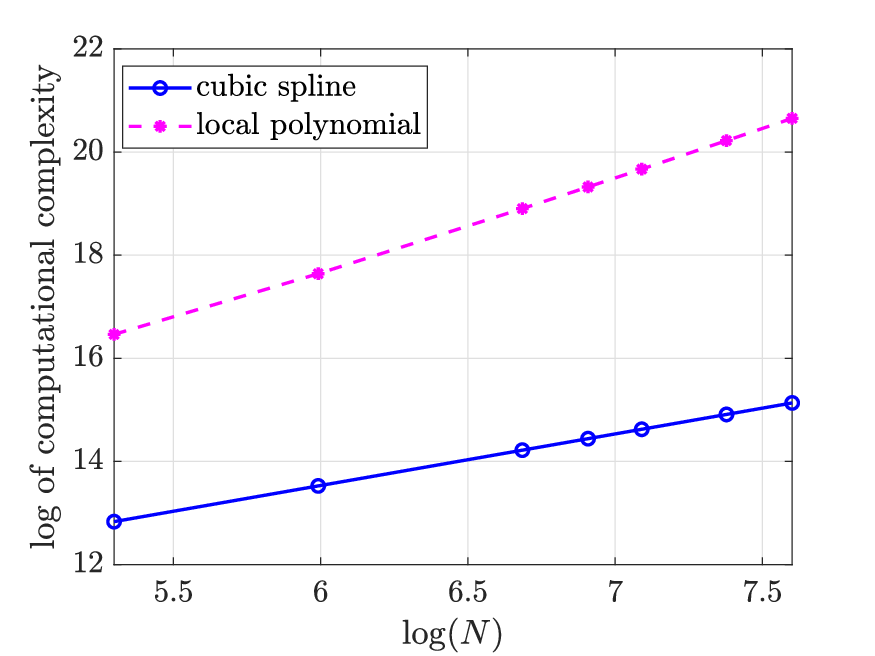}&
		\includegraphics[width=0.25\textwidth]{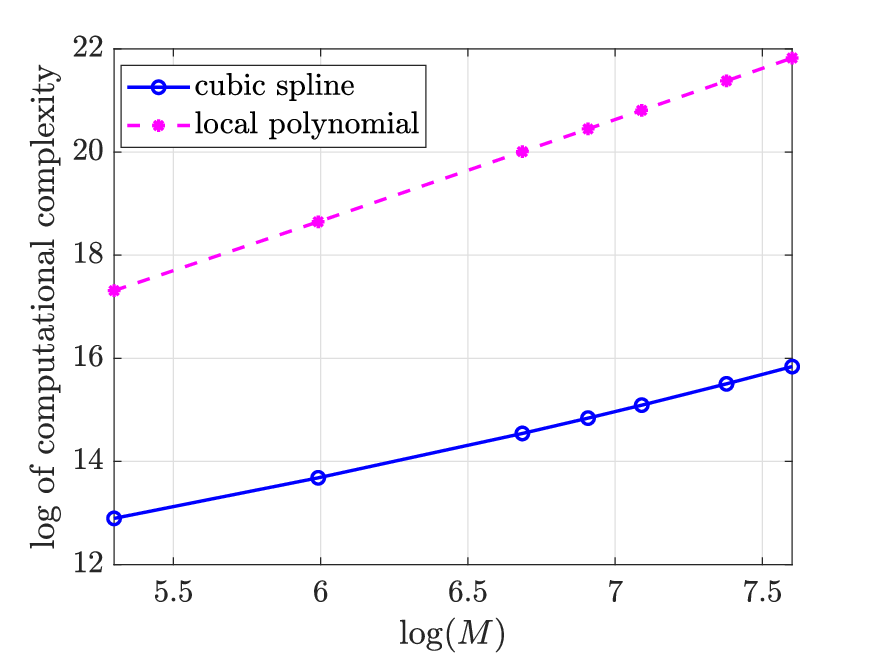}\\
		(a) fixed $M = 20$ & (b) fixed $N=20$
	\end{tabular}
	\caption{Computational complexity of the cubic spline and a local polynomial.}
	\label{fig: sim -- example 1 -- complexity}
\end{figure}

First, we consider the computational complexity of the functional estimation stage.
We select the local polynomial regression as a benchmark, and visualize the number of numerical operations of the two methods in Fig. \ref{fig: sim -- example 1 -- complexity}, where the x-axis is $\log(M)$ or $\log(N)$, and the y-axis is the logarithm of the number of numerical operations.
In Fig. \ref{fig: sim -- example 1 -- complexity}, two scenarios are discussed:
(1) $M$ is fixed as $20$, and $N$ varies from $200$ to $2000$; and
(2) $N$ is fixed as $20$, and $M$ varies from $200$ to $2000$.
Fig. \ref{fig: sim -- example 1 -- complexity} shows that, as $M$ or $N$ increases, so does the number of numerical operations in the functional estimation stage.
We find that the cubic splines method needs fewer numerical operations, compared with the local polynomial regression.
Furthermore, a simple linear regression of the four lines in Fig. \ref{fig: sim -- example 1 -- complexity} shows that in (a), the slope of the cubic spline is $0.9998$, and as $N$ goes to infinity, the slope gets get closer to one.
This validates that the computational complexity of the cubic splines-based method is of order $O(N)$ when $M$ is fixed.
The result in (b) is similar.
Thus, we numerically verify that the computational complexity of the cubic spline method is of order $O(MN)$.
Similarly, for a local polynomial,
we can numerically
validate its computational complexity, which is $\max\{O(M^2 N), O(MN^2)\}$.

\begin{figure}[htbp]
	\centering
	\begin{tabular}{ccc}
		\includegraphics[width = 0.22\textwidth]{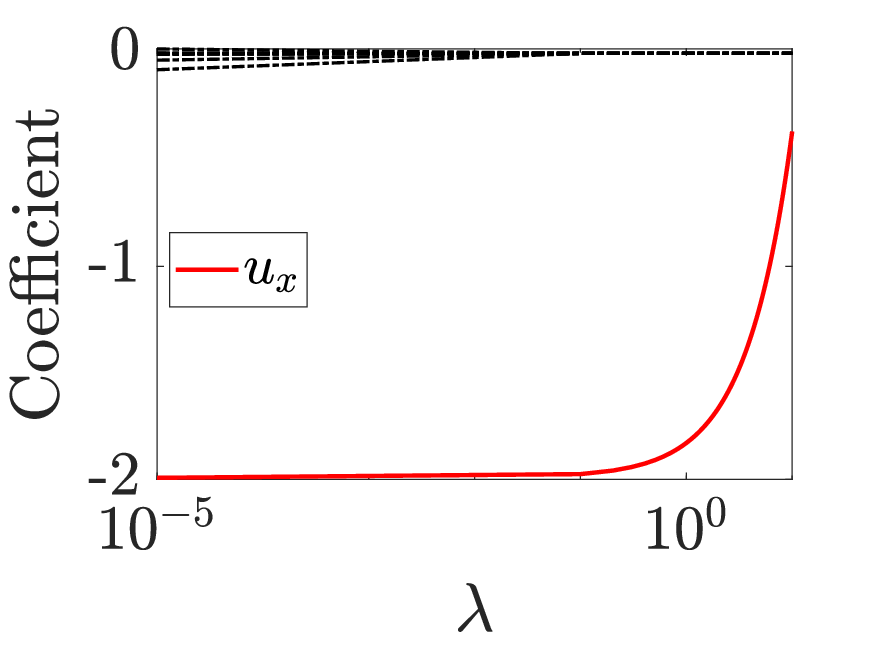} &
		\includegraphics[width = 0.22\textwidth]{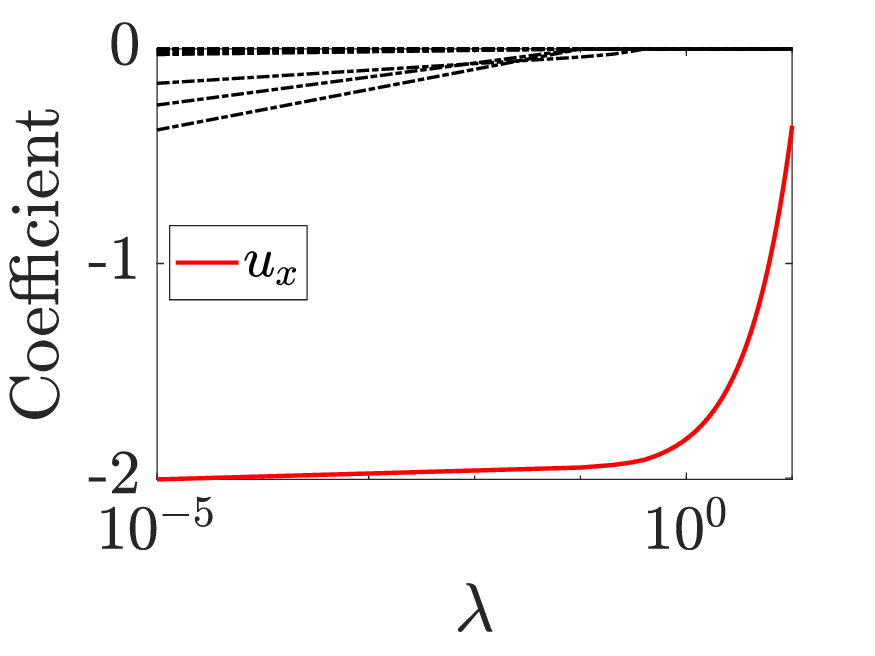}  &
		\includegraphics[width = 0.22\textwidth]{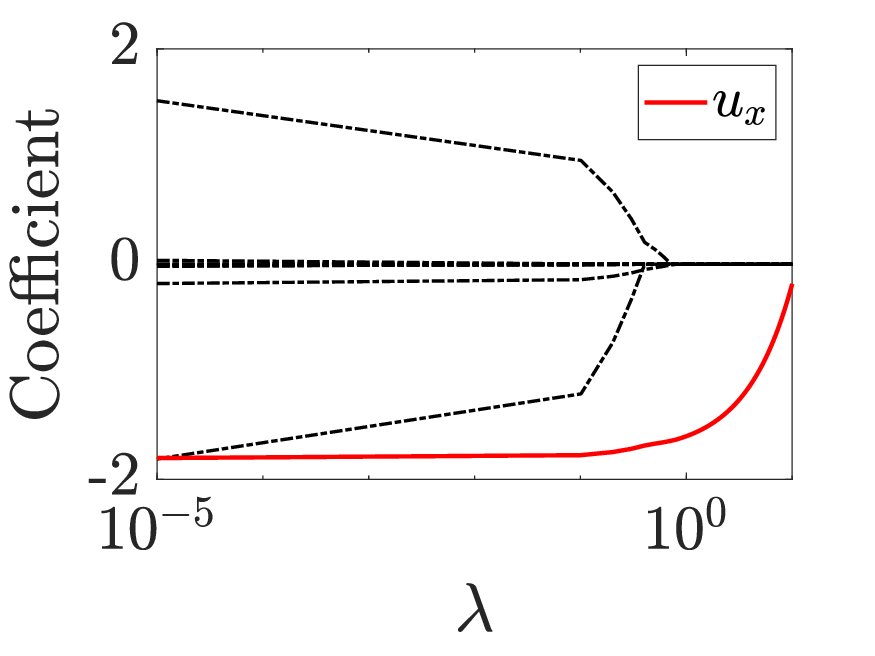}\\
		(a) $\sigma = 0.01$ & (b) $\sigma = 0.1$ & (c) $\sigma = 1$\\
	\end{tabular}
	\caption{Solution paths in the transport equation  under different $\sigma$ and $M=N=100$.
		The otation $u_x$ is a simplification of $\frac{\partial}{\partial x} u(x,t)$.
		\label{fig: sim -- example 1 -- support set recovery} }
\end{figure}

We now verify numerically that with high probability, our SAPDEMI can correctly identify the underlying PDE models.
From the formula of the transport equation in equation \eqref{equ: sim -- transport equ}, we know that the correct feature variable is $\frac{\partial}{\partial x} u(x,t)$, 
and that other feature variables
should not be identified.
We discuss the identification accuracy under different sample sizes and magnitudes of noise.
We find that the accuracy stays at 100\%.
To explain the high accuracy,  we plot the solution paths in Fig. \ref{fig: sim -- example 1 -- support set recovery} under different $\sigma$, namely, $\sigma = 0.01, 0.1, 1$.
From Fig. \ref{fig: sim -- example 1 -- support set recovery}, we can increase $\lambda$ to overcome this difficulty, and thus achieve a correct PDE identification. 

\subsection{Example 2: Inviscid Burgers' Equation}

In this section, we investigate the inviscid Burgers' equation \cite[see][Section 8.4]{olver2014introduction}, which is representative of a first-order nonlinear PDE and is used frequently in applied mathematics, such as fluid mechanics, nonlinear acoustics, gas dynamics, and traffic flow.
This PDE model was first introduced by Harry Bateman in 1915, and later studied by Johannes Martinus Burgers in 1948 \citep{whitham2011linear}.
The formula of the inviscid Burgers' equation is listed below:
\begin{equation}
	\label{equ: sim -- inviscid burgers equation}
	\left\{
	\begin{array}{rcll}
		\frac{\partial}{\partial t}u(x,t)
		& = &
		-\frac{1}{2} u(x,t)\frac{\partial}{\partial x}u(x,t) & \\
		u(x,0) & = & f(x) & 0 \leq x \leq X_{\max} \\
		u(0,t) & = & u(1,t) =0 & 0 \leq t \leq T_{\max}
	\end{array}
	\right.,
\end{equation}
where we set $f(x) = \sin(2\pi x), X_{\max} = 1$ and $T_{\max} = 0.1$.
Fig. \ref{fig: sim -- inviscid Burgers' equation -- true, noise, denoised}(a), (b), and (c) show the ground truth and noisy observations under $\sigma = 0.05$ and $0.1$, respectively.
\begin{figure}[htbp]
	\centering
	\begin{tabular}{ccc}
		\includegraphics[width = 0.22\textwidth]{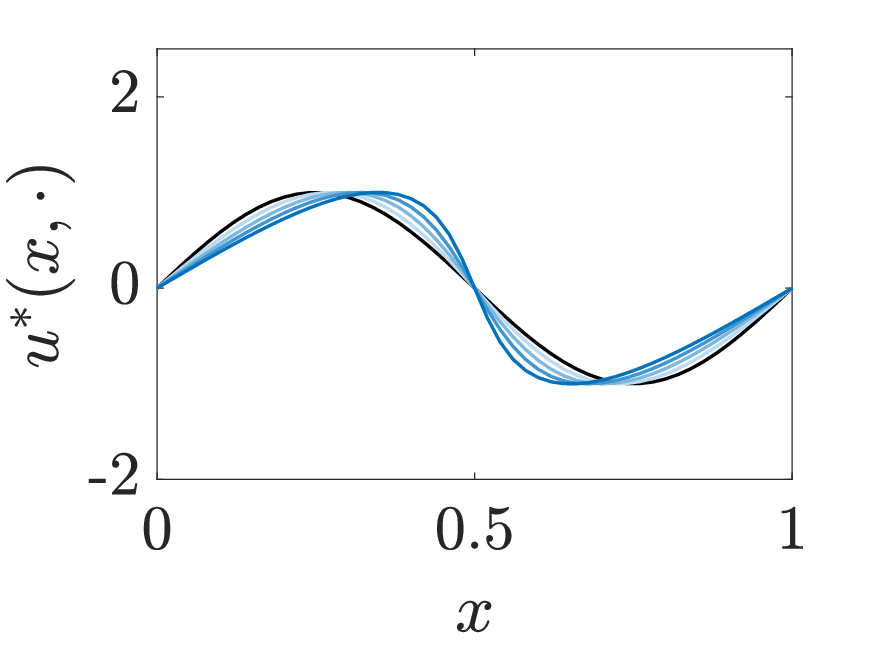} &
		\includegraphics[width = 0.22\textwidth]{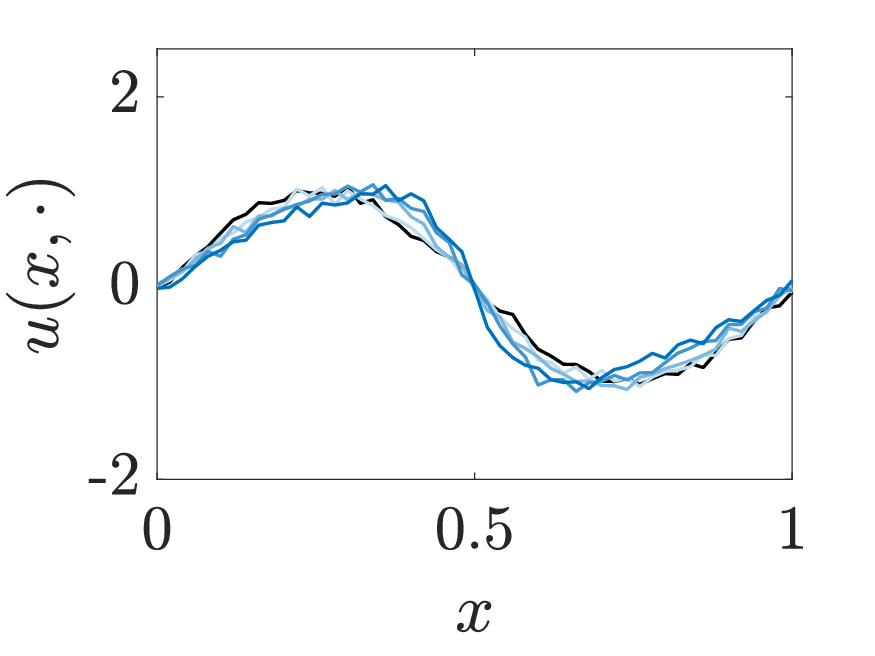} &
		\includegraphics[width = 0.22\textwidth]{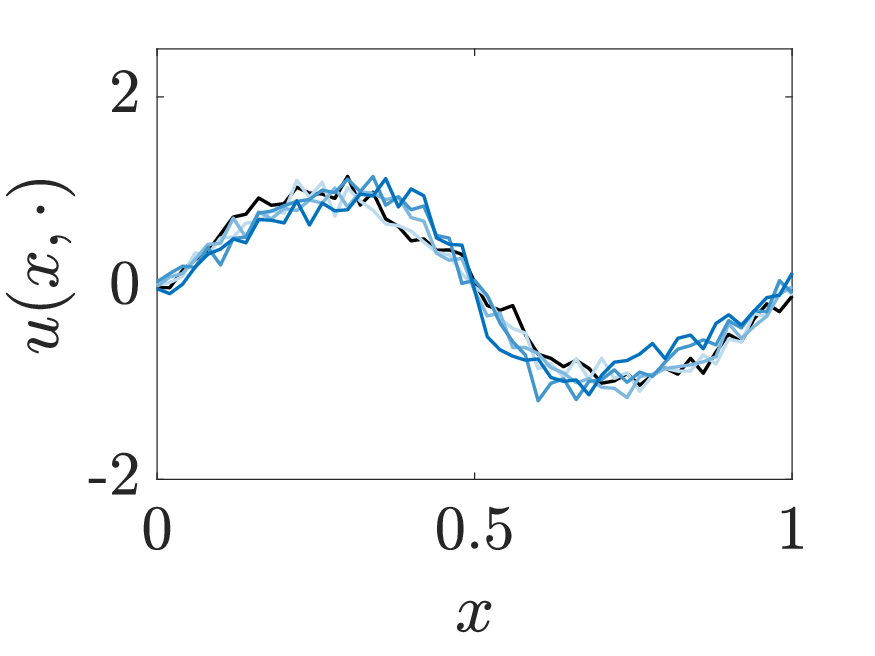} \\
		(a) truth & (b) $\sigma = 0.05$ & (c) $\sigma = 0.1$ \\
	\end{tabular}
	\caption{Noisy/True curves from \eqref{equ: sim -- inviscid burgers equation} ($M=50, N=50$).
		\label{fig: sim -- inviscid Burgers' equation -- true, noise, denoised} }
\end{figure}
Compared with our first example (transport equation in \eqref{equ: sim -- transport equ}), the inviscid Burgers’ equation can be regarded an extension from the linear transport equation to a nonlinear transport equation.
Specifically, if we set $a$ in \eqref{equ: sim -- transport equ} as 
$
  a = - \frac{1}{2} u(x, t),
$                
then \eqref{equ: sim -- transport equ} is equivalent to \eqref{equ: sim -- inviscid burgers equation}.
In the literature, this PDE model is considerably more challenging than the linear transport PDE in \eqref{equ: sim -- transport equ}: the wave speed in \eqref{equ: sim -- transport equ} depends only on the spatial variable $x$, whereas the wave speed in \eqref{equ: sim -- inviscid burgers equation} depends on both the spatial variable $x$ and the size of the disturbance $u(x,t)$.
Given the complicated wave speed in \eqref{equ: sim -- inviscid burgers equation}, it can model more complicated dynamic patterns.
For example, larger waves move faster, and overtake smaller, slow-moving waves. 

\begin{figure}[htbp]
	\centering
	\begin{tabular}{ccc}
		\includegraphics[width = 0.25\textwidth]{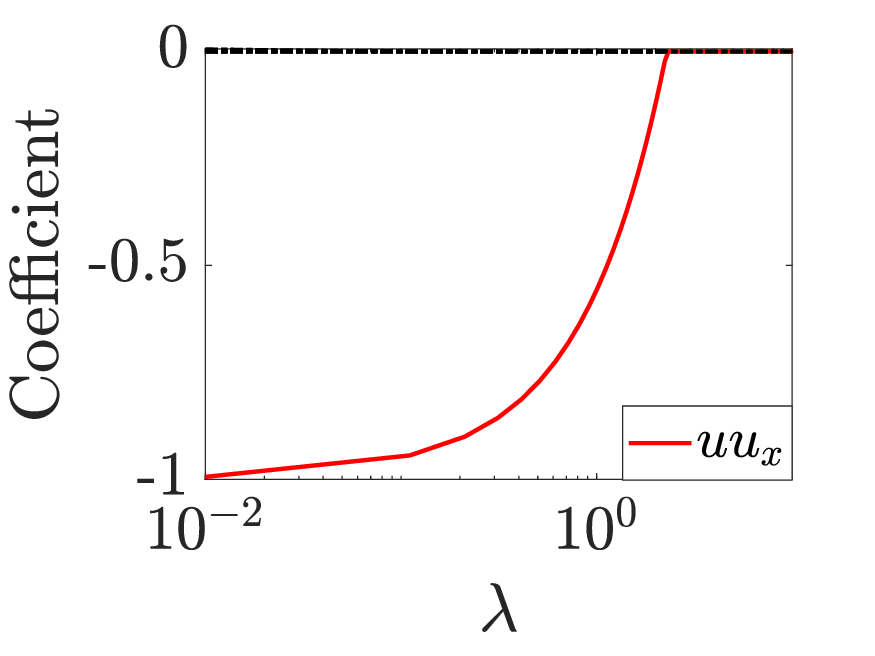} &
		\includegraphics[width = 0.25\textwidth]{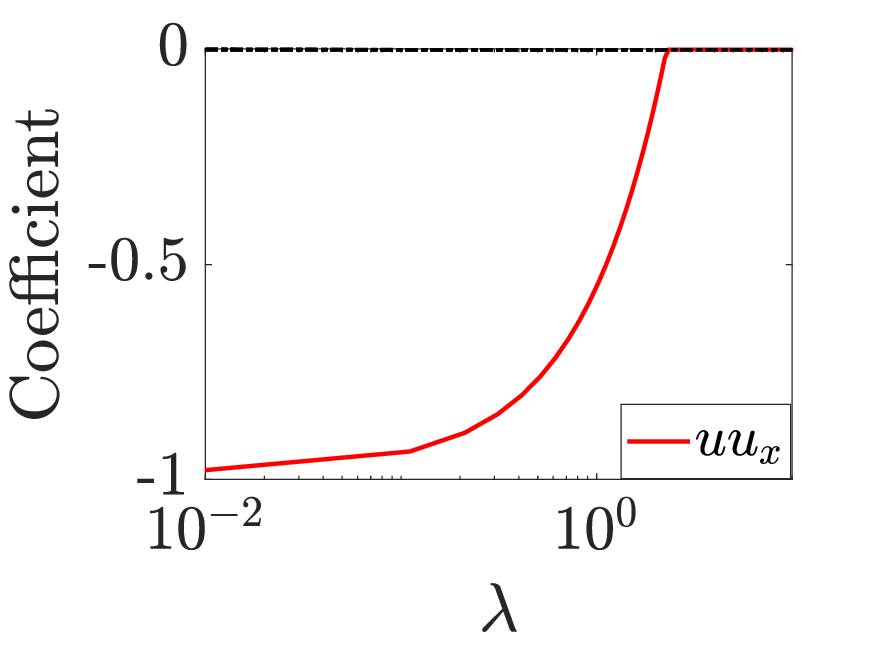}  &
		\includegraphics[width = 0.25\textwidth]{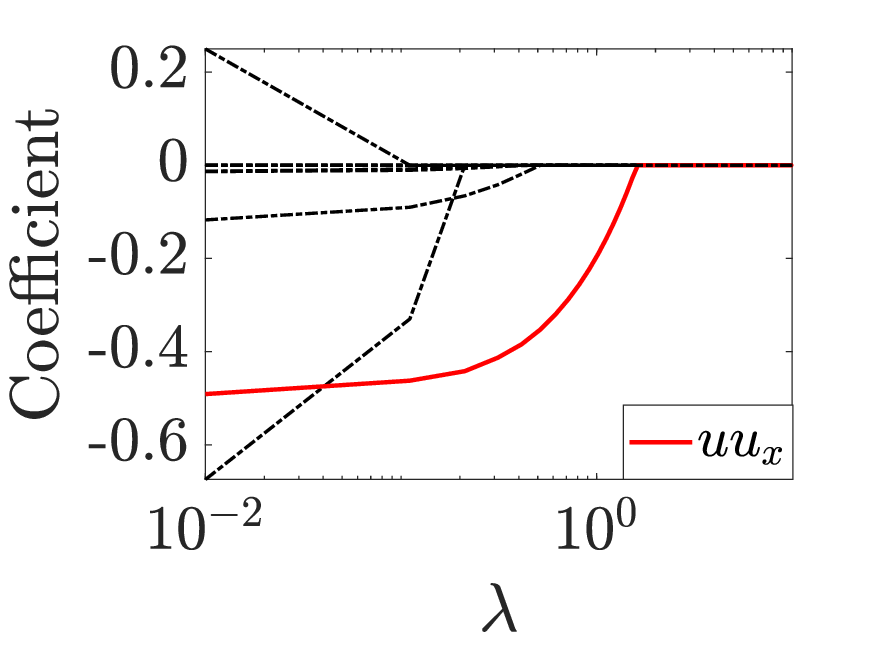}   \\
		(a) $\sigma = 0.01$ & (b) $\sigma = 0.5$ & (c) $\sigma= 1$ \\
	\end{tabular}
	\caption{Solution paths in the inviscid Burgers' equation under different $\sigma$ and $M=N=100$.
	Here $u$ and $u_x$ are simplifications of $u(x,t)$ and $\frac{\partial}{\partial x}u(x,t)$, respectively.
	\label{fig: sim -- inviscid Burgers' equation -- support set recovery}}
\end{figure}

In this example, SAPDEMI correctly identifies with an accuracy above  $99\%$ (see Fig. \ref{fig: sim -- successful identification prob curve}(a)). 
The effect of $\sigma$ is also reflected in Fig. \ref{fig: sim -- inviscid Burgers' equation -- support set recovery}, where the length of the $\lambda$-interval for correct identification decreases as $\sigma$ increases.


\subsection{Example 3: Viscous Burgers' Equation}

In this section, we investigate the more challenging viscous Burgers' equation \cite[see][Section 8.4]{olver2014introduction}, which is a fundamental second-order semilinear PDE.
It is frequently employed to model physical phenomena in fluid dynamics \citep{bonkile2018systematic, nakatsuka2025all, chen2023anomaly} and nonlinear acoustics in dissipative media \citep{rudenko1975theoretical}.
For example, in fluid and gas dynamics, we can interpret the term $\nu \frac{\partial^2}{\partial x^2} u(x, t)$ as modeling the effect of viscosity \citep[Section 8.4]{olver2014introduction}.
Thus, the viscous Burgers’ equation represents a version of the equations of the viscous fluid flows, including the celebrated and widely applied Navier-Stokes equations \citep{whitham2011linear}:
\begin{equation}
	\label{equ: sim -- viscous burgers equation}
	\left\{
	\begin{array}{rcll}
		\frac{\partial u(x,t)}{\partial t}
		& = &
		-\frac{1}{2} u(x,t)\frac{\partial}{\partial x}u(x,t) + \nu \frac{\partial^2}{\partial x^2} u(x,t) \\
		u(x,0) & = & f(x) & 0 \leq x \leq X_{\max}\\
		u(0,t) & = & u(1,t) =0 & 0 \leq t \leq T_{\max}
	\end{array}
	\right.,
\end{equation}
where we set $f(x) = \sin^2(4\pi x) + \sin^3(2\pi x), X_{\max} = 1, T_{\max} = 0.1$ and $\nu = 0.1$.
Fig. \ref{fig: sim -- viscous Burgers' equation -- true, noise, denoised} shows the corresponding curves, where (a), (b), and (c) are the ground truth and noisy observations under $\sigma = 0.05$ and $\sigma = 0.1$, respectively.

\begin{figure}[htbp]
	\centering
	\begin{tabular}{ccc}
		\includegraphics[width = 0.23\textwidth]{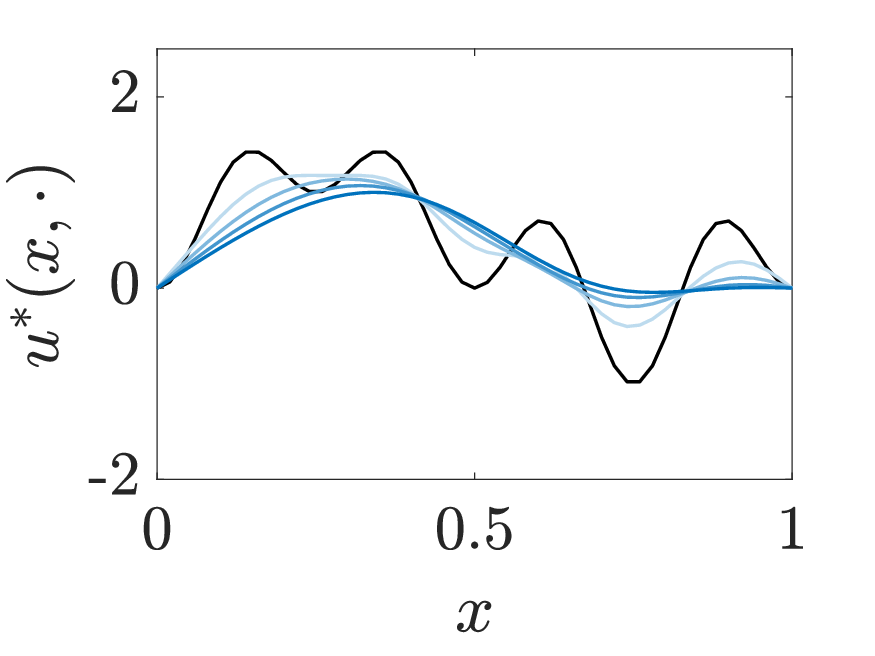} &
		\includegraphics[width = 0.23\textwidth]{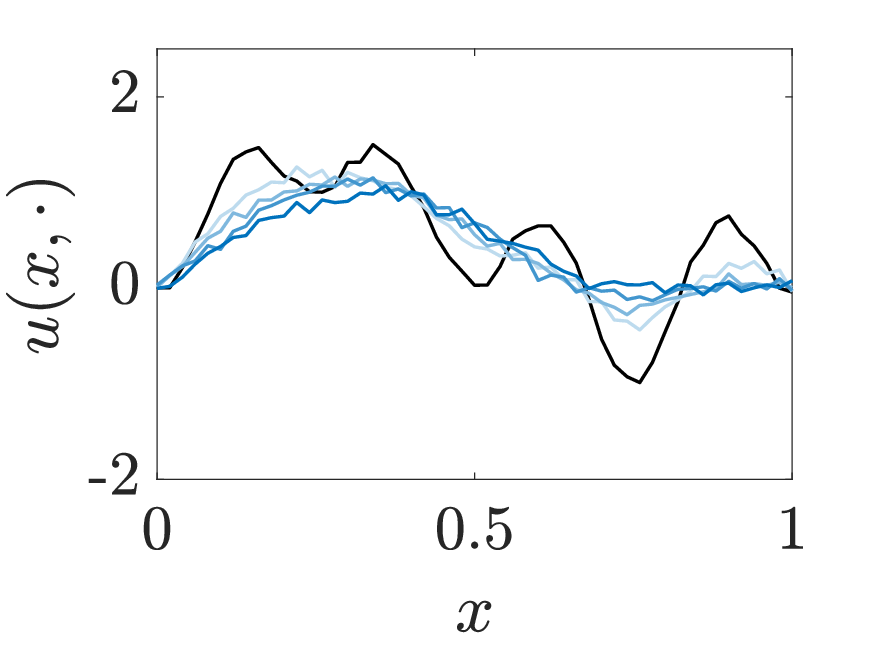} &
		\includegraphics[width = 0.23\textwidth]{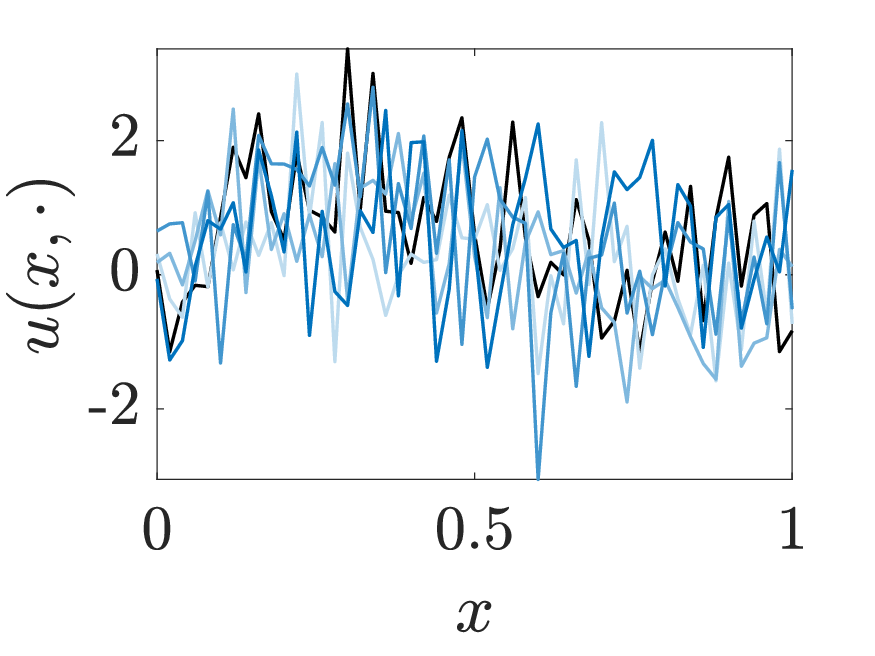} \\
		(a) true & (b) $\sigma = 0.05$ & (c) $\sigma = 1$ \\
	\end{tabular}
	\caption{Noisy/True curves from \eqref{equ: sim -- viscous burgers equation} ($M=50, N=50$).
    \label{fig: sim -- viscous Burgers' equation -- true, noise, denoised}}
\end{figure}

Compared with the previous two PDE models (transport equation in \eqref{equ: sim -- transport equ} and inviscid Burgers' equation in \eqref{equ: sim -- inviscid burgers equation}), the above PDE is more complicated and challenging.
This is because the viscous Burgers’ equation involves not only the first-order derivative, but also the second-order derivatives. Our simulations provide sufficiently complicated examples.  

\begin{figure}[htbp]
	\centering
	\begin{tabular}{ccc}
		\includegraphics[width = 0.23\textwidth]{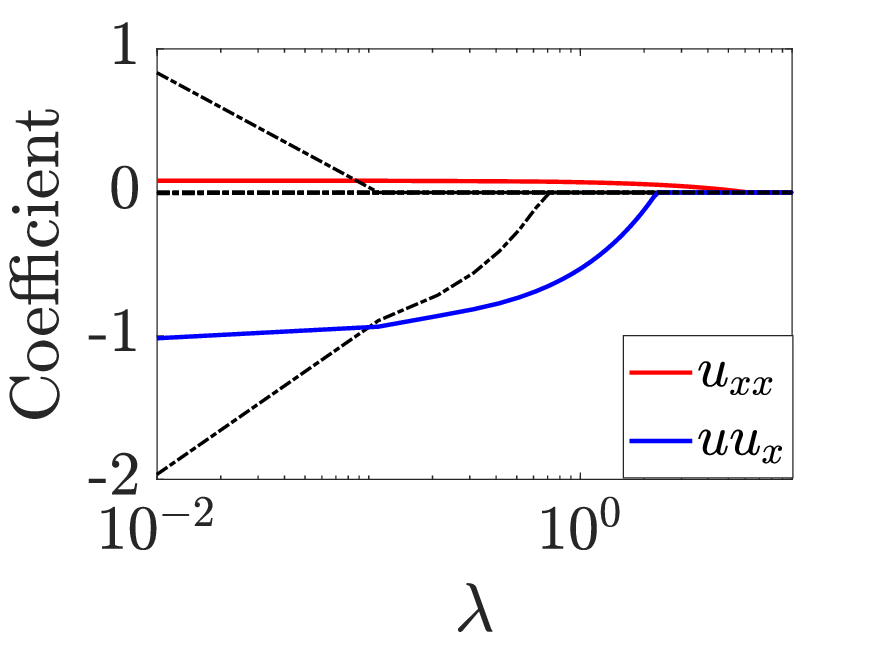} &
		\includegraphics[width = 0.23\textwidth]{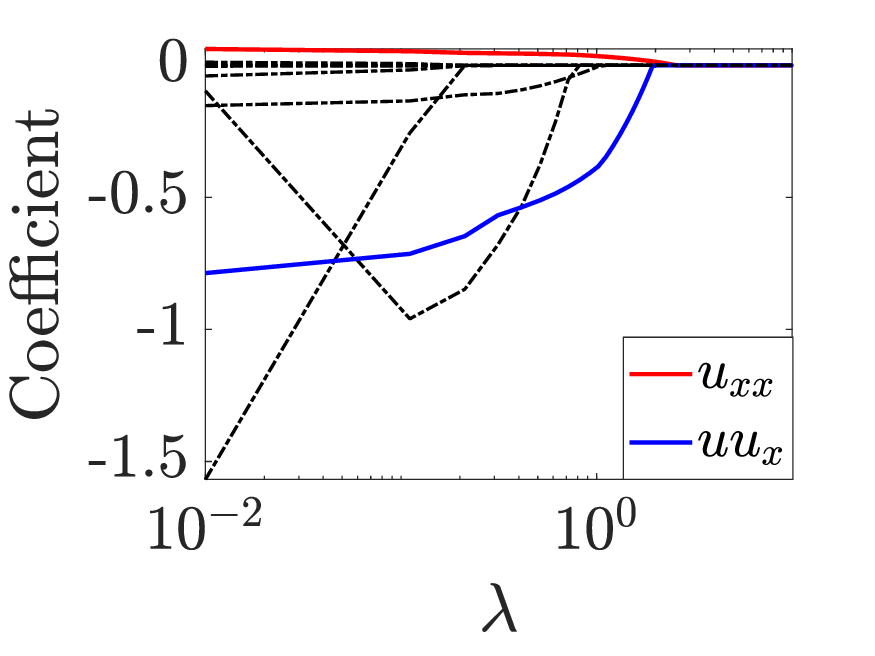} &
		\includegraphics[width = 0.23\textwidth]{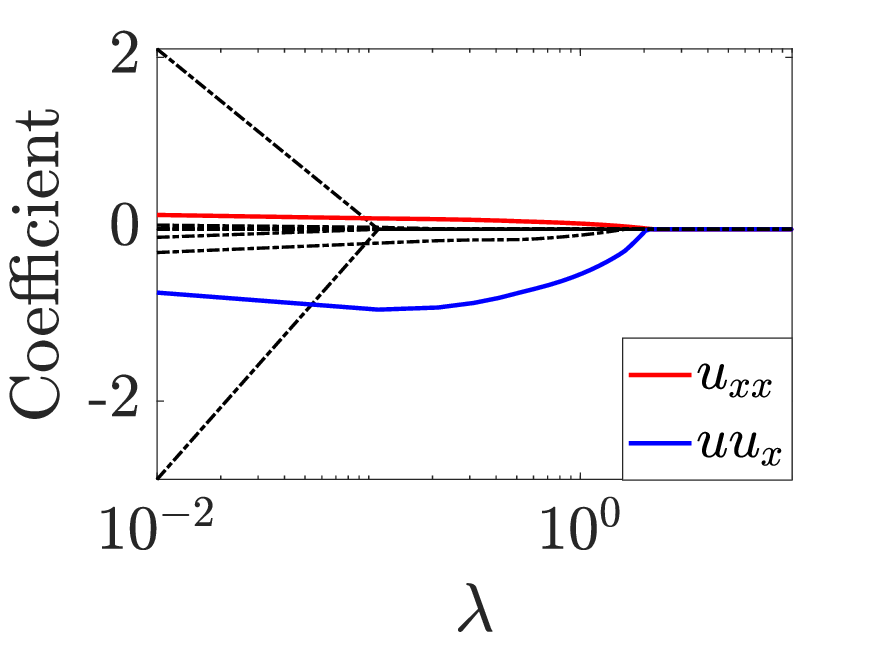} \\
		(a) $\sigma = 0.01$ & (b) $\sigma = 0.5$  & (c) $\sigma= 1$  \\
	\end{tabular}
	\caption{Solution paths in the viscous Burgers' equation under different $\sigma$ and $M=N=100$.
		The notation $u_{xx}$ and $uu_{x}$ stand for $u(x, t) \frac{\partial}{\partial x} u(x,t)$ and $\frac{\partial^2}{\partial x^2} u(x,t)$, respectively.}
		\label{fig: sim -- viscous Burgers' equation -- support set recovery}
\end{figure}

\begin{figure}[htbp]
	\centering
	\begin{tabular}{ccc}
		\includegraphics[width = 0.25\textwidth]{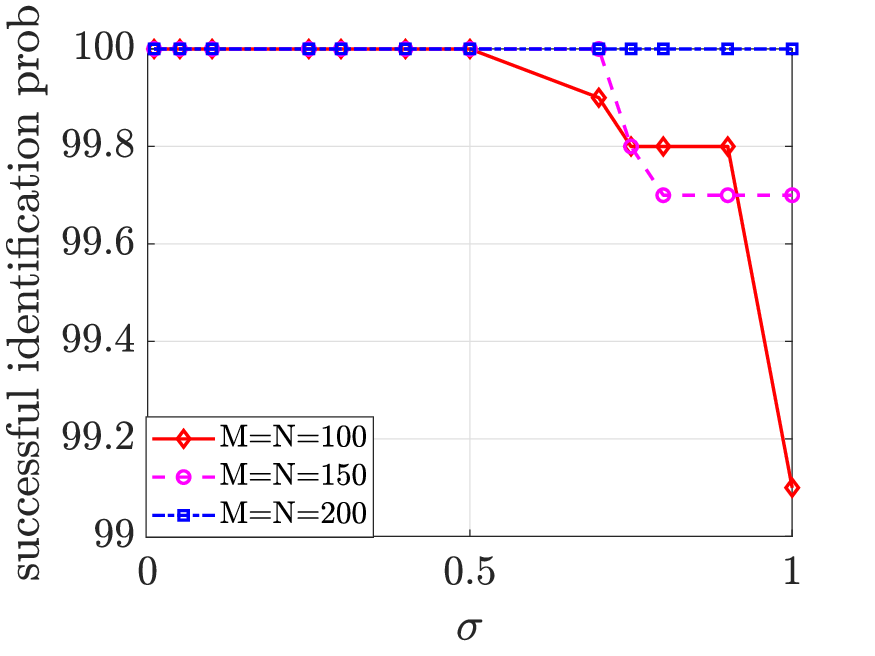}  &
		\includegraphics[width = 0.25 \textwidth]{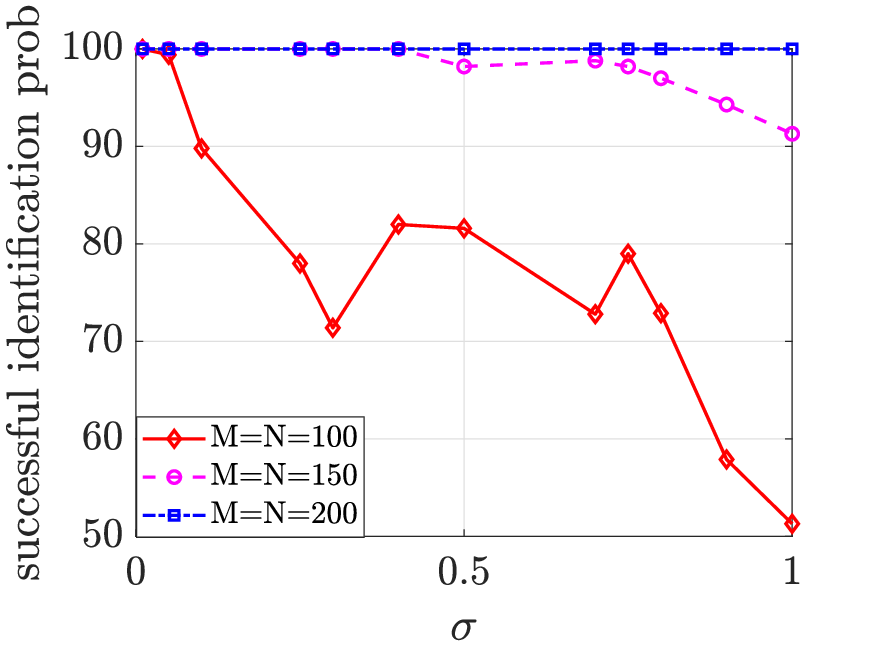}   \\
		(a) example 2 & (b) example 3 \\
	\end{tabular}
	\caption{Curves of successful identification probability.
		\label{fig: sim -- successful identification prob curve}}
\end{figure}

Based on Fig. \ref{fig: sim -- successful identification prob curve}(b), we conclude that with high probability, our proposed SAPDEMI can correctly identify the underlying viscous Burgers' equation, for the following reasons.
When $M=N=200$ or $150$, the accuracy stays above $90\%$ for all levels of $\sigma \in [0.01, 1]$.
When $M=N=100$, the accuracy is above $70\%$ when $\sigma \in [0.01, 0.5]$, and reduces to $50\%$ when $\sigma = 1$. This makes sense, because as shown in Fig. \ref{fig: sim -- viscous Burgers' equation -- support set recovery}, when $\sigma$ increases from 0.01 to 1, the length of the $\lambda$-interval for correct identification decreases, making it more difficult to realize a correct identification.
Thus, if we encounter a very noisy data set $\mathcal D$, a larger sample size is preferred.

\section{Case Study}
\label{sec: case study}
In this section, we apply SAPDEMI to a real-world data set that is a subset of the Cloud-Aerosol Lidar and Infrared Pathfinder Satellite Observations (CALIPSO) data set downloaded from NASA.
The CALIPSO reports the monthly mean of temperature in 2017 at $34^{\circ}$N and $110.9418$ meters above the Earth's surface over a uniform spatial grid from $180^{\circ}$W to $180^{\circ}$E, with equally spaced $5^{\circ}$ intervals. 
The missing data are handled either by direct imputation or by using the instrument methods \cite{chen2018pseudo, chen2021instrument, chen2019semiparametric, chen2018semiparametric}.

\begin{figure}[htbp]
	\centering
	\begin{tabular}{ccc}
		\includegraphics[width = 0.25\textwidth]{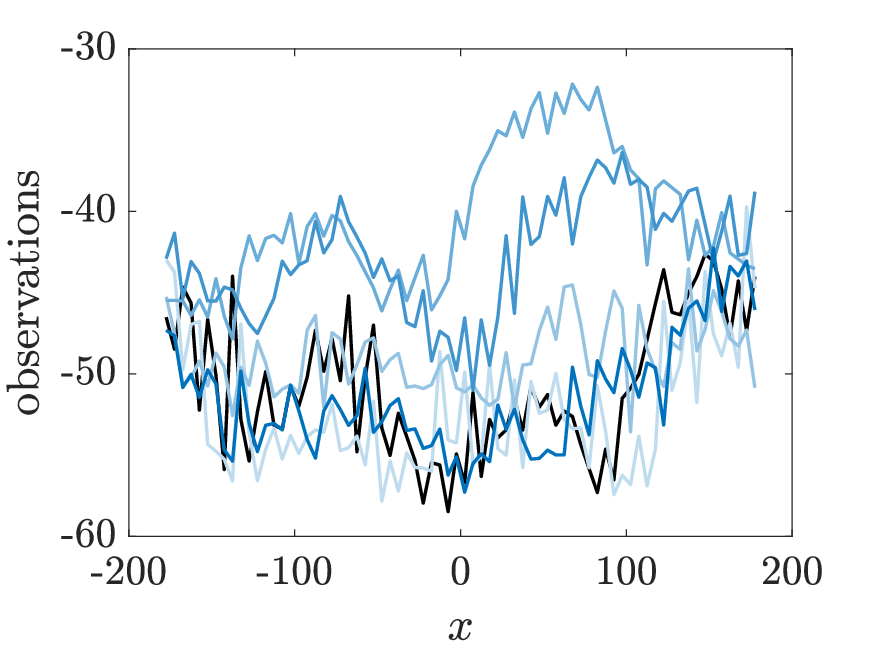}  &
		\includegraphics[width = 0.25\textwidth]{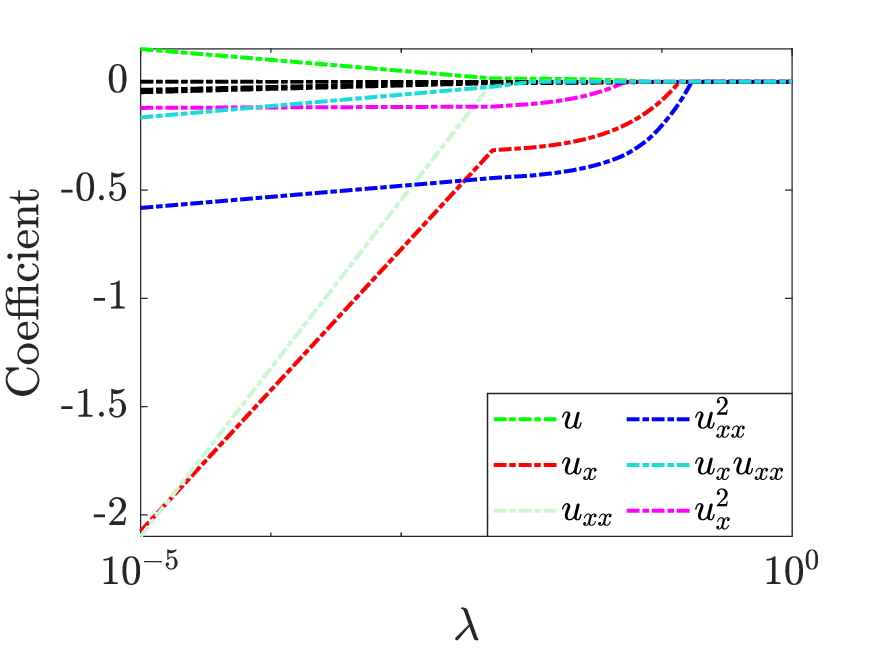}   \\
		(a)  observed temperature & (b) solution path\\
	\end{tabular}
	\caption{
		Visualization and identification of the CALIPSO data.
		\label{fig: case study -- curve and solution path}}
\end{figure}
The identified PDE model ($N = 12, M = 72$), reasonably speaking, is
\begin{equation}
	\label{equ: case study -- identified model}
	\frac{\partial}{\partial t}u(x,t)
	=
	a
	\frac{\partial}{\partial x}u(x,t)
	+
	b
	\left( \frac{\partial^2}{\partial x^2}u(x,t) \right)^2,
\end{equation}
where the values of $a$ and $b$ can be estimated using a simple linear regression on the selected derivatives, that is,
$
\frac{\partial}{\partial x}u(x,t)
$
and
$
\left( \frac{\partial^2}{\partial x^2}u(x,t) \right)^2.
$
The linear regression suggests reasonable values of $a=-0.2505$ and $b = 1.7648$.
Note that we focus on identification, that is, identifying
$
\frac{\partial}{\partial x}u(x,t)
$
and
$
\left( \frac{\partial^2}{\partial x^2}u(x,t) \right)^2
$
from many derivative candidates, rather than estimating the coefficients.
Therefore, we use $a=-0.2505$ and $b=1.7648$ as a reference.

\begin{figure}[htbp]
	\centering
	\begin{tabular}{ccc}
		\includegraphics[width=0.24 \textwidth]{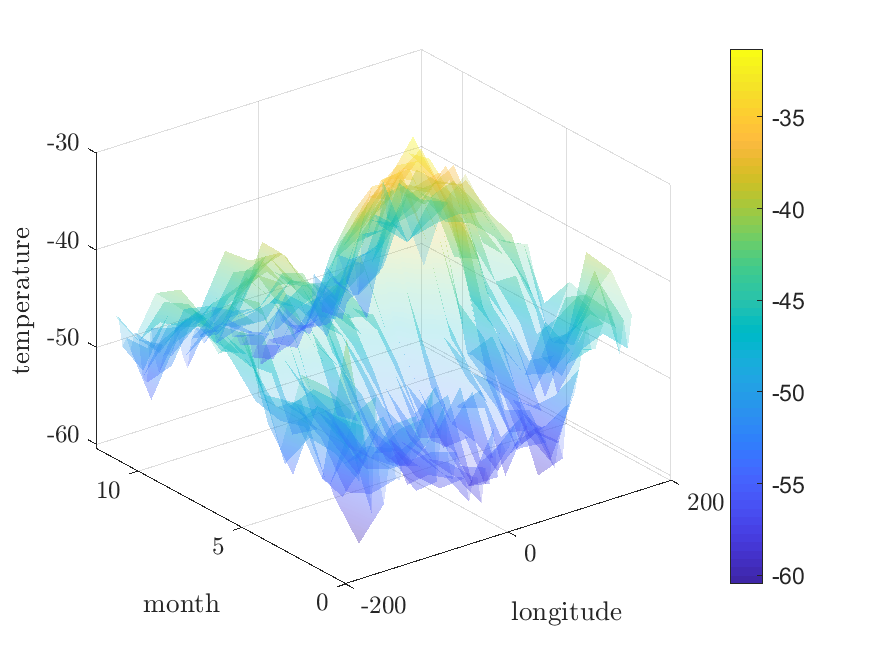} &
		\includegraphics[width=0.24 \textwidth]{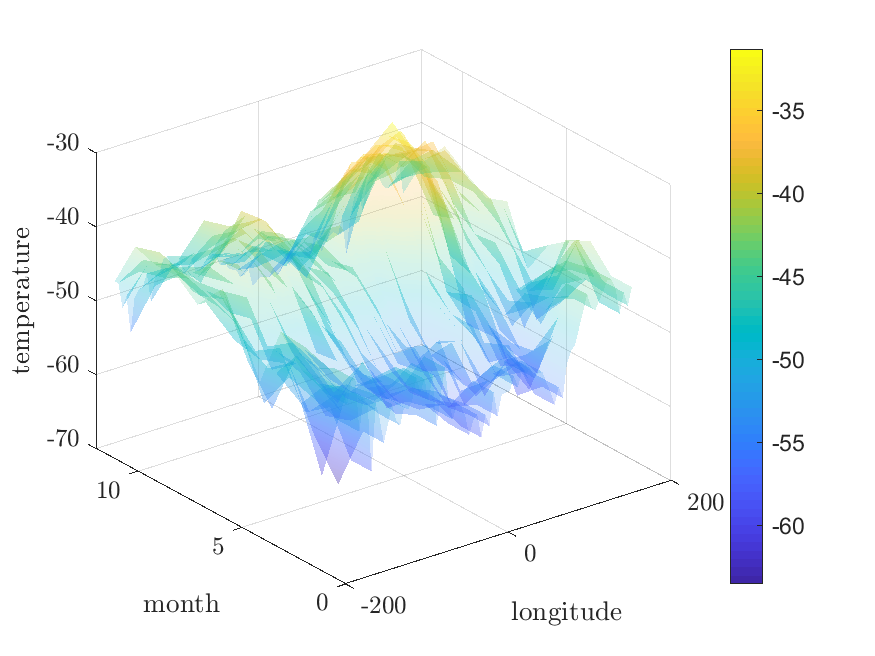}   &
		\includegraphics[width=0.24 \textwidth]{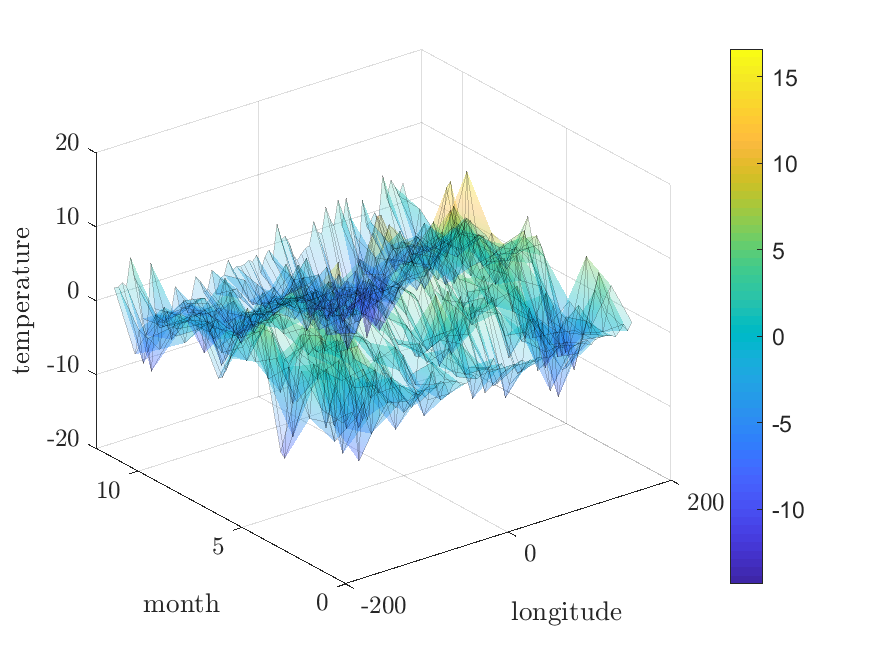} \\
		(a.1) observed 2017 temp  &
		(a.2) fitted 2017 temp&
		(a.3) 2017 residual \\
		\includegraphics[width=0.24 \textwidth]{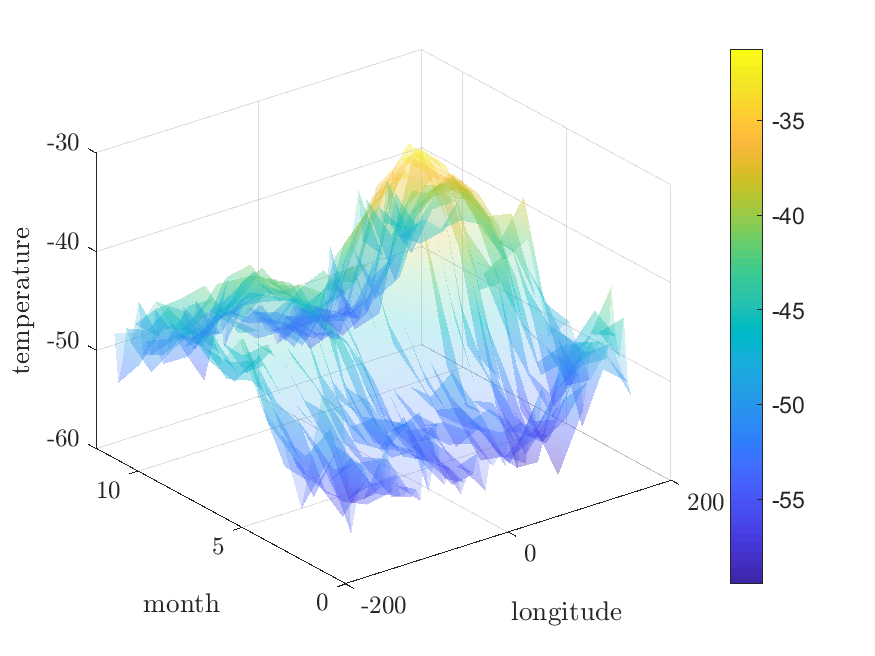}  &
		\includegraphics[width=0.24 \textwidth]{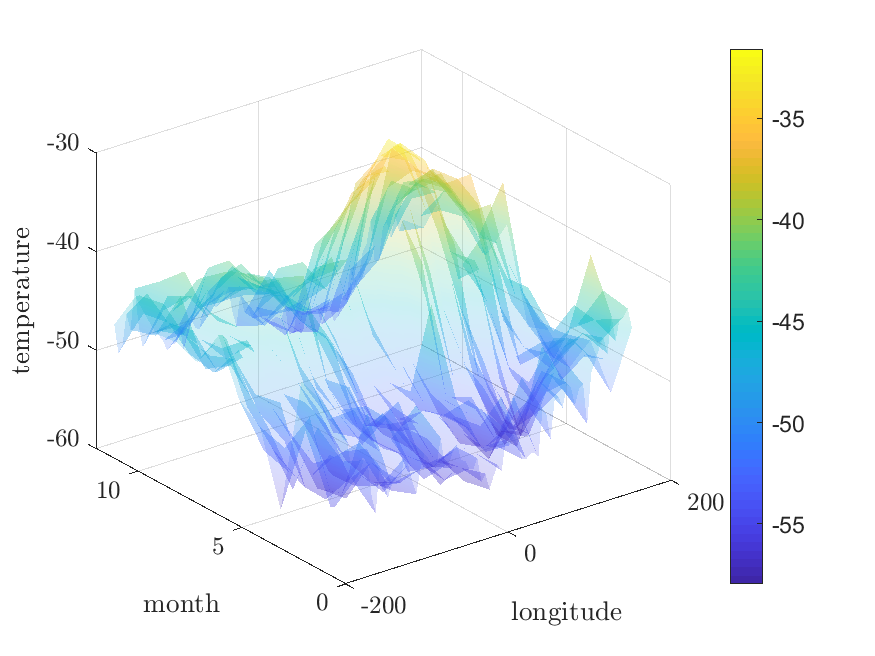} &
		\includegraphics[width=0.24 \textwidth]{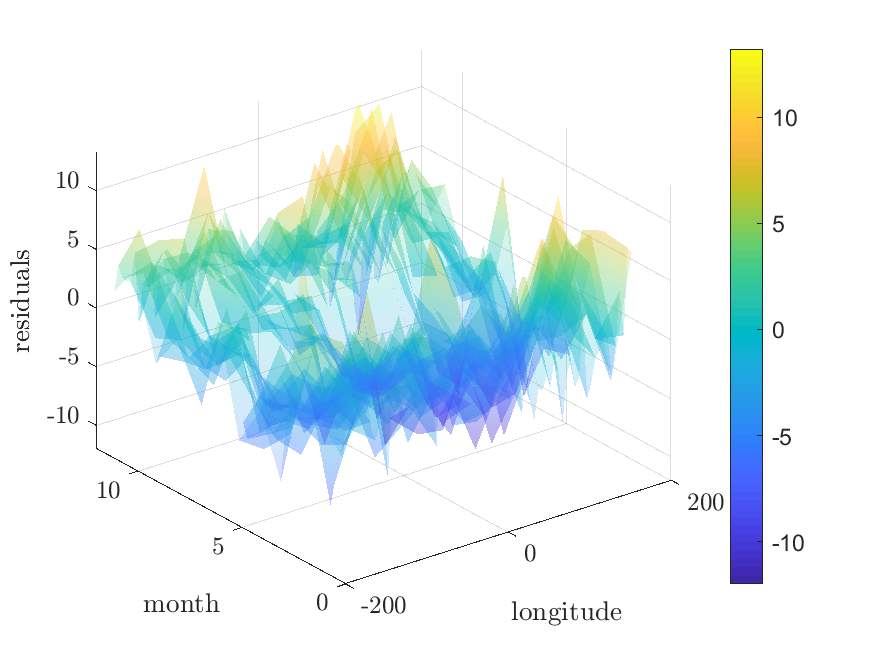}  \\
		(b.1) observed 2018 temp &
		(b.2) predicted 2018 temp &
		(b.3) 2018 residual \\
	\end{tabular}
	\caption{
		3D surface plots of the temperatures in 2017/2018. 
		\label{fig: case study -- surface}}
\end{figure}

Because the CALIPSP is a real-world data set, we do not know the ground truth of the underlying PDE model.
Here we provide some justifications. 
First, from the solution path in Fig. \ref{fig: case study -- curve and solution path}(b), the coefficients of
$
\frac{\partial}{\partial x}u(x,t)
$
and
$
\left( \frac{\partial^2}{\partial x^2}u(x,t) \right)^2
$
remain nonzeros under $\lambda = 0.05$, whereas the other coefficients are all zero.
Second, the identified PDE model in \eqref{equ: case study -- identified model} fits well to the training data (see Fig.  \ref{fig: case study -- surface} (a.1)-(a.3)).
Third, the identified PDE model in \eqref{equ: case study -- identified model} predicts well in the testing data (see Fig  \ref{fig: case study -- surface} (b.1)-(b.3)).
Thus, our proposed SAPDEMI method performs well in the CALIPSO data set, beacuse it adequately predicts the feature values in 2018.

\section{Conclusion}
\label{sec: conclusion}
We have proposed an SAPDEMI method for identifying underlying PDE models from noisy data. 
The proposed method is computationally efficient, and we derive a statistical guarantee on its performance.
We realize there are many promising future research directions, including, but not limited to, incorporating a multivariate spatial variable ($\myx \in \mathbb R^d$ with $d \geq 2$)  \citep{habermann2007multidimensional}, and the interactions between spatial and temporal variables.
In our paper, we aim at showing the methodology to solve the PDE identification, so we skip discuss the above future research and our paper should provide a good starting point for these further research.

\section*{Supplementary Material}

There is an online supplementary material for this paper, which includes (1) lemmas to derive the main theory; (2) numerical details of the figures in the simulation; (3) proofs and other technical details which is not covered in the main body of the paper due to the page limitation.

\section*{Acknowledgments}

The authors gratefully acknowledge the support of NSF grants DMS-2015405, DMS-2015363, and the TRIAD (a part of the TRIPODS program at NSF, located at Georgia Tech and enabled by the NSF grant CCF-1740776).

\par


\bibhang=1.7pc
\bibsep=2pt
\fontsize{9}{14pt plus.8pt minus .6pt}\selectfont
\renewcommand\bibname{\large \bf References}
\expandafter\ifx\csname
natexlab\endcsname\relax\def\natexlab#1{#1}\fi
\expandafter\ifx\csname url\endcsname\relax
\def\url#1{\texttt{#1}}\fi
\expandafter\ifx\csname urlprefix\endcsname\relax\def\urlprefix{URL}\fi
\bibliographystyle{chicago}            
\bibliography{PDEreference.bib}   

\vskip .65cm
\noindent

\appendix
\section{Supplementary Material}
\subsection{Overview of Our Proposed Algorithm}

We give an overview our SAPDEMI method in Algorithm \ref{alg: SAPDEMI}.

\begin{algorithm}[H]
	\caption{Overview of our proposed SAPDEMI method}
	\label{alg: SAPDEMI}
	\LinesNumbered
	\KwIn{Data from the unknown PDE model as in \eqref{equ: noisy data}; penalty parameter used in the Lasso identify model: $\lambda>0$; smoothing parameter used in the cubic spline: $\alpha, \bar\alpha \in (0,1]$.}
	\KwOut{The identified/recovered PDE model.}
	\functionalEstimation{
		Estimate $\myX, \nabla_t \myu$ by cubic spline with $\alpha, \bar\alpha \in (0,1]$.
	}
	
	\PDEidentification
	{
		The unknown PDE system is recovered as:
		$
		\frac{\partial}{\partial t}u(x,t)
		=
		\myx^\top \widehat{\mybeta},
		$
		where
		$
		\widehat{\mybeta}
		=
		\arg\min_{\mybeta}
		\frac{1}{2MN}
		\|
		\nabla_t \myu - \myX \mybeta
		\|_2^2
		+
		\lambda \| \mybeta \|_1
		$
		and
		$
		\myx
		=
		\left(
		\begin{array}{ccccccccccccc}
			1, &
			u(x,t),  &
			\frac{\partial u(x, t)}{\partial x}, &
			\frac{\partial^2 u(x, t)}{\partial x^2}, &
			\left( u(x,t) \right)^2, &
			u(x,t) \frac{\partial u(x, t)}{\partial x}, &
			\ldots, &
		\end{array}
		\right)^\top.
		$\\
		
	}
\end{algorithm}

\subsection{Derivation of the 0-th, First, Second Derivative of the Cubic Spline in Section 2.1}
\label{proof: derivative of the smoothing cubic Spline}

In this section, we focus on solving the derivatives of $u(x,t_n)$ with respective to $x$, i.e.,\\
$
\left\{
u(x_i, t_n),
\frac{\partial}{\partial x} u(x_i, t_n),
\frac{\partial^2}{\partial x^2} u(x_i, t_n)
\right\}_{i=0,1\ldots, M-1}, 
$
for any $n=0,1,\ldots, N-1$.
To realize this objective, we first fix $t$ as $t_n$ for a general $n \in \{0, 1, \ldots, N-1\}$.
Then we use cubic spline to fit data $\{(x_i, u_n^i)\}_{i=0,1,\ldots,M-1}$.

Suppose the cubic polynomial spline over the knots $\{(x_i, u_n^i)\}_{i=0,1,\ldots,M-1}$ is $s(x)$.
So under good approximation, we can regard $s(x), s'(x), s''(x)$ as the estimators of $u(x_i, t_n)$, $\frac{\partial}{\partial x} u(x, t_n)$, $\frac{\partial^2}{\partial x^2} u(x, t_n)$, where $s'(x), s''(x)$ is the first and second derivatives of $s(x)$, respectively.

Let's first take a look at the zero-order derivatives of $s(x)$.
By introducing matrix algebra, the objective function in equation \eqref{equ: smoothing cubic spline -- objective fucntion} can be rewritten as
\begin{equation}
	\label{equ: J in matrix algebra}
	J_{\alpha} (s) = \alpha (\myu^n_: -\myf)^\top \myW (\myu^n_: - \myf) + (1-\alpha) \myf^\top \myA^\top \myM^{-1} \myA \myf
\end{equation}
where vector
\begin{equation*}
	\myf
	=
	\left(
	\begin{array}{c}
		s(x_0) \\
		s(x_1) \\
		\vdots \\
		s(x_{M-1})
	\end{array}
	\right)
	\triangleq
	\left(
	\begin{array}{c}
		f_0 \\
		f_1 \\
		\vdots \\
		f_{M-1}
	\end{array}
	\right),
	\myu^n_:
	=
	\left(
	\begin{array}{c}
		u_0^n \\
		u_1^n \\
		\vdots \\
		u_{M-1}^n
	\end{array}
	\right)
\end{equation*}
and matrix $\myW = \text{diag}(w_0, w_1, \ldots, w_{M-1})$ and matrix $\myA$ is defined in \eqref{equ: proof -- cubic spline -- derivative estimation -- matrix A}.
By taking the derivative of \eqref{equ: J in matrix algebra} with respective to $\myf$ and set it as zero, we have
\begin{equation}
	\label{equ: cubic spline -- zero order derivative estimation -- appendix}
	\widehat{\myf}
	=
	[\alpha \myW +(1-\alpha) \myA^\top \myM \myA]^{-1} \alpha \myW \myu_:^n.
\end{equation}

Then we solve the second-order derivative with respective to $x$.
Let us first suppose that the cubic spline $s(x)$ in $[x_i, x_{i+1}]$ is denoted $s_i(x)$, and we denote
$
s_i''(x_i) = \sigma_i,
s_i''(x_{i+1}) = \sigma_{i+1}.
$
Then we have $\forall x \in [x_i, x_{i+1}] \; (0 \leq i \leq M-2)$,
\begin{equation*}
	s_i''(x) = \sigma_i \frac{x_{i+1} -x}{h_i} + \sigma_{i+1} \frac{x-x_i}{h_i},
\end{equation*}
where matrix $\myM$ is defined in \eqref{equ: proof -- cubic spline -- derivative estimation -- matrix M}.
This is because $s_i''(x)$ with $x \in [x_i, x_{i+1}]$ is a linear function.
By taking a double integral of the above equation, we have
\begin{equation}
	\label{equ: spline s(x)}
	s_i(x)   = \frac{\sigma_i}{6h_i} (x_{i+1} - x)^3 + \frac{\sigma_{i+1}}{6h_i} (x-x_i)^3 + c_1(x-x_1) +c_2(x_{i+1}-x),\\
\end{equation}
where $c_1, c_2$ is the unknown parameters to be estimated.
Because  $s_i(x)$ interpolates two endpoints $(x_i, f_i)$ and $(x_{i+1}, f_{i+1})$ , if we plug $x_i, x_{i+1}$ into the above $s_i(x)$, we have
\begin{equation*}
	\left\{
	\begin{array}{ccl}
		f_i     & = & s_i(x_i)     = \frac{\sigma_i}{6} h_i^2 + c_2 h_i\\
		f_{i+1} & = & s_i(x_{i+1}) = \frac{\sigma_{i+1}}{6} h_i^2 + c_1 h_i,
	\end{array}
	\right.
\end{equation*}
where we can solve $c_1, c_2$ as
\begin{equation*}
	\left\{
	\begin{array}{ccc}
		c_1 & = & (f_{i+1} - \frac{\sigma_{i+1}}{6} h_i^2)/h_i, \\
		c_2 & = & (f_i - \frac{\sigma_i}{6} h_i^2)/h_i.
	\end{array}
	\right.
\end{equation*}
By plugging in the value of $c_1,c_2$ into equation \eqref{equ: spline s(x)}, we have ( $0 \leq i \leq M-2$)
\begin{equation*}
	s_i(x) =
	\frac{\sigma_i}{6h_i} (x_{i+1} -x)^3
	+
	\frac{\sigma_{i+1}}{6h_i} (x-x_i)^3
	+
	\left( \frac{f_{i+1}}{h_i} - \frac{\sigma_{i+1} h_i}{6} \right) (x-x_i)
	+
	\left( \frac{f_i}{h_i} - \frac{\sigma_i h_i}{6}  \right)(x_{i+1} -x), 
\end{equation*}
with its first derivative as
\begin{equation}
	\label{equ: spline s'(x)}
	s_i'(x) = -\frac{\sigma_i}{2h_i} (x_{i+1} -x)^2 + \frac{\sigma_{i+1}}{2h_i} (x-x_i)^2 + \frac{f_{i+1} - f_i}{h_i} - \frac{h_i}{6} (\sigma_{i+1} -\sigma_i).
\end{equation}
Because $s'_{i-1}(x_i) = s'_i(x_i)$, for $1 \leq i \leq M-2$, we have 
\begin{equation}
	\label{equ: spline continue first-order derivative}
	\frac{1}{6} h_{i-1} \sigma_{i-1} + \frac{1}{3} (h_{i-1} + h_i)\sigma_i + \frac{1}{6} h_i \sigma_{i+1}
	=
	\frac{f_{i+1} - f_i}{h_i} - \frac{f_i - f_{i-1}}{h_{i-1}}.
\end{equation}
Equation \eqref{equ: spline continue first-order derivative} gives $M-2$ equations. Recall  $\sigma_0  = \sigma_{M-1} =0$, so totally we get $M$ equations, which is enough to solve $M$ parameters, i.e., $\sigma_0, \sigma_1, \ldots, \sigma_{M-1}$.
We write out the above system of linear equations, where we hope to identify a fast numerical approach to solve it.
The system of linear equations is:
{\small \begin{equation*}
		\left\{
		\begin{array}{ccccccccc}
			& &\frac{1}{3}(h_0+h_1)\sigma_1
			& + &
			\frac{1}{6}h_1\sigma_2
			& = &
			\frac{u_2^n - u_1^n}{h_1}
			-
			\frac{f_1 - u_0}{h_0} \\
			\frac{1}{6} h_1\sigma_1
			& + &
			\frac{1}{3}(h_1+h_2)\sigma_2
			& + &
			\frac{1}{6}h_2\sigma_3
			& = &
			\frac{f_3 - f_2}{h_1}
			-
			\frac{f_2 - f_1}{h_0}\\
			&  &  & \vdots &  &  \\
			\frac{1}{6} h_{M-4}\sigma_{M-4}
			& + &
			\frac{1}{3}(h_{M-4}+h_{M-3})\sigma_{M-3}
			& + &
			\frac{1}{6}h_{M-3} \sigma_{M-2}
			& = &
			\frac{f_{M-2} - f_{M-3}}{h_{M-3}}
			-
			\frac{f_{M-3} - f_{M-4}}{h_{M-4}}\\
			\frac{1}{6} h_{M-3}\sigma_{M-3}
			& + & \frac{1}{3}(h_{M-3}+h_{M-2})\sigma_{M-2}
			& &
			& = &
			\frac{f_{M-1} - f_{M-2}}{h_{M-2}}
			-
			\frac{f_{M-2} - f_{M-3}}{h_{M-3}}
		\end{array}
		\right..
\end{equation*}}
From the above system of equation, we can see that the second derivative of cubic spline $s(x)$ can be solved by the above system of linear equation, i.e.,
\begin{equation}
	\label{equ: cubic spline -- second order derivative estimation}
	\widehat{\mysigma} = \myM^{-1} \myA \widehat{\myf} 
\end{equation}
where vector $\widehat\myf$ is defined in \eqref{equ: cubic spline -- zero order derivative estimation -- appendix}, matrix $\myA \in \mathbb R^{(M-2) \times M}$ is defined in \eqref{equ: proof -- cubic spline -- derivative estimation -- matrix A}, and matrix $\myM \in \mathbb R^{(M-2) \times (M-2)}$ is defined as \eqref{equ: proof -- cubic spline -- derivative estimation -- matrix M}.

Finally, we focus on solving the first derivative of cubic spline $s(x)$.
Let $\theta_i = s'(x_i)$ for $i = 0,1, \ldots, M-1$, then we have
\begin{equation*}
	\begin{array}{rcl}
		s_i(x)
		& = &
		\theta_i \frac{(x_{i+1}-x)^2(x-x_i)}{h_i^2}
		-
		\theta_{i+1} \frac{(x-x_i)^2(x_{i+1} -x)}{h_i^2}
		+
		f_i \frac{(x_{i+1} - x)^2 [2(x-x_i)+h_i]}{h_i^3}
		+ \\
		& &
		f_{i+1} \frac{(x-x_i)^2[2(x_{i+1} -x) + h_i]}{h_i^3}, \\
		s_i'(x)
		&=&
		\theta_i\frac{(x_{i+1}-x)(2x_i +x_{i+1} -3x)}{h_i^2} - \theta_{i+1} \frac{(x-x_i)(2x_{i+1}+x_i-3x)}{h_i^2} + 6 \frac{u_{i+1}^n - u_i^n}{h_i^3} (x_{i+1}-x) (x-x_i), \\
		s_i''(x)
		&=&
		-2\theta_i \frac{2x_{i+1} + x_i -3x}{h_i^2} - 2\theta_{i+1} \frac{2x_i + x_{i+1} -3x}{h_i^2} + 6\frac{u_{i+1}^n - u_i^n}{h_i^3} (x_{i+1} + x_i -2x). \\
	\end{array}
\end{equation*}

By plugging $x_i$ into $s_i''(x)$ and $s_{i-1}''(x)$, we have
\begin{equation*}
	\left\{
	\begin{array}{ccll}
		s_i''(x)
		& = &
		-2\theta_i \frac{2x_{i+1} + x_i -3x}{h_i^2} - 2\theta_{i+1} \frac{2x_i + x_{i+1} -3x}{h_i^2} + 6\frac{f_{i+1} - f_i}{h_i^3} (x_{i+1} + x_i -2x)\\
		s_{i-1}''(x)
		& = &
		-2\theta_{i-1} \frac{2x_{i} + x_{i-1} -3x}{h_{i-1}^2} - 2\theta_{i} \frac{2x_{i-1} + x_{i} -3x}{h_{i-1}^2} + 6\frac{f_{i} - f_{i-1}}{h_{i-1}^3} (x_{i} + x_{i-1} -2x)\\
	\end{array}
	\right.
\end{equation*}
which gives
\begin{equation*}
	\left\{
	\begin{array}{ccll}
		& s_i''(x)
		& = &
		\frac{-4}{h_i}\theta_i + \frac{-2}{h_i} \theta_{i+1} + 6\frac{f_{i+1} - f_i}{h_i^2} \\
		& s_{i-1}''(x)
		& = &
		\frac{2}{h_{i-1}}\theta_{i-1} + \frac{4}{h_{i-1}} \theta_{i} - 6\frac{f_i - f_{i-1}}{h_{i-1}^2}. \\
	\end{array}
	\right.
\end{equation*}
Because $s_i''(x_i) = s_{i-1}''(x_i)$, we have ($\forall i=1,2,\ldots, M-2$)
\begin{equation*}
	\begin{array}{cc}
		&
		\frac{-4}{h_i}\theta_i + \frac{-2}{h_i} \theta_{i+1} + 6\frac{f_{i+1} - f_i}{h_i^2}
		=
		\frac{2}{h_{i-1}}\theta_{i-1} + \frac{4}{h_{i-1}} \theta_{i} - 6\frac{f_i - f_{i-1}}{h_{i-1}^2}\\
		\Leftrightarrow &
		\frac{2}{h_{i-1}}\theta_{i-1} + (\frac{4}{h_{i-1}} + \frac{4}{h_i}) \theta_i + \frac{2}{h_i} \theta_{i+1}
		=
		6\frac{f_{i+1} - f_i}{h_i^2} + 6\frac{f_i - f_{i-1}}{h_{i-1}^2} \\
		\Leftrightarrow &
		\frac{1}{h_{i-1}}\theta_{i-1} + (\frac{2}{h_{i-1}} + \frac{2}{h_i}) \theta_i + \frac{1}{h_i} \theta_{i+1}
		=
		3\frac{f_{i+1} - f_i}{h_i^2} + 3\frac{f_i - f_{i-1}}{h_{i-1}^2}.
	\end{array}
\end{equation*}

By organizing the above system of equation into matrix algebra, we have

\begin{equation*}
	\begin{array}{l}
		\left(
		\begin{array}{ccccccccc}
			\frac{1}{h_0} & \frac{2}{h_0}+\frac{2}{h_1} & \frac{1}{h_1}               & 0     & \ldots & 0     & 0     & 0 \\
			0             & \frac{1}{h_1}               & \frac{2}{h_1}+\frac{2}{h_2} & 0     & \ldots & 0     & 0     & 0\\
			\vdots        & \vdots                      & \vdots                      &\vdots & \ddots &\vdots &\vdots &\vdots\\
			0             & 0                           & 0                           & 0     & \ldots & \frac{1}{h_{M-3}} & \frac{2}{h_{M-3}}+\frac{2}{h_{M-2}} & \frac{1}{h_{M-2}} \\
		\end{array}
		\right)
		\left(
		\begin{array}{c}
			\theta_0 \\
			\theta_1 \\
			\theta_2 \\
			\vdots \\
			\theta_{M-1}
		\end{array}
		\right)\\
		=
		\left(
		\begin{array}{c}
			3\frac{f_2-f_1}{h_1^2} + 3\frac{f_1 - f_0}{h_0^2} \\
			3\frac{f_3-f_2^n}{h_2^2} + 3\frac{f_2 - f_1}{h_1^2} \\
			\vdots \\
			3\frac{f_{M-1}-f_{M-2}}{h_{M-2}^2} + 3\frac{f_{M-2}^n - f_{M-3}}{h_{M-3}^2}
		\end{array}
		\right).
	\end{array}
\end{equation*}

For the endpoint $\theta_0$, because $s_0''(x_0) = 0$, we have
\begin{equation*}
	s_0''(x)
	=
	-2\theta_0 \frac{2x_{1} + x_0 -3x}{h_0^2} - 2\theta_{1} \frac{2x_0 + x_{1} -3x}{h_0^2} + 6\frac{f_1 - f_0}{h_0^3} (x_1 + x_0 -2x).
\end{equation*}
When we take the value of $x$ as $x_0$, we have
\begin{equation*}
	\begin{array}{crl}
		s_0''(x_0)
		& = &
		-2\theta_0 \frac{2x_{1} + x_0 -3x_0}{h_0^2} - 2\theta_{1} \frac{2x_0 + x_{1} -3x_0}{h_0^2} + 6\frac{f_1 - f_0}{h_0^3} (x_1 + x_0 -2x_0)\\
		&=&
		\frac{-4}{h_0} \theta_0 + \frac{- 2}{h_0}\theta_{1} + 6\frac{f_{1} - f_0}{h_0^2} \\
		&=&
		0. \\
	\end{array}
\end{equation*}
For the two endpoint $\theta_{M-1}$, because $s_{M-2}''(x_{M-1}) = 0$, we have
\begin{equation*}
	\begin{array}{crcl}
		s_{M-2}''(x)
		&=&
		-2\theta_{M-2} \frac{2x_{M-1} + x_{M-2} -3x}{h_{M-2}^2}
		-
		2\theta_{M-1} \frac{2x_{M-2} + x_{M-1} -3x}{h_{M-2}^2}
		+ \\
		& &
		6\frac{f_{M-1} - f_{M-2}}{h_{M-2}^3} (x_{M-1} + x_{M-2} -2x). 
	\end{array}
\end{equation*}
When we take the value of $x$ as $x_{M-1}$, we have
\begin{equation*}
	\begin{array}{crl}
		s_{M-2}''(x_{M-1})
		& = &
		-2\theta_{M-2} \frac{2x_{M-1} + x_{M-2} -3x_{M-1}}{h_{M-2}^2}
		-
		2\theta_{M-1} \frac{2x_{M-2} + x_{M-1} -3x_{M-1}}{h_{M-2}^2}
		+ \\
		& &
		6\frac{f_{M-1} - f_{M-2}}{h_{M-2}^3} (x_{M-1} + x_{M-2} -2x_{M-1})\\
		& = &
		\frac{2}{h_{M-2}}\theta_{M-2}  + \frac{4}{h_{M-2}}\theta_{M-1} - 6\frac{f_{M-1} - f_{M-2}}{h_{M-2}^2} \\
		& = &
		0. \\
	\end{array}
\end{equation*}
So the first order derivative $\mytheta = (\theta_0, \theta_1, \ldots, \theta_{M-1})^\top$ can be solved by
\begin{equation*}
	\begin{array}{l}
		\underbrace{
			\left(
			\begin{array}{ccccccccc}
				\frac{2}{h_0} & \frac{1}{h_0}               & 0                           & 0     & \ldots & 0     & 0     & 0 \\
				\frac{1}{h_0} & \frac{2}{h_0}+\frac{2}{h_1} & \frac{1}{h_1}               & 0     & \ldots & 0     & 0     & 0 \\
				0             & \frac{1}{h_1}               & \frac{2}{h_1}+\frac{2}{h_2} & 0     & \ldots & 0     & 0     & 0\\
				\vdots        & \vdots                      & \vdots                      &\vdots & \ddots &\vdots &\vdots &\vdots\\
				0             & 0                           & 0                           & 0     & \ldots & \frac{1}{h_{M-3}} & \frac{2}{h_{M-3}}+\frac{2}{h_{M-2}} & \frac{1}{h_{M-2}} \\
				0             & 0                           & 0                           & 0     & \ldots & 0     &\frac{1}{h_{M-2}} & \frac{2}{h_{M-2}}
			\end{array}
			\right)
		}_{\myQ \in \mathbb R^{M \times M}}
		\underbrace{
			\left(
			\begin{array}{c}
				\theta_0 \\
				\theta_1 \\
				\theta_2 \\
				\theta_3 \\
				\theta_4 \\
				\vdots \\
				\theta_{M-1}
			\end{array}
			\right)
		}_{\mytheta}\\
		=
		\underbrace{
			\left(
			\begin{array}{c}
				3 \frac{f_1 - f_0}{h_0^2} \\
				3\frac{f_2-f_1}{h_1^2} + 3\frac{f_1 - f_0}{h_0^2} \\
				3\frac{f_3-f_2}{h_2^2} + 3\frac{f_2 - f_1}{h_1^2} \\
				\vdots \\
				3\frac{f_{M-1}-f_{M-2}}{h_{M-2}^2} + 3\frac{f_{M-2} - f_{M-3}}{h_{M-3}^2} \\
				3 \frac{f_{M-1} - f_{M-2}}{h_{M-2}^2}
			\end{array}
			\right)
		}_{\myq}. 
	\end{array}
\end{equation*}
In matrix notation, the first order derivative $\mytheta = (\theta_0, \theta_1, \ldots, \theta_{M-1})^\top$ can be solved by
\begin{equation}
	\label{equ: cubic spline -- first order derivative estimation}
	\widehat{\mytheta} = \myQ^{-1}\widehat{\myq} = \myQ^{-1} \myB \widehat{\myf},
\end{equation}
where $\widehat \myf$ is defined in \eqref{equ: cubic spline -- zero order derivative estimation -- appendix}, and matrix $\myB \in \mathbb R^{M \times M}$ is defined as
\begin{equation*}
	\myB=
	\left(
	\begin{array}{ccccccccc}
		\frac{-3}{h_0^2} & \frac{3}{h_0^2}                 & 0                & 0                & \ldots & 0     & 0     & 0 \\
		\frac{-3}{h_0^2} & \frac{3}{h_0^2}-\frac{3}{h_1^2} & \frac{3}{h_1^2}  & 0                & \ldots & 0     & 0     & 0 \\
		0                & \frac{-3}{h_1^2} & \frac{3}{h_1^2}-\frac{3}{h_2^2} & \frac{3}{h_2^2}  & \ldots & 0     & 0     & 0\\
		\vdots        & \vdots                      & \vdots                      &\vdots & \ddots &\vdots &\vdots &\vdots \\
		& 0     & 0     & 0 & 0 & \frac{-3}{h_{M-3}^2} & \frac{3}{h_{M-3}^2}-\frac{3}{h_{M-2}^2} & \frac{3}{h_{M-2}^2} \\
		& 0     & 0     & 0 & 0 & 0                    & \frac{-3}{h_{M-2}^2}  &\frac{3}{h_{M-2}^2}
	\end{array}
	\right).
\end{equation*}


\subsection{Computational Complexity of Local Polynomial Regression Method}\label{sec: local polynomial reg comp}
In the functional estimation stage, the computational complexity of the local polynomial regression method is stated in the following proportion.

\begin{myprop}
	\label{theo: computation complexity of local polynomial}
	Given data $\mathcal D$ in \eqref{equ: noisy data}, if we use the local polynomial regression in the functional estimation stage, i.e., estimate
	$
	\myX \in \mathbb R^{MN \times K}, \nabla_t \myu \in \mathbb R^{MN}
	$
	via the local polynomial regression described as in this online supplementary material, then the computation complexity of this stage is of order
	$$
	\max\{O( q_{\max}^2 M^2 N), O(M N^2), O(q_{\max}^3 MN), O(p_{\max}MN), O(K^3)\},
	$$
	where $p_{\max}$ is the highest polynomial order in \eqref{equ: temporal evolutionary PDE}, $q_{\max}$ is the highest order of derivatives in \eqref{equ: temporal evolutionary PDE}, $M$ is the spatial resolution, $N$ is the temporal resolution, and $K$ is the number of columns of $\myX$.
	
	If we set $q_{\max} = 2$ to match the derivative order of the local polynomial regression to the cubic spline, then the computation complexity is of order
	$$
	\max\{ O(M^2 N), O(M N^2) , O(p_{\max} MN), O(K^3) \}.
	$$
\end{myprop}

See a proof in Section \ref{appendix: Computational Complexity of Local Polynomial Regression}. 

As suggested by Proposition \ref{theo: computation complexity of local polynomial}, the computational complexity of local polynomial regression is much higher than that in the cubic spline.
But the advantage of local polynomial regression is that it can derive any order of derivatives, i.e., $q_{\max} \geq 0$ in \eqref{equ: temporal evolutionary PDE}, while for the cubic spline, $q_{\max} = 2$.
In applications, $q_{\max} = 2$ should be sufficient because most of the PDE models are governed by derivatives up to the second derivative, for instance, heat equation, wave equation, Laplace's equation, Helmholtz equation, Poisson's equation, and so on.
In our paper, we mainly use cubic spline as an illustration example due to its simplification and computational efficiency.
Readers can extend our proposed SAPDEMI method to the higher-order spline with $q_{\max} > 2$ if they are interested in higher-order derivatives.



\subsection{Coordinate Gradient Descent to Solve the Optimization problem in Section \ref{sec: use LASSO to identify the PDE model}.}
\label{appendix: Coordinate Gradient Descent}

In this section, we briefly review the implement of the coordinate descent algorithm in \cite{friedman2010regularization} to solve \eqref{equ: LASSO model - matrix algebra}.
The main idea of the coordinate descent is to update the estimator in a coordinate-wise fashion, which is the main difference between the coordinate descent and regular gradient descent.
For instance, in the $k$-th iteration, the coordinate descent updates the iterative estimator $\mybeta^{(k)}$ by using partial of the gradient information, instead of the whole gradient information.
Mathematically speaking, in the $k$-th iteration, the coordinate descent optimizes
$
F(\mybeta)
=
\frac{1}{2MN} \| \nabla_t \myu - \myX \mybeta \|_2^2 + \lambda \| \mybeta \|_1
$
with respective to $\mybeta$ by
$$
\beta_j^{(k+1)}
=
\arg\min_{\beta_j}
F(
(
\beta_{1}^{(k)},
\beta_{2}^{(k)},
\ldots,
\beta_{j-1}^{(k)},
\beta_j,
\beta_{j+1}^{(k)},
\ldots,
\beta_{K}^{(k)}
)
)
$$
for all $j= 1, 2, \ldots, K$.
To minimize the above optimization problem, we can derive the first derivative and set it as $0$:
$$
\frac{\partial}{\partial \beta_j} F(\mybeta^{(k)})
=
\frac{1}{MN} \left( \mye^\top_j \myX^\top \myX \mybeta^{(k)} - \nabla_t \myu^\top \myX \mye_j \right) + \lambda\text{sign}(\mybeta_j)
=
0,
$$
where $\mye_j$ is a vector of length $K$ whose entries are all zero expect the $j$-th entry is $1$.
By solving the above equation, we can solve $\beta_j^{(k+1)}$ by
$$
\beta_j^{(k+1)}
=
S
\left(
\nabla_t \myu^\top \myX \mye_j
-
\sum_{l \neq j} (\myX^\top \myX)_{jl} \mybeta_l^{(k)}, MN\lambda
\right)
\bigg/
(\myX^\top \myX)_{jj},
$$
where $S(\cdot)$ is the soft-thresholding function defined as
$$
S(x,\alpha) =
\left\{
\begin{array}{cl}
	x-\alpha & \text{if}\; x\geq  \alpha \\
	x+\alpha & \text{if}\; x\leq -\alpha \\
	0 & \text{otherwise}
\end{array}
\right..
$$
The detailed procedure of this algorithm is summarized in Algorithm \ref{alg: coordinate descent to lasso}.

\begin{algorithm}[htbp]
	\caption{Algorithm for the coordinate descent to minimize $F(\mybeta)$ }
	\label{alg: coordinate descent to lasso}
	\LinesNumbered
	\KwIn{response vector $\nabla_t \myu$, design matrix $\myX$, and number of iterations $M$}
	\KwOut{coefficient estimation $\widehat{\mybeta}$}
	\textbf{Initialize} $\mybeta^{(0)}$  \label{algLine: coordinate descent initial point}\\
	\For{$\mathcal \ell = 1, \ldots, \mathcal L$}{
		\For{$j = 1, \ldots, K$}{
			$  \mybeta^{(\ell)}_j
			=
			S\left(
			\nabla_t \myu^\top \myX \mye_j
			-
			\sum_{l \neq j} (\myX^\top \myX)_{jl} \mybeta_l^{(\ell - 1)}
			,
			MN\lambda
			\right)
			\bigg/
			(\myX^\top \myX)_{jj}
			$
		}
	}
	$\widehat{\mybeta} = \mybeta^{(\mathcal L)}$
\end{algorithm}




\subsection{A review of methods to select the smoothing parameter in the cubic spline literature}
\label{sec: review of smooth parameter in cubic spline}
We consider a noisy data $ \{(x_i, y_i)\}_{i = 1, \ldots, n}$, where $$y_i = g(x_i) + \epsilon_i,$$ with $\epsilon_i \sim N(0, \sigma^2)$. 
To fit this noisy data, the cubic spline use a spline function $s(x)$ to approximate $g(x)$.
And the function $s(x)$ can be solved as the minimizer of the following optimization problem:
$$
  J_{\lambda} (s)
  =
  \frac{1}{n}\sum_{i=1}^{n}  [y_i - s(x_i)]^2
  +
  \lambda \int_{x_1}^{x_{n}} s''(x)^2 dx,
$$
where the first term $\sum_{i=1}^{n} [y_i - s(x_i)]^2$ is the sum of squares for residuals.  
And this term is commonly called \textit{infidelity} of the data. 
In the second term $\lambda \int_{x_1}^{x_{n}} s''(x)^2 dx$, the function $s''(x)$ is the second derivative of $s(x)$, and this term is the penalty of the smoothness.
In the above optimization problem, the parameter $\lambda > 0$ controls the trade off between the goodness of fit and the smoothness of the cubic spline.
            
We will discuss the selection of $\lambda$ under two scenarios: $\sigma$ is known and $\sigma$ is unknown.
            
\begin{itemize}
    \item \textbf{Scenario 1: $\sigma$ is known.}
    As suggested by \cite{reinsch1967smoothing}, a good value of $\lambda$ should be the one make the infidelity ($\frac{1}{n}\sum_{i=1}^{n}  [y_i - s(x_i)]^2$) equals to $\sigma^2$, i.e.,
    $$
      \lambda^* 
      = 
      \left\{
      \lambda: 
      \frac{1}{n}
      \sum_{i = 1}^n
      \left[y_i - s(x_i)\right]^2 = \sigma^2
      \right\}.
    $$
              
    An alternative way to select $\lambda$ is to choose the optimal $\lambda$ which minimizes the true mean square error averaged over the data points \citep{wahba1975smoothing, craven1978smoothing}.
    And the true mean square error $R(\lambda)$ is defined as
    $$
      R(\lambda) 
      =
      \frac{1}{n}
      \sum_{i = 1}^n
      \left[g(x_i) - s(x_i)\right]^2.
    $$
    So the optimal $\lambda$ is
    $$
      \lambda^* = \arg\min_{\lambda}R(\lambda).
    $$
              
    In practice, the above two approaches from \cite{reinsch1967smoothing, wahba1975smoothing, craven1978smoothing} are not feasible, because $\sigma$ is commonly unknown.
              
    \item \textbf{Scenario 2: $\sigma$ is unknown.} 
              
    The first representative method is from \cite{mallows2000some, hudson1974empirical}, where the optimal $\lambda^*$ is
    \begin{eqnarray*}
      \lambda^*
      & = &
      \arg\min_{\lambda} E\left( \frac{1}{n} A(\lambda) y - g \right) \\
      & = &
      \arg\min_{\lambda} \frac{1}{n} 
      \|[I - A(\lambda)] g \|^2
      +
      \frac{\sigma^2}{n} \text{tr}(A^2(\lambda)).
    \end{eqnarray*}
    Here the matrix $A(\lambda) \in \mathbb R^{n \times n}$ depends on $\lambda$ and is defined by the following equation:
    $$
      \left(
      \begin{array}{c}
      s(x_1) \\ s(x_2) \\ \vdots \\ s(x_n) \\
      \end{array}
      \right)
      = 
      A(\lambda) 
      \left(\begin{array}{c}
      y_1 \\y_2 \\\vdots \\y_n \\
      \end{array}
      \right).
    $$
    In the above equation, the vectors $y$ and $g$ are defined $y = (y_1, \ldots, y_n)^\top$ and $g = (g(x_1), \ldots, g(x_n))^\top$.
    And the norm $\|\cdot\|^2$ is the Euclidean norm.
              
    The second representative method is generalized cross-validation (GCV) \citep{craven1978smoothing, aydin2013smoothing}.
    Mathematically, it takes the optimal $\lambda$ as the minimizer of $V(\lambda)$, i.e.,
    \begin{eqnarray*}
      \lambda^*
      & = &
      \arg\min_{\lambda} V(\lambda) \\
      & = &
      \arg\min_{\lambda} \frac{1}{n} 
      \|[I - A(\lambda)] y \|^2
      \bigg /
      \left[\frac{1}{n} \text{tr}(I - A(\lambda)) \right]^2.
      \end{eqnarray*}
     \end{itemize}
            

\subsection{Some Important Lemmas}
\label{sec: Some Preliminaries}

In this section, we present some important preliminaries, which are important blocks for the proofs of the main theories.
To begin with, we first give the upper bound of $\widehat{u(x,t_n)} -  u(x,t_n)$ for $x \in \{ x_0, x_1, \ldots, x_{M-1} \}$, which is distance between the ground truth $u(x,t_n)$ and the estimated zero-order derivatives by cubic spline $\widehat{u(x,t_n)}$.
\begin{mylemma}
	\label{lemma: u_x and u_x hat difference bound}
	Assume that
	\begin{enumerate}
		\item for any fixed $n = 0,1,\ldots, N-1$, we have the spatial variable $x$ is sorted in nondecreasing order, i.e., $x_0 < x_1 \ldots <x_{M-1}$;
		\item for any fixed $n = 0,1,\ldots, N-1$, we have the ground truth function $ f^*(x) := u(x,t_n) \in C^4$, where $C^4$ refers to the set of functions that is forth-time differentiable;
		\item for any fixed $n = 0,1,\ldots, N-1$, we have
		$
		\frac{\partial^2}{\partial x^2} u(x_0,t_n)
		=
		\frac{\partial^2}{\partial x^2} u(x_{M-1},t_n) = 0,
		$
		and
		$
		\frac{\partial^3}{\partial x^3} u(x_0,t_n) \neq 0,
		\frac{\partial^3}{\partial x^3} u(x_{M-1},t_n) = 0;
		$
		\item for any fixed $n = 0,1, \ldots, N-1$, the value of third order derivative of function $f^*(x) := u(x, t_n)$ at point $x = 0$ is bounded, i.e., $\frac{d^3}{dx^3} f^*(0) < + \infty$;
		\item for any $U_i^n$ generated by the underlying PDE system
		$
		U_i^n = u(x_i, t_n) + w_i^n
		$
		with
		$
		w_i^n \stackrel{i.i.d}
		{\sim}
		N(0,\sigma^2),
		$
		we have
		$
		\eta^2 : = \max_{i = 0,\ldots,M-1, n = 0, \ldots, N-1} E(U_i^n)^2
		$
		is bounded;
		\item for function
		$
		K(x)
		=
		\frac{1}{2}
		e^{-|x|/\sqrt{2}}
		\left[
		\sin(|x|\sqrt{2}) + \pi/4
		\right],
		$
		we assume that it is uniformly continuous with modulus of continuity $w_K$ and of bounded variation $V(K)$ and we also assume that
		$
		\int |K(x)| dx,
		$
		$
		\int |x|^{1/2} |dK(x)|,
		$
		$
		\int |x \log |x||^{1/2} |d K(x)|
		$
		are bounded and denote
		\begin{equation*}
			K_{\max}
			:=
			\max_{x \in \max_{x \in [0, X_{\max}] \cup [0, T_{\max}] }}
			K( x );
		\end{equation*}
		\item the smoothing parameter in \eqref{equ: smoothing cubic spline -- objective fucntion} is set as $\alpha = \left( 1 + M^{-4/7} \right)^{-1}$;
		\item the Condition \ref{assumption -- spline -- convergence cdf} - Condition \ref{assumption -- spline -- bounded pdf}  hold.
	\end{enumerate}
	Then there exist finite positive constant
	$
	\mathscr C_{ (\sigma, \|u\|_{L^\infty(\Omega)}) } > 0,
	C_{ (\sigma, \|u\|_{L^\infty(\Omega)}) } > 0,
	\widetilde C_{ (\sigma, \|u\|_{L^\infty(\Omega)}) } > 0 ,
	Q_{ (\sigma, \|u\|_{L^\infty(\Omega)}) } > 0,
	\gamma_{ (M) } > 0,
	\omega_{ (M) } >1,
	$
	such that for any $\epsilon$ satisfying
	\begin{equation*}
		\begin{array}{ccll}
			\epsilon
			& > &
			\mathscr C_{  (\sigma, \|u\|_{L^\infty(\Omega)}) } \;
			\max&
			\left\{
			4 K_{\max} M^{- 3/7},
			4 A M^{-3/7},
			4 \sqrt{2} \frac{d^3}{dx^3} f^*(0) M^{-3/7},
			\right. \\
			&  & &
			\left.
			\frac{
				16
				\left[
				C_{ (\sigma, \|u\|_{L^\infty(\Omega)}) } \log (M) + \gamma_{ (M) }
				\right]
				\log(M)
			}{ M^{3/7}} ,
			\right. \\
			&  & &
			\left.
			16 \sqrt{\frac{\omega_{ (M) } }{7}} \widetilde{C}_{ (\sigma, \|u\|_{L^\infty(\Omega)})}
			\frac{\sqrt{\log(M)}}{M^{3/7}}
			\right\},
		\end{array}
	\end{equation*}
	there exist a $\dot M >0$, such that when $M > \dot M$, we have
	\begin{eqnarray*}
		&&
		P
		\left[
		\sup_{x\in[0,X_{\max}]}
		\left|
		\frac{\partial^k}{\partial x^k} \widehat{ u(x,t_n)} -  \frac{\partial^k}{\partial x^k} u(x,t_n)
		\right|
		> \epsilon
		\right] \\
		&< &
		2M e^{-\frac{( M^{3/7} - \| u \|_{L^{\infty}(\Omega)})^2 }{2\sigma^2}} +
		Q_{ (\sigma, \|u\|_{L^\infty(\Omega)}) } e^{-L \gamma_{(M)}  }
		+
		4\sqrt{2} \eta^4 M^{-\omega_{(M)} / 7 }
	\end{eqnarray*}
	for $k = 0, 1, 2$.
	Here
	$
	A = \sup_{\alpha} \int |u|^s f_M(\alpha, u) du \times \int_{x \in [0, X_{\max}]} |K(x)| dx.
	$
\end{mylemma}

\begin{proof}
	See in Section  \ref{proof: Bound of zero-order derivative} in this file.
\end{proof}

In the above lemma, we add $(\sigma, \|u\|_{L^\infty(\Omega)})$ as the subscript of constants $\mathscr C, C, \widetilde C, Q$ to emphasize that these constant are independent of the temporal resolution $N$ and spatial resolution $M$, and only depends on the noisy data $\mathcal D$ in \eqref{equ: noisy data} itself.
We add $M$ as the subscript of constants $\gamma, \omega$ to emphasize that $\gamma, \omega$ are function of the spatial resolution $M$, and we will discuss the value of $\gamma, \omega$ in Lemma \ref{lemma: bound y-Xb*}.

The above lemma show the closeness between
$
\frac{\partial^k}{\partial x^k} \widehat{ u(x,t_n)}
$
and
$
\frac{\partial^k}{\partial x^k} u(x,t_n)
$
for $k = 0,1,2$.
This results can be easily extend of the closeness between
$
\frac{\partial}{\partial t} \widehat{ u(x_i, t)}
$
and
$
\frac{\partial}{\partial t} u(x_i, t),
$
which is shown in the following corollary.

\begin{mycorollary}
	\label{lemma: u_t and u_t hat difference bound}
	Assume that
	\begin{enumerate}
		\item for any fixed $i = 0,1,\ldots, M-1$, we have the spatial variable $t$ is sorted in nondecreasing order, i.e., $t_0 < t_1 \ldots <t_{N-1}$;
		\item for any fixed $i = 0,1,\ldots, M-1$, we have the ground truth function $ f^*(t) := u(x_i,t) \in C^4$, where $C^4$ refers to the set of functions that is forth-time differentiable;
		\item for any fixed $i = 0,1,\ldots, M-1$, we have
		$
		\frac{\partial^2}{\partial t^2} u(x_i,t_0)
		=
		\frac{\partial^2}{\partial t^2} u(x_i,t_{N-1}) = 0,
		$
		and
		$
		\frac{\partial^3}{\partial t^3} u(x_i,t_0) \neq 0,
		\frac{\partial^3}{\partial t^3} u(x_i,t_{N-1}) = 0;
		$
		\item for any fixed $i = 0,1, \ldots, M-1$, the value of third order derivative of function $\bar f^*(x) := u(x_i, t)$ at point $t = 0$ is bounded, i.e., $\frac{d^3}{dt^3} \bar f^*(0) < + \infty$;
		\item for any $U_i^n$ generated by the underlying PDE system
		$
		U_i^n = u(x_i, t_n) + w_i^n
		$
		with
		$
		w_i^n \stackrel{i.i.d}
		{\sim}
		N(0,\sigma^2),
		$
		we have
		$
		\max_{i=0,\ldots,M-1, n = 0, \ldots, N-1} E(U_i^n)^2
		$
		is bounded;
		\item for function
		$
		K(x)
		=
		\frac{1}{2}
		e^{-|x|/\sqrt{2}}
		\left[
		\sin(|x|\sqrt{2}) + \pi/4
		\right],
		$
		we have $K(x)$ is uniformly continuous with modulus of continuity $w_K$ and of bounded variation $V(K)$, and we also assume that
		$
		\int_{x \in [0, X_{\max}]} |K(x)| dx,
		$
		$
		\int |x|^{1/2} |dK(x)|,
		$
		$
		\int |x \log |x||^{1/2} |d K(x)|
		$
		are bounded
		and denote
		$
		K_{\max}:= \max_{x \in [0, X_{\max}] \cup [0, T_{\max}]} K\left( x \right);
		$
		\item the smoothing parameter in \eqref{equ: smoothing cubic spline -- objective fucntion} is set as $\bar\alpha = O\left( \left( 1 + N^{-4/7} \right)^{-1} \right)$;
		\item the Condition \ref{assumption -- spline -- convergence cdf} - Condition \ref{assumption -- spline -- bounded pdf}  hold.
	\end{enumerate}
	then there exist finite positive constant
	$
	\mathscr C_{ (\sigma, \|u\|_{L^\infty(\Omega)}) } > 0,
	C_{ (\sigma, \|u\|_{L^\infty(\Omega)}) } > 0,
	\widetilde C_{ (\sigma, \|u\|_{L^\infty(\Omega)}) } > 0 ,
	Q_{ (\sigma, \|u\|_{L^\infty(\Omega)}) } > 0,
	\gamma_{ (N) } > 0,
	\omega_{ (N) } >1,
	$
	such that for any $\epsilon$ satisfying
	\begin{equation*}
		\begin{array}{ccll}
			\epsilon
			& > &
			\mathscr C_{  (\sigma, \|u\|_{L^\infty(\Omega)}) } \;
			\max&
			\left\{
			4 K_{\max} N^{-3/7},
			4 \bar A N^{-3/7},
			4 \sqrt{2} \frac{d^3}{dx^3} f^*(0) N^{-3/7},
			\right. \\
			&  & &
			\left.
			\frac{
				16
				\left[
				C_{ (\sigma, \|u\|_{L^\infty(\Omega)}) } \log(N) + \gamma_{(N)}
				\right]
				\log(N)
			}{ N^{3/7}} ,
			\right. \\
			&  & &
			\left.
			16 \sqrt{\frac{\omega_{(N)} }{7}}
			\widetilde{C}_{ (\sigma, \|u\|_{L^\infty(\Omega)})  }
			\frac{\sqrt{\log(N)}}{N^{3/7}}
			\right\},
		\end{array}
	\end{equation*}
	there exist a $\dot N >0$, such that when $N > \dot N$, we have
	\begin{eqnarray*}
		P
		\left[
		\sup_{t\in[0,T_{\max}]}
		\left|
		\frac{\partial}{\partial t} \widehat{ u(x_i, t)} -  \frac{\partial}{\partial t} u(x_i, t)
		\right|
		> \epsilon
		\right]
		& < &
		2N e^{-\frac{( N^{3/7} - \| u \|_{L^{\infty}(\Omega)})^2 }{2\sigma^2}} +\\
		&   &
		Q_{ (\sigma, \|u\|_{L^\infty(\Omega)}) } e^{-L \gamma_{(N)}  }
		+
		4\sqrt{2} \eta^4 N^{-\omega_{(N)} / 7 }.
	\end{eqnarray*}
	Here
	$
	\bar A =
	\sup_{\alpha} \int |u|^s \bar f_N(\alpha, u) du
	\times
	\int_{t \in [0, T_{\max}]} |K(x)| dx.
	$
\end{mycorollary}

After bounding the error of all the derivatives, we then aim to bound $\| \nabla_t \myu - \myX \mybeta^* \|_\infty$.
It is important to bound $\| \nabla_t \myu - \myX \mybeta^* \|_\infty$, with the reason described as follows in Lemma \ref{lemma: bound y-Xb*}.

\begin{mylemma}
	\label{lemma: bound y-Xb*}
	Suppose the conditions in Lemma \ref{lemma: u_x and u_x hat difference bound} and Corollary \ref{lemma: u_t and u_t hat difference bound} hold and we set $M = O(N)$, then there exist finite positive constant
	$
	\mathscr C_{ (\sigma, \|u\|_{L^\infty(\Omega)}) } > 0
	$
	such that for any $\epsilon$ satisfying
	$$
	\epsilon
	>
	\mathscr C_{ (\sigma, \|u\|_{L^\infty(\Omega)}) }
	\frac{\log(N)}{N^{3/7 - r}},
	$$
	and any $r \in \left(0, \frac{3}{7} \right)$, there exist $\dot N > 0$, such that when $ N > \dot N$, we have
	\begin{equation*}
		P
		\left(
		\| \nabla_t \myu - \myX \mybeta^* \|_\infty
		> \epsilon
		\right)
		<
		N e^{-N^r},
	\end{equation*}
	where $\mathscr C_{ (\sigma, \|u\|_{L^\infty(\Omega)}) }$ is a constant which do not depend on the temporal resolution $M$ and spatial resolution $N$.
\end{mylemma}

\begin{proof}
	See Section \ref{sec: proof}.
\end{proof}


\subsection{Justification of $\alpha, \bar \alpha$ in Theorem 3.1}
\label{sec: supp -- theorem 1 -- alpha}

We acknowledge that our way to select the smoothing parameters $\alpha, \bar \alpha$ is different from that in the cubic spline literature (see a detailed literature review in the supplementary material). 
The root cause of the difference lies in the different objectives in theory. For the existing methods in the cubic spline literature, the objective is to minimize the fitting error when one fits the data (similar to \textit{single objective optimization}). 
However, for our proposed SAPDEMI method, the objective is to maximize the accuracy when one identifies the underlying PDE models. To build a path to this objective, we apply the cubic spline as an important block. And the selection of the smoothing parameter is required to, on the one hand, have a relatively small fitting error; on the other hand, leads to a high identification accuracy (similar to \textit{multiple objective optimizations}).

\subsection{Tables to draw the curve in Fig. \ref{fig: sim -- example 1 -- complexity} and Fig. \ref{fig: sim -- successful identification prob curve}}
\label{appendix: table for successful identification curves}

In this section, we present the table to draw the curves in Fig. \ref{fig: sim -- example 1 -- complexity},\ref{fig: sim -- successful identification prob curve} in Table \ref{table: sim -- example 1 -- complexity}, \ref{table: sim -- correct identification prob}, respectively.

\begin{table}[htbp]
	\caption{Computational complexity of the functional estimation by cubic spline and local polynomial regression in transport equation
		\label{table: sim -- example 1 -- complexity}}
	\centering
	\begin{adjustbox}{max width=0.95\textwidth}
		\centering
		\begin{threeparttable}
			\begin{tabular}{c|cccccccc}
				\hline
				& \multicolumn{7}{c}{M = 20}\\
				\hline        & N=200  & N=400  & N=800  &N=1000 & N=1200 & N=1600 & N=2000\\
				\cline{2-8}
				cubic spline  & 374,389     & 748,589    &  1,496,989   & 1,871,189   & 2,245,389   & 2,993,789   & 3,742,189\\
				local poly    & 14,136,936  & 45,854,336 &  162,089,136 & 246,606,536 & 348,723,936 & 605,758,736 & 933,193,536\\
				\hline
				& \multicolumn{7}{c}{N = 20}\\
				\cline{2-8}    & M=200 & M=400 & M=800 & M=1000 & M=1200& M=1600 & M=2000\\
				\cline{2-8}
				cubic spline   & 398,573    & 875,773    & 207,0173   & 2,787,373  & 3,584,573    & 5,418,973    & 7,573,373\\
				local poly     & 33,046,336 & 125,596,136 & 489,255,736 &760,365,536 & 1,090,995,336 & 1,930,814,936 & 3,008,714,536\\
				\hline
			\end{tabular}
		\end{threeparttable}
	\end{adjustbox}
\end{table}

\begin{table}[htbp]
	\caption{Correct identification probability of transport equation, inviscid Burgers equation and viscous Burgers's equation
		\label{table: sim -- correct identification prob}}
	\centering
	\begin{adjustbox}{max width=0.95\textwidth}
		\centering
		\begin{threeparttable}
			\begin{tabular}{c|cccccccccccc}
				\hline
				&\multicolumn{12}{c}{$\sigma$}\\
				& 0.01 & 0.05 & 0.1 & 0.25 & 0.3 & 0.4& 0.5 & 0.7& 0.75 & 0.8 & 0.9 & 1\\
				\hline
				& \multicolumn{11}{c}{transport equation}\\
				\cline{2-13}
				$M=N=100$ &100\% & 100\% & 100\% & 100\% & 100\% &100\% &100\% & 100\% & 100\% & 100\% & 100\% & 100\%\\
				$M=N=150$ &100\% & 100\% & 100\% & 100\% & 100\% &100\% &100\% & 100\% & 100\% & 100\% & 100\% & 100\%\\
				$M=N=200$ &100\% & 100\% & 100\% & 100\% & 100\% &100\% &100\% & 100\% & 100\% & 100\% & 100\% & 100\%\\
				\cline{2-13}
				& \multicolumn{11}{c}{inviscid Burgers equation}\\
				\cline{2-13}
				$M=N=100$ &100\% & 100\% & 100\% &100\% &100\%& 100\%& 100\% & 99.9\%&99.8\% &99.8\% & 99.8\% & 99.1\% \\
				$M=N=150$ &100\% & 100\% & 100\% &100\% &100\%& 100\%& 100\% & 100\% &99.8\% &99.7\% & 99.7\% & 99.7\% \\
				$M=N=200$ &100\% & 100\% & 100\% &100\% &100\%& 100\%    & 100\% & 100\% &100\%     &100\%  & 100\%  & 100\% \\
				\cline{2-13}
				& \multicolumn{11}{c}{viscous Burgers equation}\\
				\cline{2-13}
				$M=N=100$&100\% & 99.4\%&89.8\%&78.0\%& 71.4\% &82.0\% &91.6\%& 72.8\%&79.0\% &72.9\%&57.9\%& 51.3\%\\
				$M=N=150$&100\% & 100\% &100\% &97.3\%& 96.5\% &96.2\% &97.6\%& 95.6\%&93.3\% &86.6\%&79.9\%& 73.6\%\\
				$M=N=200$&100\% & 100\% & 100\%& 100\%&  99.6\%&99.6\% &98.2\%& 98.8\%&98.2\% & 97.0\%&94.3\%& 91.3\%\\
				\hline
			\end{tabular}
			\begin{tablenotes}
				\footnotesize
				\item[1] The simulation results are based on 1000 times of simulations.
			\end{tablenotes}
		\end{threeparttable}
	\end{adjustbox}
\end{table}


\subsection{The reasons why the RK4 is not feasible.}
\label{app: Runge-Kutta}

In this section, we discuss the reasons why the RK4 is not feasible.
Generally speaking,  RK4 is used to approximate solutions of ordinary differential equations.
In our content, it aims at solveing the solution of the following differential equation with fixed $i \in \{0,\ldots, M-1\}$:
\begin{equation}
	\label{equ: case study -- RK4 -- problem formula}
	\left\{
	\begin{array}{l}
		\frac{\partial}{\partial t}u(x_i,t) = p(u,t) \\
		u(x_i, t_0) = u_i^0
	\end{array}
	\right.,
\end{equation}
where $p(u,t)$ is the function interpolated through data set
\begin{equation*}
	\left\{
	t_n, u_i^n,  \frac{u(x_i, t_n + \Delta t) - u(x_i, t_n)}{\Delta t}
	\right\}_{n = 0, \ldots, N-2}.
\end{equation*}
Then as shown by Chapter 5 in \cite{lambert1991numerical}, the solution can be approximate by
$$
u(x_i, t_{n+1}) = u(x_i, t_n) + \frac{\Delta t}{6}(k_1 + k_2 + k_3 + k_4),
$$
where $\Delta t = t_{n+1} - t_n$ and $k_1, k_2, k_3, k_4$ are defined as
\begin{equation}
	\label{equ: RK4 -- 4k define}
	\left\{
	\begin{array}{l}
		k_1 = p(t_n, u_i^n) \\
		k_2 = p(t_n + \Delta t/2, u_i^n + k_1 \Delta t/2 ) \\
		k_3 = p(t_n + \Delta t/2, u_i^n + k_2 \Delta t/2 ) \\
		k_4 = p(t_n + \Delta t, u_i^n + k_3 \Delta t ).
	\end{array}
	\right.
\end{equation}

Given the above implementation of RK4, we find it is infeasible to be used in our case study due to the following two reasons.
First, it is infeasible to obtain $p(u,t)$ in \eqref{equ: case study -- RK4 -- problem formula}, even though the interpolation methods.
Second, it is infeasible to get the value of $k_3$ in \eqref{equ: RK4 -- 4k define}.
Because the calculation of $k_3$ depends on the value of $k_2$ and $k_2$ needs to be obtained (at least) by interpolation, it is complicated to calculate $k_1, k_2, k_3, k_4$ through one-time operation.
Given the complicated implementation of RK4, we use the  explicit Euler method in our case study.


\subsection{Proofs}\label{sec: proof}
\subsubsection{Proof of Proposition \ref{theo: computation complexity of cubic spline}}
\label{proof: computation complexity in cubic spline (without interaction)}
\begin{proof}
	
	The computational complexity in the functional estimation stage lies in calculating all elements in matrix $\myX$ and vector $\nabla_t \myu$, including
	$$
	\left\{
	\widehat{ u(x_i, t_n)  },
	\widehat{ \frac{\partial}{\partial x} u(x_i, t_n) },
	\widehat{ \frac{\partial^2}{\partial x^2} u(x_i, t_n)  },
	\widehat{ \frac{\partial}{\partial t} u(x_i, t_n) }
	\right\}_{i = 0, \ldots, M-1, n = 0,\ldots, N-1}.
	$$
	by cubic spline in \eqref{equ: smoothing cubic spline -- objective fucntion}.
	
	We divide our proof into two scenarios:
	(1) $\alpha = 1$ and
	(2) $\alpha \in (0,1)$.
	
	\begin{itemize}
		\item First of all, we discuss a very simple case, i.e., $\alpha = 1$.
		When $\alpha = 1$, we call the cubic spline as \textit{interpolating cubic spline} since there is no penalty on the smoothness.
		
		For the zero-order derivative, i.e.,
		$
		\left\{
		\widehat{u(x_i,t_n)}
		\right\}_{i=0,\ldots, M-1, n = 0,\ldots, N-1},
		$
		it can be estimated as $\widehat{u(x_i,t_n)} = u_i^n$ for $i = 0,1,\ldots, M-1, n = 0,1,\ldots, N-1$.
		So there is no computational complexity involved.
		
		For the second order derivatives, i.e.,
		$
		\left\{
		\frac{\partial^2}{\partial x^2} \widehat{u(x_i,t_n)}
		\right\}_{i=0,\ldots, M-1},
		$
		with $n \in \{ 0,\ldots, N-1 \} $ fixed, it can be solved in a closed-form, i.e.,
		$$
		\widehat\mysigma = \myM^{-1} \myA \myu_:^n
		$$
		where
		$
		\widehat\mysigma
		=
		\left(
		\widehat{ \frac{\partial^2}{\partial x^2} u(x_0,t_n) },
		\widehat{ \frac{\partial^2}{\partial x^2} u(x_1,t_n) },
		\ldots,
		\widehat{ \frac{\partial^2}{\partial x^2} u(x_{M-1},t_n) }
		\right)^\top.
		$
		So the main computational load lies in the calculation of $\myM^{-1}$.
		Recall $\myM \in \mathbb R^{(M-2) \times (M-2)}$ is a tri-diagonal matrix:
		\begin{equation*}
			\myM
			=
			\left(
			\begin{array}{cccccc}
				\frac{h_0+h_1}{3} & \frac{h_1}{6}     & 0                 & \ldots & 0 & 0 \\
				\frac{h_1}{6}     & \frac{h_1+h_2}{3} & \frac{h_2}{6}     & \ldots & 0 & 0 \\
				0                 & \frac{h_2}{6}     & \frac{h_2+h_3}{3} & \ddots & 0 & 0 \\
				\vdots            & \vdots            & \ddots            & \ddots & \ddots & \vdots \\
				0                 & 0                 & 0                 & \ddots & \frac{h_{M-4}+h_{M-3}}{3} & \frac{h_{M-3}}{6} \\
				0                 & 0                 & 0                 & \ldots & \frac{h_{M-3}}{6} & \frac{h_{M-3} + h_{M-2}}{3}
			\end{array}
			\right).
		\end{equation*}
		For this type of tri-diagonal matrix, there exist a fast algorithm to calculate its inverse.
		The main idea of this fast algorithm is to decompose $\myM$ through Cholesky decomposition as
		$$
		\myM = \myL \myD \myL^\top,
		$$
		where
		$
		\myL \in \mathbb R^{(M-2) \times (M-2)},
		\myD \in \mathbb R^{(M-2) \times (M-2)}
		$
		has the form of
		\begin{equation*}
			\myL
			=
			\left(
			\begin{array}{cccccc}
				1      & 0      & 0     & \ldots & 0  \\
				l_1    & 1      & 0     & \ldots & 0  \\
				0      & l_2    & 1     & \ldots & 0  \\
				\vdots & \vdots & \ddots& \ddots & \vdots \\
				0      & 0      & 0     & l_{M-3}& 1
			\end{array}
			\right),
			\myD
			=
			\left(
			\begin{array}{cccc}
				d_1 & 0 & \ldots & 0\\
				0 & d_2 & \ldots &0\\
				\vdots & \vdots &\ddots & \vdots \\
				0 & 0 & \ldots &d_{M-2}
			\end{array}
			\right).
		\end{equation*}
		After decomposing matrix $\myM$ into $\myL \myD \myL^\top$, the second derivatives $\widehat \mysigma$ can be solved as
		$$
		\widehat \mysigma
		=
		(\myL^\top)^{-1} \myD^{-1} \myL^{-1}
		\underbrace{
			\myA \myu_:^n
		}_{\myxi}.
		$$
		In the remaining of the proof in this scenario, we will verify the following two issues:
		\begin{enumerate}
			\item the computational complexity to decompose $\myM$ into $\myL \myD \myL^\top$ is $O(M)$ with $n \in \{0,\ldots, N-1\}$ fixed;
			\item the computational complexity to compute $\widehat \mysigma = (\myL^\top)^{-1} \myD^{-1} \myL^{-1} \myxi$ is $O(M)$ with $n \in \{0,\ldots, N-1\}$ fixed and $\myL, \myD$ available.
		\end{enumerate}
		
		For the decomposition of $\myM = \myL \myD \myL^\top$, its essence is to derive $l_1, \ldots l_{M-3}$ in matrix $\myL$ and $d_1, \ldots, d_{M-2}$ in matrix $\myD$.
		By utilizing the method of undetermined coefficients to inequality $\myM = \myL \myD \myL^\top$, we have:
		\begin{equation*}
			\begin{array}{ccl}
				&&
				\left[
				\begin{array}{cccccc}
					d_1     & d_1 l_1 & 0      & \ldots & 0                           & 0\\
					d_1 l_1 & d_2     & d_2 l_2& \ldots & 0                           & 0\\
					0       & d_2 l_2 & d_3    & \ldots & 0                           & 0\\
					\vdots  & \vdots  & \vdots & \ddots & \vdots                      & \vdots\\
					0       & 0       & 0      & \ldots & d_{M-3} l_{M-3}             & d_{M-3} l_{M-3}^2 + d_{M-2}
				\end{array}
				\right]\\
				& = &
				\left[
				\begin{array}{cccc}
					M_{11} & M_{12} & \ldots      & 0  \\
					M_{21} & M_{22} & \ldots      & 0  \\
					0      & M_{32} & \ldots      & 0  \\
					\vdots & \vdots & \ddots      & \vdots \\
					0      & 0      & \ldots      & M_{M-2,M-2}
				\end{array}
				\right],
			\end{array}
		\end{equation*}
		where $M_{i,j}$ is the $(i,j)$th entry in matrix $\myM$.
		Through the above method of undetermined coefficients, we can solve the exact value of the entries in matrix $\myL, \myD$, which is summarized in Algorithm \ref{alg: solve L & D}.
		It can be seen from Algorithm \ref{alg: solve L & D} that, the computational complexity of solve $\myL, \myD$ is of order $O(M)$.
		
		For the calculation of
		$
		\widehat \mysigma
		=
		(\myL^\top)^{-1} \myD^{-1} \myL^{-1} \myxi
		$
		with matrix $\myL, \myD$ available, we will first verify that the computational complexity to solve
		$
		\bar{\myxi} = \myL^{-1} \myxi
		$
		is $O(M)$.
		Then, we will verify that the computational complexity to solve
		$
		\bar{ \bar{\myxi} }= \myD^{-1} \bar{\myxi}
		$
		is $O(M)$.
		Finally, we will verify that the computational complexity to solve
		$
		\bar{\bar{ \bar{\myxi} }}= (\myL^\top)^{-1} \bar{ \bar{\myxi} }
		$
		is $O(M)$.
		First, the computational complexity of calculating
		$
		\bar{\myxi} = \myL^{-1} \myxi
		$
		is $O(M)$, this is because by $\myL \bar{\myxi} = \myxi$, we have the following system of equations:
		\begin{equation*}
			\left\{
			\begin{array}{l}
				\xi_1  = \bar{\xi}_1\\
				\xi_2  = \bar{\xi}_2 + l_1  \bar{\xi}_1\\
				\vdots \\
				\xi_{M-2} = \bar{\xi}_{M-2} + l_{M-3}  \bar{\xi}_{M-3}, 
			\end{array}
			\right.
		\end{equation*}
		where $\xi_i, \bar{\xi}_i$ is the $i$-th entry in $\myxi, \bar{\myxi}$ ,respectively.
		Through the above system of equations, we can solve the values of all entries in $\bar{\myxi}$, which is summarized in Algorithm \ref{alg: solve L inverse y}.
		From Algorithm \ref{alg: solve L inverse y}, we know that the computational complexity of solving $\myL^{-1} \myxi$ is $O(M)$.
		Next, it is obvious that the computational complexity of
		$
		\bar{\bar{\myxi}}
		=
		\myD^{-1} \bar{\myxi}
		$
		is $O(M)$, because $\myD$ is a diagonal matrix.
		Finally, with the similar logic flow, we can verify that the computational complexity of
		$
		\bar{\bar{\bar{\myxi}}}
		=
		(\myL^\top)^{-1} \bar{\bar{\myxi}}
		$
		is still $O(M)$.
		So, the computational complexity is to calculate
		$
		\widehat \mysigma
		=
		(\myL^\top)^{-1} \myD^{-1} \myL^{-1} \myxi,
		$
		with known $\myL, \myD$ is $O(M)$.
		
		As a summary, the computational complexity is to calculate
		$
		\left\{
		\widehat{ \frac{\partial^2}{\partial x^2} u(x_i, t_n)}
		\right\}_{i = 0,\ldots, M-1}
		$
		with a fixed $n \in \{0,1, \ldots, N-1 \}$ is $O(M)$.
		Accordingly, the computational complexity to solve
		$
		\left\{
		\widehat{ \frac{\partial^2}{\partial x^2} u(x_i, t_n)}
		\right\}_{i = 0,\ldots, M-1, n = 0,\ldots, N-1}
		$
		is $O(MN)$.
		
		For the first order derivatives, i.e.,
		$
		\left\{
		\frac{\partial}{\partial x} u(x_i, t_n),
		\frac{\partial}{\partial x} u(x_i, t_n)
		\right\}_{i = 0,\ldots, M-1, n = 0,\ldots, N-1},
		$
		we can verify the computational complexity to solve them is also $O(MN)$ with the similar logic as that in the second order derivatives.
		
		\begin{algorithm}[H]
			\caption{Pseudo code to solve $\myL, \myD$ }
			\label{alg: solve L & D}
			\LinesNumbered
			\KwIn{matrix $\myM$}
			\KwOut{matrix $\myL, \myD$}
			\textbf{Initialize} $d_1 = M_{1,1}$  \\
			\For{$i = 1,2, \ldots, M-3$}{
				$l_i = M_{i,i+1}/d_i$\\
				$d_{i+1} = M_{i+1,i+1} - d_i l_i^2$
			}
		\end{algorithm}
		
		\begin{algorithm}[H]
			\caption{Pseudo code to solve $\myL^{-1} \myxi$ }
			\label{alg: solve L inverse y}
			\LinesNumbered
			\KwIn{matrix $\myL,\myxi$}
			\KwOut{matrix $\bar{\myxi}$}
			\textbf{Initialize} $\bar{\xi}_1 = \xi_1$  \\
			\For{$i = 2, \ldots, M-2$}{
				$\bar{\xi}_i = \xi_i - l_{i-1} \bar{\xi}_{i-1}$
			}
		\end{algorithm}

		\item Next, we discuss the scenario when $\alpha \in (0,1)$.
		
		Since all the derivatives has similar closed-form formulation as shown in
		\eqref{equ: cubic spline -- zero order derivative estimation},
		\eqref{equ: cubic spline -- first order derivative estimation},
		\eqref{equ: cubic spline -- second order derivative estimation},
		we take the zero-order derivative
		$
		\left\{
		u(x_i, t_n)
		\right\}_{i = 0, \ldots, M-1, n = 0,\ldots, N-1}
		$
		as an illustration example, and other derivatives can be derived similarly.
		
		Recall that in Section \ref{sec: solve derivatives by cubic spline}, the zero-order derivative
		$
		\left\{
		u(x_i, t_n)
		\right\}_{i = 0, \ldots, M-1}
		$
		with $n \in \{0,1,\ldots, N-1\}$ fixed can be estimated through cubic spline as in equation \eqref{equ: cubic spline -- zero order derivative estimation}:
		\begin{equation*}
			\widehat{\myf}
			=
			[
			\underbrace{
				\alpha \myW +(1-\alpha) \myA^\top \myM \myA
			}_{\myZ}
			]^{-1}
			\underbrace{
				\alpha \myW \myu_:^n
			}_{\myy},
		\end{equation*}
		where $\alpha \in (0,1)$ trades off the fitness of the cubic spline and the smoothness of the cubic spline, vector
		$
		\widehat{\myf}
		=
		\left(
		\widehat{u(x_0, t_n)},
		\widehat{u(x_1, t_n)},
		\ldots,
		\widehat{u(x_{M-1}, t_n)}
		\right)^\top,
		$
		vector
		$
		\myu_:^{n} = \left(u_0^n, \ldots,  u_{M-1}^n \right)^\top,
		$
		matrix
		$
		\myW = \text{diag}(w_0, w_1, \ldots, w_{M-1}),
		$
		matrix $\myA \in \mathbb R^{(M-2) \times M}, \myM \in \mathbb R^{(M-2) \times (M-2)}$ are defined as
		\begin{equation*}
			\myA=
			\left(
			\begin{array}{ccccccccc}
				\frac{1}{h_0} & -\frac{1}{h_0} - \frac{1}{h_1} & \frac{1}{h_1}                  & 0             & \ldots & 0 & 0& 0 \\
				0             & \frac{1}{h_1}                  & -\frac{1}{h_1} - \frac{1}{h_2} & \frac{1}{h_2} & \ldots & 0 & 0& 0  \\
				\vdots  & \vdots & \vdots & \vdots  & \ddots & \vdots & \vdots  & \vdots \\
				0             & 0                              & 0                              & 0             & \ldots & \frac{1}{h_{M-3}} & -\frac{1}{h_{M-3}} - \frac{1}{h_{M-2}} & \frac{1}{h_{M-2}}\\
			\end{array}
			\right),
		\end{equation*}
		\begin{equation*}
			\myM
			=
			\left(
			\begin{array}{cccccc}
				\frac{h_0+h_1}{3} & \frac{h_1}{6}     & 0                 & \ldots & 0 & 0 \\
				\frac{h_1}{6}     & \frac{h_1+h_2}{3} & \frac{h_2}{6}     & \ldots & 0 & 0 \\
				0                 & \frac{h_2}{6}     & \frac{h_2+h_3}{3} & \ldots & 0 & 0 \\
				\vdots            & \vdots            & \vdots            & \ddots & \vdots & \vdots \\
				0                 & 0                 & 0                 & \ldots & \frac{h_{M-4}+h_{M-3}}{3} & \frac{h_{M-3}}{6} \\
				0                 & 0                 & 0                 & \ldots & \frac{h_{M-3}}{6} & \frac{h_{M-3} + h_{M-2}}{3}
			\end{array}
			\right)
		\end{equation*}
		with $h_i = x_{i+1} - x_{i}$ for $i = 0, 1,\ldots, M-2$.
		
		By simple calculation, we know that matrix
		$
		\myZ
		=
		\alpha \myW +(1-\alpha) \myA^\top \myM \myA
		\in
		\mathbb R^{M \times M}
		$
		is a symmetric seventh-diagonal matrix:
		\begin{equation*}
			\myZ
			=
			\left(
			\begin{array}{ccccccc}
				z_{11} & z_{12} & z_{13} & z_{14} & 0      & \ldots &  \\
				z_{21} & z_{22} & z_{23} & z_{24} & z_{25} & \ldots & \\
				z_{31} & z_{32} & z_{33} & z_{34} & z_{35} & \ddots &  \\
				z_{41} & z_{42} & z_{43} & z_{44} & z_{45} & \ddots &  \\
				0      & z_{52} & z_{53} & z_{54} & z_{55} & \ddots &  \\
				\vdots & \vdots & \ddots & \ddots & \ddots & \ddots &  \\
			\end{array}
			\right). 
		\end{equation*}
		By applying Cholesky decomposition to matrix $\myZ$ as $\myZ = \myP \mySigma \myP^\top$, we can calculate $\widehat \myf$ as
		$$
		\widehat \myf
		=
		\myZ^{-1} \myy
		=
		(\myP^\top)^{-1} \mySigma^{-1} \myP^{-1} \myy,
		$$
		where $\myP \in \mathbb R^{M \times M}, \mySigma \in \mathbb R^{M \times M}$ has the form of
		\begin{equation*}
			\myP
			=
			\left(
			\begin{array}{ccccccc}
				1        & 0        & 0        & 0          & \ldots       & 0 & 0 \\
				\ell_1   & 1        & 0        & 0          & \ldots       & 0 & 0 \\
				\gamma_1 & \ell_2   & 1        & 0          & \ldots       & 0 & 0 \\
				\eta_1   & \gamma_2 & \ell_3   & 1          & \ldots       & 0 & 0 \\
				0        & \eta_2   & \gamma_3 & \ell_4     & \ddots       & 0 & 0 \\
				\vdots   & \vdots   & \ddots   & \ddots     & \ddots       & 1 & 0 \\
				0        & 0        & \ldots   & \eta_{M-3} & \gamma_{M-2} & \ell_{M-1} & 1
			\end{array}
			\right)
			,
			\mySigma
			=
			\left(
			\begin{array}{ccccc}
				s_1    & 0      & 0      & \ldots & 0 \\
				0      & s_2    & 0      & \ldots & 0 \\
				0      & 0      & s_3    & \ldots & 0 \\
				\vdots & \vdots & \vdots & \ddots & 0 \\
				0      & 0      & 0      & \ldots & s_{M}
			\end{array}
			\right).
		\end{equation*}
		In the remaining of the proof in this scenario, we will verify the following two issues:
		\begin{enumerate}
			\item the computational complexity to decompose $\myZ$ into $\myP \mySigma \myP^\top$ is $O(M)$ with $n \in \{0,\ldots, N-1\}$ fixed;
			\item the computational complexity to compute $(\myP^\top)^{-1} \mySigma^{-1} \myP^{-1} \myy$ is $O(M)$ with $n \in \{0,\ldots, N-1\}$ fixed.
		\end{enumerate}
		
		First of all, we verify that the computational complexity to decompose $\myZ$ into $\myP \mySigma \myP^\top$ is $O(M)$ when $n \in \{0,\ldots, N-1\}$ fixed
		By applying method of undetermined coefficients to equality $\myZ = \myP \mySigma \myP^\top$, we have
		\begin{equation*}
			\left[
			\begin{array}{ccccccc}
				s_1          & s_1 \ell_1                       & s_1 \gamma_1                                           & \ldots & 0      \\
				s_1 \ell_1   & s_1 \ell_1^2 + s_2               & s_1 \ell_1 \gamma_1 + s_2 \ell_2                       & \ldots & 0      \\
				s_1 \gamma_1 & s_1 \ell_1 \gamma_1 + s_2 \ell_2 & s_1 \gamma_1^2 + s_2 \ell_2^2 + s_3                    & \ldots & 0      \\
				s_1 \eta_1   & s_1 \eta_1 \ell_1 + s_2 \gamma_2 & s_1 \eta_1 \gamma_1 + s_2 \gamma_2 \ell_2 + s_3 \ell_3 & \ldots & 0      \\
				0            & s_2 \eta_2                       & s_2 \eta_2 \ell_2 + s_3 \gamma_3                       & \ldots & 0      \\
				0            & 0                                & s_3 \eta_3                                             & \ldots & 0      \\
				\vdots       & \vdots                           & \vdots                                                 & \ddots & \vdots \\
				0            & 0                                & 0                                                      & \ldots & s_{M-3} \eta_{M-3}^2 + s_{M-2} \gamma_{M-2}^2       \\
				&                                  &                                                        &        & + s_{M-1} \gamma_{M-1} \ell_{M-1}^2 + s_M      \\
			\end{array}
			\right]
			=
			[z_{i,j}],
		\end{equation*}
		where $[z_{ij}]$ denotes matrix $\myZ$ with its $(i,j)$th entry as $z_{i,j}$.
		Through the above method of undetermined coefficients, we can solve the explicit value of all entries in matrix $\myP, \mySigma$, i.e., $\ell_1, \ldots, \ell_{M-1}, \gamma_1, \ldots, \gamma_{M-2}, \eta_1, \ldots, \eta_{M-3}$ in matrix $\myP$ and $s_1, \ldots, s_{M}$ in matrix $\mySigma$, which is summarized in Algorithm \ref{alg: solve P & Sigma}.
		From Algorithm \ref{alg: solve P & Sigma}, we can see that the computational complexity to decompose $\myZ$ into $\myP \mySigma \myP^\top$ is $O(M)$ with $n \in \{0,\ldots, N-1\}$ fixed.
		
		Second, we verify the computational complexity to compute $(\myP^\top)^{-1} \mySigma^{-1} \myP^{-1} \myy$ is $O(M)$ with $n \in \{0,\ldots, N-1\}$ fixed and matrix $\myP, \mySigma$ available.
		To realize this objective, we will first verify that the computational complexity to calculate $\bar{\myy} = \myP^{-1} \myy$ is $O(M)$.
		Then, we will first verify that the computational complexity to calculate $\bar{\bar{\myy}} = \mySigma^{-1} \bar{\myy}$ is $O(M)$.
		Finally, we will first verify that the computational complexity to calculate $\bar{\bar{\bar{\myy}}} = (\myP^\top)^{-1} \bar{\bar{\myy}}$ is $O(M)$.
		First of all, let us verify the computational complexity to compute $\bar{\myy} = \myP^{-1} \myy$ is $O(M)$ with $n \in \{0,\ldots, N-1\}$ fixed.
		Because we have a system of equations derived from $\myP \bar{\myy} = \myy$:
		\begin{equation*}
			\left\{
			\begin{array}{ccl}
				\bar{y}_1 & = & y_1 \\
				\bar{y}_2 & = & y_2 - \ell_1   \bar{y}_1 \\
				\bar{y}_3 & = & y_3 - \gamma_1 \bar{y}_1 - \ell_2   \bar{y}_2 \\
				\bar{y}_4 & = & y_4 - \eta_1   \bar{y}_1 - \gamma_2 \bar{y}_2 - \ell_3 \bar{y}_3\\
				\bar{y}_5 & = & y_5 - \eta_2   \bar{y}_3 - \gamma_3 \bar{y}_3 - \ell_4 \bar{y}_4\\
				\vdots    &&  \\
				\bar{y}_M & = & y_M - \eta_{M-3}\bar{y}_{M-3} - \gamma_{M-2} \bar{y}_{M-2} - \ell_{M-1} \bar{y}_{M-1}, \\
			\end{array}
			\right.
		\end{equation*}
		we can solve vector
		$
		\bar{\myy} = (\bar{y}_1, \bar{y}_2, \ldots, \bar{y}_M)^\top
		$
		explicitly through Algorithm \ref{alg: solve P inverse y}, which only requires $O(M)$ computational complexity.
		After deriving
		$
		\bar{\myy} = \myP^{-1} \myy,
		$
		we can easily verify that the computational complexity to derive
		$
		\bar{\bar{\myy}} = \mySigma^{-1} \bar{\myy}
		$
		is still $O(M)$ because $\mysigma$ is a diagonal matrix.
		Finally, after deriving
		$
		\bar{\bar{\myy}} = \mySigma^{-1} \bar{\myy},
		$
		we can verify that the computational complexity to derive
		$
		\bar{\bar{\bar{\myy}}} = (\myP^\top)^{-1} \bar{\bar{\myy}}
		$
		is still $O(M)$ with the similar logic as that in
		$
		\bar{\myy} = \myP^{-1} \myy.
		$
		
		From the above discussion, we know that the computational complexity to calculate
		$
		\widehat{\myf}
		=
		(\widehat{u(x_0, t_n)}, \widehat{u(x_1, t_n)}, \ldots, \widehat{u(x_{M-1}, t_n)})^\top,
		$
		is $O(M)$ with $n \in \{0,1,\ldots, N-1\}$ fixed.
		In other words, the computational complexity to derive
		$
		\left\{ \widehat{u(x_i, t_n)} \right\}_{i=0,\ldots,M-1}
		$
		is $O(M)$.
		According, the computational complexity to derive
		$
		\left\{ \widehat{u(x_i, t_n)} \right\}_{i=0,\ldots, M-1, n=0,\ldots, N-1}
		$
		is $O(MN)$.
		
		\begin{algorithm}[H]
			\caption{Pseudo code to solve $\myP, \mySigma$ }
			\label{alg: solve P & Sigma}
			\LinesNumbered
			\KwIn{matrix $\myZ$}
			\KwOut{matrix $\myP, \mySigma$}
			\textbf{Initialize} $s_j = \eta_j = \gamma_j = \ell_j = 0 \; \forall j \leq 0$  \\
			\For{$i = 1,2, \ldots, M$}{
				$s_i = z_{ii} - s_{i-3} \eta_{i-3}^2 - s_{i-2} \gamma_{i-2}^2 - s_{i-1} \ell_{i-1}^2$ \\
				$\ell_i = (z_{i,i+1} - s_{i-2} \gamma_{i-2} \eta_{i-2} - s_{i-1} \gamma_{i-1} \ell_{i-1}) / s_{i}$ \\
				$\eta_i = a_{i, i+3} / s_i$
			}
		\end{algorithm}
		
		\begin{algorithm}[H]
			\caption{Pseudo code to solve $\myP^{-1} \myy$ }
			\label{alg: solve P inverse y}
			\LinesNumbered
			\KwIn{matrix $\myP,\myy$}
			\KwOut{vector $\bar{\myy}$}
			\textbf{Initialize} $\eta_i = \gamma_i = \ell_i = 0 \; \forall i\leq 0$  \\
			\For{$i = 1, \ldots, M$}{
				$
				\bar{y}_i
				= y_i - \eta_{i-3}\bar{y}_{i-3} - \gamma_{i-2} \bar{y}_{i-2} - \ell_{i-1} \bar{y}_{i-1}
				$
			}
		\end{algorithm}

	\end{itemize}

\end{proof}

\subsubsection{Proof of Proposition \ref{theo: computation complexity of local polynomial}}
\label{appendix: Computational Complexity of Local Polynomial Regression}

\begin{proof}
	We have discussed how to use cubic spline to derive derivatives of $u(x,t)$.
	In this section, we discuss how to use local polynomial regression to derive derivatives, as a benchmark method.
	
	Recall that the derivatives can be estimated by local polynomial regression includes
	\linebreak[4]{
		$
		u(x_i, t_n),
		\frac{\partial}{\partial x} u(x_i,t_n),
		\frac{\partial^2}{\partial x^2} u(x_i,t_n),
		\ldots.
		$
	}
	And here we take the derivation $\frac{\partial^l}{\partial x^l} u(x,t_n)$ as an example ($l=0,1,2,\ldots$), and the other derivatives can be derived with the same logic flow.
	To derive the estimation of $\frac{\partial^l}{\partial x^l} u(x,t_n)$, we fix the temporal variable $t_n$ for a general $n \in \{0, 1, \ldots, N-1 \}$.
	Then we locally fit a degree $\check p$ polynomial over the data
	$
	\left\{
	(x_i, u_i^n)
	\right\}_{i=0,\ldots, M-1},
	$
	i.e.,
	\begin{equation*}
		\left\{
		\begin{array}{ccl}
			u(x_0, t_n)
			& = &
			u(x,t_n)
			+
			\frac{\partial}{\partial x}  u(x,t_n) (x_0 - x)
			+
			\ldots
			+
			\frac{\partial^{\check p}}{\partial x^{\check p}} u(x,t_n) (x_0-x)^{\check p} \\
			u(x_1, t_n)
			& = &
			u(x,t_n)
			+
			\frac{\partial}{\partial x} u(x,t_n) (x_1-x)
			+
			\ldots
			+
			\frac{\partial^{\check p}}{\partial x^{\check p}} u(x,t_n) (x_1-x)^{\check p} \\
			\vdots          &   & \vdots\\
			u(x_{M-1}, t_n)
			& = &
			u(x,t_n)
			+
			\frac{\partial}{\partial x} u(x,t_n) (x_{M-1} - x)
			+
			\ldots
			+
			\frac{\partial^{\check p}}{\partial x^{\check p}} u(x,t_n) (x_{M-1} - x)^{\check p}
		\end{array}
		\right..
	\end{equation*}
	For the choice of $\check p$, we choose $\check{p} = l+3$ to realize minmax efficiency \cite[see][]{fan1997local}.
	If we denote
	$
	\myb(x)
	=
	\left( u(x,t_n),
	\frac{\partial}{\partial x} u(x,t_n),
	\ldots,
	\frac{\partial^{\check p}}{\partial x^{\check p}} u(x,t_n) \right)^\top,
	$
	then $\frac{\partial^l}{\partial x^l} u(x,t_n)$ can be obtained as the $(l+1)$-th entry of the vector $\widehat{\myb(x)}$, and $\widehat{\myb(x)}$ is obtained by the following optimization problem:
	\begin{equation}
		\label{equ: solve derivative w.r.t x -- local polynomial}
		\widehat{ \myb(x) }
		=
		\arg\min_{\myb(x)}
		\sum_{i = 0}^{M-1}
		\left[
		u_i^n
		-
		\sum_{j=0}^{{\check p}}
		\frac{\partial^j}{\partial x^j} u(x,t_n) (x_i - x)^j
		\right]^2
		\mathcal{K} \left( \frac{x_i - x}{h} \right),
	\end{equation}
	where $h$ is the bandwidth parameter, and $\mathcal K$ is a kernel function, and in our paper, we use the Epanechikov kernel $\mathcal K(x) = \frac{3}{4} \max\{0, 1-x^2\}$ for $x \in \mathbb R$.
	Essentially, the optimization problem in equation \eqref{equ: solve derivative w.r.t x -- local polynomial} is a weighted least squares model, where $\myb(x)$ can be solved in a close form:
	\begin{equation}
		\label{equ: solve derivative w.r.t x -- local polynomial (weighted LS)}
		\myb(x)
		=
		\left( \myX_{\text{spa}}^\top \myW_{\text{spa}} \myX_{\text{spa}} \right)^{-1}
		\myX_{\text{spa}}^\top \myW_{\text{spa}} \myu_{:}^n,
	\end{equation}
	where
	\begin{equation*}
		\myX_{\text{spa}}
		=
		\left[
		\begin{array}{cccc}
			1 & (x_0-x)     & \ldots & (x_0-x)^{\check p} \\
			1 & (x_1-x)     & \ldots & (x_1-x)^{\check p} \\
			\vdots & \vdots      & \ddots & \vdots   \\
			1 & (x_{M-1}-x) & \ldots & (x_{M-1}-x)^{\check p}
		\end{array}
		\right],
		\myu_{:}^n
		=
		\left[
		\begin{array}{c}
			u_0^n \\
			u_1^n \\
			\vdots \\
			u_{M-1}^n
		\end{array}
		\right]
	\end{equation*}
	and
	$
	\myW_{\text{spa}}
	=
	\text{diag}
	\left(
	\mathcal{K} \left( \frac{x_0 - x}{h} \right),
	\ldots,
	\mathcal{K} \left( \frac{x_{M-1} - x}{h} \right)
	\right)
	$.
	
	By implementing the local polynomial in this way, the computational complexity is much higher than our method, and we summarize its computational complexity in the following proposition.
	
	Following please find the proof.
	
	Similar to the proof of the computational complexity in cubic spline, the proof of the computational complexity of local polynomial regression in the funcational estimation stage lies in calculating all elements in matrix $\myX$ and vector $\nabla_t \myu$, including
	$$
	\left\{
	\widehat{ u(x_i, t_n)  },
	\widehat{ \frac{\partial}{\partial x} u(x_i, t_n) },
	\widehat{ \frac{\partial^2}{\partial x^2} u(x_i, t_n)  },
	\widehat{ \frac{\partial}{\partial t} u(x_i, t_n) }
	\right\}_{i = 0, \ldots, M-1, n = 0,\ldots, N-1}.
	$$
	
	We will take the estimation of
	$
	\widehat{ \frac{\partial^p}{\partial x^p} u(x_i, t_n)}
	$
	with a general $p \in \mathbb N$ as an example.
	To solve
	$
	\left\{
	\widehat{ \frac{\partial^p}{\partial x^p} u(x_i, t_n)}
	\right\}_{i=0,\ldots,M-1, n = 0,\ldots, N-1},
	$
	we first focus on
	$
	\left\{
	\widehat{ \frac{\partial^p}{\partial x^p} u(x_i, t_n)}
	\right\}_{i=0,\ldots,M-1},
	$
	with $n \in \{ 0,\ldots, N-1\}$ fixed.
	To solve it, the main idea of local polynomial regression is to do Taylor expansion:
	\begin{equation*}
		\left\{
		\begin{array}{ccl}
			u(x_0, t_n)
			& = &
			u(x,t_n)
			+
			\frac{\partial}{\partial x}  u(x,t_n) (x_0 - x)
			+
			\ldots
			+
			\frac{\partial^ {\check p}}{\partial x^{\check p}}  u(x,t_n) (x_0-x)^{\check p} \\
			u(x_1, t_n)
			& = &
			u(x,t_n)
			+
			\frac{\partial}{\partial x} u(x,t_n) (x_1-x)
			+
			\ldots
			+
			\frac{\partial^ {\check p}}{\partial x^{\check p}} u(x,t_n) (x_1-x)^{\check p} \\
			\vdots          &   & \vdots\\
			u(x_{M-1}, t_n)
			& = &
			u(x,t_n)
			+
			\frac{\partial}{\partial x} u(x,t_n) (x_{M-1} - x)
			+
			\ldots
			+
			\frac{\partial^ {\check p}}{\partial x^{\check p}} u(x,t_n) (x_{M-1} - x)^{\check p}
		\end{array}
		\right.,
	\end{equation*}
	where $\check p$ is usually set as $\check p = p + 3$ to obtain asymptotic minimax efficiency \cite[see][]{fan1997local}.
	In the above system of equations, if we denote
	$$
	\myb(x)
	=
	\left(
	u(x,t_n),
	\frac{\partial}{\partial x}  u(x,t_n),
	\ldots,
	\frac{\partial^ {\check p}}{\partial x^{\check p}}  u(x,t_n)
	\right)^\top,
	$$
	then we can solve $\myb(x)$ through the optimization problem in \eqref{equ: solve derivative w.r.t x -- local polynomial} with a closed-form solution shown in \eqref{equ: solve derivative w.r.t x -- local polynomial (weighted LS)}:
	\begin{equation}
		\myb(x)
		=
		\left( \myX_{\text{spa}}^\top \myW_{\text{spa}} \myX_{\text{spa}} \right)^{-1}
		\myX_{\text{spa}}^\top \myW_{\text{spa}} \myu_{:}^n,
	\end{equation}
	where
	\begin{equation*}
		\myX_{\text{spa}}
		=
		\left[
		\begin{array}{cccc}
			1 & (x_0-x)     & \ldots & (x_0-x)^{\check p} \\
			1 & (x_1-x)     & \ldots & (x_1-x)^{\check p} \\
			\vdots & \vdots      & \ddots & \vdots   \\
			1 & (x_{M-1}-x) & \ldots & (x_{M-1}-x)^{\check p}
		\end{array}
		\right],
		\myu_{:}^n
		=
		\left[
		\begin{array}{c}
			u_0^n \\
			u_1^n \\
			\vdots \\
			u_{M-1}^n
		\end{array}
		\right]
	\end{equation*}
	and
	$
	\myW_{\text{spa}}
	=
	\text{diag}
	\left(
	\mathcal{K} \left( \frac{x_0 - x}{h} \right),
	\ldots,
	\mathcal{K} \left( \frac{x_{M-1} - x}{h} \right)
	\right)
	$.
	
	The main computational complexity to derive $\myb(x)$ lies in the computation of inverse of matrix
	$
	\myX_{\text{spa}}^\top \myW_{\text{spa}} \myX_{\text{spa}}
	\in
	\mathbb R^{(\check p + 1) \times (\check p + 1)},
	$
	where
	\begin{equation*}
		\begin{array}{ccl}
			& &
			\myX_{\text{spa}}^\top \myW_{\text{spa}} \myX_{\text{spa}} \\
			& = &
			\left[
			\begin{array}{cccccc}
				\sum\limits_{i=0}^{M-1} w_i
				& \sum\limits_{i=0}^{M-1} w_i(x_i-x)
				& \sum\limits_{i=0}^{M-1} w_i(x_i-x)^2
				& \ldots
				& \sum\limits_{i=0}^{M-1} w_i(x_i-x)^{\check p} \\
				\sum\limits_{i=0}^{M-1} w_i(x_i-x)
				& \sum\limits_{i=0}^{M-1} w_i(x_i-x)^2
				& \sum\limits_{i=0}^{M-1} w_i(x_i-x)^3
				& \ldots
				& \sum\limits_{i=0}^{M-1} w_i(x_i-x)^{\check p + 1} \\
				\sum\limits_{i=0}^{M-1} w_i(x_i-x)^2
				& \sum\limits_{i=0}^{M-1} w_i(x_i-x)^3
				& \sum\limits_{i=0}^{M-1} w_i(x_i-x)^4
				& \ldots
				& \sum\limits_{i=0}^{M-1} w_i(x_i-x)^{\check p + 2} \\
				\vdots
				& \vdots
				& \vdots
				& \ddots
				& \vdots \\
				\sum\limits_{i=0}^{M-1} w_i(x_i-x)^{\check p}
				& \sum\limits_{i=0}^{M-1} w_i(x_i-x)^{\check p + 1}
				& \sum\limits_{i=0}^{M-1} w_i(x_i-x)^{\check p + 2}
				& \ldots
				& \sum\limits_{i=0}^{M-1} w_i(x_i-x)^{2\check p}
			\end{array}
			\right]
		\end{array}
	\end{equation*}
	and
	\begin{equation*}
		\myX_{\text{spa}}^\top \myW_{\text{spa}} \myu_:^n
		=
		\left(
		\begin{array}{c}
			\sum\limits_{i=0}^{M-1} w_i u_i^n \\
			\sum\limits_{i=0}^{M-1} w_i(x_i-x) u_i^n \\
			\sum\limits_{i=0}^{M-1} w_i(x_i-x)^2 u_i^n \\
			\sum\limits_{i=0}^{M-1} w_i(x_i-x)^3 u_i^n \\
			\sum\limits_{i=0}^{M-1} w_i(x_i-x)^4 u_i^n
		\end{array}
		\right),
	\end{equation*}
	we know that for a fixed $n \in \{0, \ldots, N-1\}$ and $x \in \{x_0, \ldots, x_{M-1}\}$, the computational complexity of computing
	$
	\myX_{\text{spa}}^\top \myW_{\text{spa}} \myX_{\text{spa}}
	$
	and
	$
	\myX_{\text{spa}}^\top \myW_{\text{spa}} \myu_:^n
	$
	is $O(\check p^2 M).$
	Besides, the computational complexity to derive $( \myX_{\text{spa}}^\top \myW_{\text{spa}} \myX_{\text{spa}})^{-1}$ is $O(\check p^3)$.
	So we know that for a fixed $n \in \{0, \ldots, N-1\}$ and $x \in \{x_0, \ldots, x_{M-1}\}$, the computational complexity of computing
	$
	\frac{\partial^p}{\partial x^p} u(x_i,t_n)
	$
	is
	$
	\max\{ O(\check p^2 M), O(\check p^3) \}
	$
	with $\check p$ usually set as $\check p = p + 3$.
	Accordingly, the computational complexity of computing
	$
	\left\{
	\frac{\partial^p}{\partial x^p} u(x_i,t_n)
	\right\}_{i= 0,\ldots, M-1, n = 0, \ldots, N-1}
	$
	is
	$
	\max\{ O(\check p^2 M^2 N), O(\check p^3 MN) \}.
	$
	Because $p \leq q_{\max}$, we know that the computational complexity of computing all derivatives with respective to $x$ with highest order as $q_{\max}$ is
	$
	\max\{ O(q_{\max}^2 M^2 N), O(q_{\max}^3 MN) \}.
	$
	Similarly, the computational complexity of computing the first order derivatives with respective to $t$ is
	$
	\max\{ O( M N^2), O( MN) \}.
	$
	In conclusion, the computational complexity to derive all elements in matrix $\myX$ and vector $\nabla_t \myu$, including
	$$
	\left\{
	\widehat{ u(x_i, t_n)  },
	\widehat{ \frac{\partial}{\partial x} u(x_i, t_n) },
	\widehat{ \frac{\partial^2}{\partial x^2} u(x_i, t_n)  },
	\widehat{ \frac{\partial}{\partial t} u(x_i, t_n) }
	\right\}_{i = 0, \ldots, M-1, n = 0,\ldots, N-1}.
	$$
	by local polynomial regression in \eqref{equ: smoothing cubic spline -- objective fucntion} is
	$
	\max\{ O(q_{\max}^2 M^2 N), O(M N^2) , O(q_{\max}^3 MN) \},
	$
	where $q_{\max}$ is the highest order of derivatives desired in \eqref{equ: temporal evolutionary PDE}.
\end{proof}

\subsubsection{Proof of Lemma \ref{lemma: u_x and u_x hat difference bound}}
\label{proof: Bound of zero-order derivative}

\begin{proof}
	In this proof, we take $k=0$ as an illustration example, i.e., prove that when
	\begin{equation*}
		\begin{array}{ccll}
			\epsilon
			& > &
			\mathscr C(\sigma, \|u\|_{L^\infty(\Omega)}) \;
			\max&
			\left\{
			\frac{4 K_{\max} }{ M^{3/7} },
			4 A M^{-3/7},
			4 \sqrt{2} \frac{d^3}{dx^3} f^*(0) M^{-3/7},
			\frac{16 (C \log M + \gamma) \log(M)}{ M^{3/7}} ,
			\right. \\
			&  & &
			\left.
			16 \sqrt{\frac{\omega}{7}} \widetilde{C}(\sigma, \|u\|_{L^\infty(\Omega)})
			\frac{\sqrt{\log(M)}}{M^{3/7}}
			\right\},
		\end{array}
	\end{equation*}
	we have
	\begin{equation*}
		P
		\left[
		\sup_{x\in[0,X_{\max}]}
		\left|
		\widehat{ u(x,t_n)} -  u(x,t_n)
		\right|
		> \epsilon
		\right]
		<
		2 M e^{-\frac{M^{2/7}}{2\sigma^2}}
		+
		Q e^{-L\gamma}
		+
		4\sqrt{2} \eta^4 M^{-\frac{2}{7}\omega}
	\end{equation*}
	for a fixed $t_n$ with $n \in \{0,1, \ldots, N-1 \}$.
	For $k=1,2$, it can be derived with the same logic flow.
	
	Recall in Section \ref{sec: solve derivatives by cubic spline}, the fitted value of the smoothing cubic spline $s(x)$ is the minimizer of the optimization problem in \eqref{equ: smoothing cubic spline -- objective fucntion}.
	From Theorem A in \cite{silverman1984spline} (also mentioned by \cite{messer1991comparison} in the Section 1, and equation (2.2) in \cite{craven1978smoothing}) that when Condition \ref{assumption -- spline -- convergence cdf} - Condition \ref{assumption -- spline -- bounded pdf}  hold and for large $M$ and small $\widetilde\lambda = \frac{1-\alpha}{\alpha}$, we have
	\begin{equation*}
		\widehat{f_i}
		=
		\frac{1}{M \widetilde\lambda^{1/4}} \sum_{j=0}^{M-1} K \left(\frac{x_i - x_j}{ \widetilde\lambda^{1/4}} \right) u_j^n,
	\end{equation*}
	where $\widehat{f_i} = \widehat{u(x_i, t_n)}$, $\widetilde\lambda$ trades off the goodness-of-fit and smoothness of the cubic spline in \eqref{equ: smoothing cubic spline -- objective fucntion} and $K(\cdot)$ is a fixed kernel function defined as
	$$
	K(x)
	=
	\frac{1}{2}
	e^{-|x|/\sqrt{2}}
	\left[ \sin(|x|/\sqrt{2} + \pi/4) \right].
	$$
	
	For a general spatial variable $x$ and fixed $n \in \{0,1,\ldots,N-1 \}$, we denote
	$$
	f^*(x) = u(x,t_n),
	$$
	which is the ground truth of the underlying dynamic function $u(x,t_n)$ with $t_n$ fixed.
	Besides, we denote
	$
	\widehat f(x) = \widehat{u(x,t_n)},
	$
	which is an estimation of the ground truth of
	$
	f^*(x) = u(x,t_n)
	$
	with $t_n$ fixed.
	Accordingly to the above discussion, this estimation of $\widehat f(x)$ can be written as
	$$
	\widehat f(x)
	=
	\frac{1}{M \widetilde\lambda^{1/4}}
	\sum_{j=0}^{M-1}
	K \left(\frac{x - x_j}{ \widetilde\lambda^{1/4} } \right) u_j^n,
	$$
	where $\widehat f_i = \widehat{f}(x_i)$ for $ i \in \{0,1, \ldots, M-1 \}$
	
	In order to bound $P\left(\sup |\widehat f(x) - f^*(x)| > \epsilon\right)$ for a general $x$, we decompose it as follows:
	\begin{equation}
		\label{equ: proof-- decompose the error bound}
		\begin{array}{ccl}
			& &
			P
			\left(
			\sup |\widehat f(x) - f^*(x)| > \epsilon
			\right)\\
			& = &
			P
			\left(
			\sup |\widehat f(x) - \widehat f^B(x) + \widehat f^B(x) - \widehat f^*(x)| > \epsilon
			\right)\\
			& = &
			P
			\left(
			\sup |
			\widehat f(x)
			-
			\widehat f^B(x)
			-
			E(\widehat f(x) - \widehat f^B(x))
			+
			E(\widehat f(x) - \widehat f^B(x))
			+
			\widehat f^B(x) - \widehat f^*(x)
			| > \epsilon
			\right)\\
			& = &
			P
			\left(
			\sup |
			\underbrace{ \widehat f(x) - \widehat f^B(x)  }_{\mathcal{A} } -
			\underbrace{E(\widehat f(x) - \widehat f^B(x))  }_{\mathcal{B} } +
			\right.\\
			&  &
			\left.
			\underbrace{E(\widehat f(x) ) - f^*(x)  }_{\mathcal{C} } +
			\underbrace{\widehat f^B(x) - E(\widehat f^B(x))  }_{\mathcal{D} }
			| > \epsilon
			\right) \\
			& \leq &
			P\left(\sup |\mathcal{A} | > \frac{\epsilon}{4} \right) +
			P\left(\sup |\mathcal{B} | > \frac{\epsilon}{4} \right) +
			P\left(\sup |\mathcal{C} | > \frac{\epsilon}{4} \right) +
			P\left(\sup |\mathcal{D} | > \frac{\epsilon}{4} \right)
		\end{array},
	\end{equation}
	where  the $\widehat{f^B}(x)$ in (\ref{equ: proof-- decompose the error bound}) the truncated estimator defined as
	$$
	\widehat{f^B}(x)
	=
	\frac{1}{M \widetilde\lambda^{1/4}}
	\sum_{j=0}^{M-1} K
	\left(\frac{x - x_j}{\widetilde \lambda^{1/4}} \right) u_j^n \mathbbm{1}\{u_j^n < B_M\}.
	$$
	Here $\{ B_M \}$ is an increasing sequence and $B_M \to +\infty$ as $M \to +\infty$, i.e., $B_M = M^b$ with constant $b>0$, and we will discuss the value of $b$ at the end of this proof.
	
	In the remaining of the proof, we work on the upper bound of the four decomposed terms, i.e.,
	$
	P\left(\sup |\mathcal{A} | > \frac{\epsilon}{4} \right),
	P\left(\sup |\mathcal{B} | > \frac{\epsilon}{4} \right),
	P\left(\sup |\mathcal{C} | > \frac{\epsilon}{4} \right),
	P\left(\sup |\mathcal{D} | > \frac{\epsilon}{4} \right).
	$
	
	\bigskip
	\textbf{First, let us discuss the upper bound of $P\left(\sup |\mathcal{A} | > \frac{\epsilon}{4} \right)$.}
	
	Because
	\begin{eqnarray*}
		P\left(\sup |\mathcal{A} | > \frac{\epsilon}{4} \right)
		& = &
		P\left( \sup \left| \widehat f(x) - \widehat f^B(x)  \right| > \frac{\epsilon}{4} \right) \\
		& = &
		P
		\left(
		\sup \left| \frac{1}{M \widetilde\lambda^{1/4}}
		\sum_{j=0}^{M-1}
		K
		\left(
		\frac{x - x_j}{ \widetilde\lambda^{1/4}}
		\right)
		u_j^n \mathbbm{1}\{u_j^n \geq B_M\} \right| > \frac{\epsilon}{4}
		\right)\\
		& \leq &
		P
		\left(
		\sup
		\left|
		\frac{K_{\max}}{M \widetilde\lambda^{1/4}}
		\sum_{j=0}^{M-1}
		u_j^n
		\mathbbm{1}\{u_j^n \geq B_M\}
		\right| > \frac{\epsilon}{4}
		\right),\\
	\end{eqnarray*}
	where
	$
	K_{\max} = \max_{x \in [0, X_{\max}] \cup [0, T_{\max}] } K (x).
	$
	If we let $\frac{\epsilon}{4}  > \frac{K_{\max}}{M \widetilde\lambda^{1/4}} B_M$, then we have
	\begin{eqnarray*}
		P\left( \sup \left|\mathcal A  \right| > \frac{\epsilon}{4} \right)
		&\leq&
		P\left( \exists \; i=0,\ldots,M-1, s.t. \; |u_i^n| \geq B_M \right)\\
		&=&
		P\left( \max_{ i=0,\ldots,M-1} |u_i^n| \geq B_M \right). 
	\end{eqnarray*}
	Let $C_M = B_M -\|U\|_{L^\infty(\Omega)}$, where $U$ is the random variable generated from the unknown dynamic system, i.e., $U = u(x,t) + \epsilon$ with $\epsilon \sim N(0, \sigma^2)$.
	Then we have
	\begin{eqnarray*}
		P\left(\sup |\mathcal{A} | > \frac{\epsilon}{4} \right)
		& = &
		P\left( \sup \left| \widehat f(x) - \widehat f^B(x)  \right| > \frac{\epsilon}{4} \right) \\
		&\leq&
		P\left( \max_{ i=0,\ldots,M-1} |U_i^n - u_i^n| \geq C_M \right)\\
		&\leq&
		2Me^{-C_M^2/(2\sigma^2)}. 
	\end{eqnarray*}

	\bigskip
	\paragraph{Next, let us discuss the upper bound of $P\left(\sup |\mathcal{B} | > \frac{\epsilon}{4} \right)$.}
	
	\begin{eqnarray}
		\mathcal{B}
		& = & \nonumber
		E\left(|\widehat f(x) - \widehat f^B(x)| \right)  \\
		& = & \nonumber
		E\left(
		\left|
		\frac{1}{M \widetilde\lambda^{1/4}}
		\sum_{j=0}^{M-1}
		K
		\left(
		\frac{x - x_j}{\widetilde\lambda^{1/4}}
		\right)
		u_j^n
		\mathbbm{1}\{u_j^n \geq B_M\}
		\right| \right) \\
		& \leq & \nonumber
		E\left(
		\frac{1}{M \widetilde\lambda^{1/4}} \sum_{j=0}^{M-1}
		\left|
		K \left(\frac{x - x_j}{\widetilde\lambda^{1/4}} \right)
		\right| |u_j^n|  \mathbbm{1}\{u_j^n \geq B_M\}
		\right)  \\
		& = & \label{equ: proof -- bound u and u hat in part B empirical distribution}
		\frac{1}{\widetilde\lambda^{1/4}}
		\int \int_{|u| \geq B_M}
		\left|
		K \left(\frac{x - a}{\widetilde\lambda^{1/4}} \right)
		\right|
		|u|
		d F_M(a, u)     \\
		& \leq &   \label{equ: proof -- bound u and u hat in part B empirical distribution pdf}
		\int |K(\xi)| d\xi
		\times
		\underbrace{
			\sup_{\alpha} \int_{|u| \geq B_M} |u| f_M( \alpha, u) du
		}_{\mathcal V}. 
	\end{eqnarray}
	Here in \eqref{equ: proof -- bound u and u hat in part B empirical distribution}, $F_M(\cdot, \cdot)$ is the empirical c.d.f. of $(x, u)$'s, and in \eqref{equ: proof -- bound u and u hat in part B empirical distribution pdf}, $f_M(\cdot, \cdot)$ is the empirical p.d.f. of $(x, u)$'s.
	
	Now let us take a look at the upper bound of $\mathcal V$.
	For any $s>0$, we have
	\begin{eqnarray}
		\sup_{\alpha}
		\int_{|u| \geq B_M} \frac{|u|}{B_M} f_M(\alpha, u) du
		& \leq & \nonumber
		\sup_{\alpha}
		\int_{|u| \geq B_M} \left( \frac{|u|}{B_M} \right)^s f_M(\alpha, u) du \\
		& \leq & \nonumber
		\sup_{\alpha}
		\int \left( \frac{|u|}{B_M} \right)^s f_M(\alpha, u) du,
	\end{eqnarray}
	which gives
	$$
	\mathcal V
	:=
	\sup_{\alpha}
	\int_{|u| \geq B_M} |u| f_M(\alpha, u) du
	\leq
	B_M^{1-s}
	\underbrace{
		\sup_{\alpha}
		\int |u|^s f_M(\alpha, u) du
	}_{\pi_s}.
	$$
	From the lemma statement we know that when $s = 2$, we have
	$
	\pi_s:= \sup_{\alpha} \int |u|^s f_M(\alpha, u) du  < +\infty.
	$
	If we set
	$
	A = \pi_s \int |K(\xi)| d\xi,
	$
	then we have
	$$
	\mathcal B
	\leq A B_M^{1-s}. 
	$$
	So when $\frac{\epsilon}{4}  >  AB_M^{1-s}$ , we have
	$$
	P\left(\sup |\mathcal{B} | > \frac{\epsilon}{4} \right)
	=
	P\left( E\left( |\widehat f(x) - \widehat f^B(x)| \right)
	\geq
	\frac{\epsilon}{4}  \right)
	= 0.
	$$

	\bigskip
	\textbf{ Then, let us discuss the upper bound of $P\left(\sup |\mathcal{C} | > \frac{\epsilon}{4} \right)$.}
	According to Lemma 5 in \cite{rice1983smoothing}, when
	
	$f^*(x) \in C^4$,
	$
	\frac{d^2}{dx^2}f^*(x_0) = \frac{d^2}{dx^2}f^*(x_{M-1}) = 0
	$
	and
	$
	\frac{d^3}{dx^3}f^*(x_0) \neq 0,  \frac{d^3}{dx^3}f^*(x_{M-1}) = 0,
	$
	we have
	\begin{eqnarray*}
		&  &
		E(\widehat f(x)) - f^*(x)\\
		& = &
		\sqrt{2}
		\frac{d^3}{dx^3} f^*(0)
		\widetilde\lambda^{3/4}
		\exp\left( \frac{-x}{\sqrt{2}} \widetilde\lambda^{-1/4} \right)
		\cos\left( \frac{x}{\sqrt{2}} \widetilde\lambda^{-1/4} \right)
		+
		\ell(x), \\
	\end{eqnarray*}
	where the error term $\ell(x)$ satisfies
	$$
	\int [\ell(x)]^2 dx
	=
	o
	\left(
	\int
	\left[
	E(\widehat f(x)) - f^*(x)
	\right]^2
	dx
	\right)   .
	$$
	So when
	$
	\frac{\epsilon}{4}
	>
	\sqrt{2}
	\frac{d^3}{dx^3} f^*(0)
	\widetilde\lambda^{3/4}
	$
	and $M$ is sufficiently large then we have
	$$
	P\left( \sup |\mathcal C| > \frac{\epsilon}{4} \right) = 0. 
	$$

	\textbf{ Finally, let us discuss the upper bound of $P\left(\sup |\mathcal{D} | > \frac{\epsilon}{4} \right)$.}
	\bigskip
	
	In order to bound $P\left( \sup|D| > \frac{\epsilon}{4}\right)$, we further decompose $\mathcal D$ into two components, i.e.,
	$$
	\mathcal D
	:=
	\widehat f^B(x) - E(\widehat f^B(x))
	=
	e_M(x,t_n) +  \frac{1}{\sqrt{M}} \rho_M(x,t_n).
	$$
	The decomposition procedure and the definition of $e_M(x,t_n), \rho_M(x,t_n)$ are described in the following system of equations \cite[see][Proposition 2]{mack1982KernelRegression}:
	\begin{eqnarray}
		\mathcal{D}
		& = & \nonumber
		\widehat{f}^B(x) - E(\widehat{f}^B(x)) \\
		& = & \nonumber
		\frac{1}{M \widetilde\lambda^{1/4}}
		\sum_{j=0}^{M-1}
		K \left(\frac{x - x_j}{\widetilde\lambda^{1/4}} \right)
		u_j^n \mathbbm{1}\{u_j^n < B_M\}
		-\\
		& & \nonumber
		E
		\left(
		\frac{1}{M \widetilde\lambda^{1/4}}
		\sum_{j=0}^{M-1}
		K
		\left(
		\frac{x - x_j}{\widetilde\lambda^{1/4}}
		\right)
		u_j^n
		\mathbbm{1}\{u_j^n < B_M\}
		\right) \\
		& = & \label{equ: proof -- bound u - u hat in part D meaning of FM}
		\frac{1}{\sqrt{M} \widetilde\lambda^{1/4}}
		\int_{a \in \mathbb{R}}
		\int_{|u| < B_M}
		K
		\left(
		\frac{x - a}{\widetilde\lambda^{1/4} }
		\right)
		u\;
		d \underbrace{\left( \sqrt{M} (F_M(a,u) - F(a,u) )   \right) }_{Z_M(a,u)}\\
		& = & \nonumber
		\frac{1}{\sqrt{M} \widetilde\lambda^{1/4}}
		\int_{a\in \mathbb{R}}  K\left( \frac{x - a}{\widetilde\lambda^{1/4} }\right)
		\int_{|u| < B_M}
		u \;
		d (Z_M(a,u))  \\
		& = & \nonumber
		\frac{1}{\sqrt{M} \widetilde\lambda^{1/4}}
		\int_{a \in \mathbb{R}}
		K \left( \frac{x - a}{\widetilde\lambda^{1/4} }\right)
		\left[
		\int_{|u| < B_M} u \; d (Z_M(a,u) - B_0(T(a,u)))  +
		\right.\\
		& & \label{equ: proof -- bound u - u hat in part D meaning of B0}
		\left.
		\int_{|u| < B_M} u \; d B_0(T(a,u))
		\right] \\
		& = & \nonumber
		\underbrace{
			\frac{1}{ \sqrt{M} \widetilde\lambda^{1/4}}
			\int_{a \in \mathbb{R}}
			\int_{|u| < B_M}
			K\left( \frac{x - a}{\widetilde\lambda^{1/4} }\right)
			u\;
			d (Z_M(a,u) - B_0(T(a,u)))
		}_{e_M(x,t_n)}\\
		& & \nonumber
		+ \frac{1}{\sqrt{M}}
		\underbrace{
			\frac{1}{ \widetilde\lambda^{1/4}}
			\int_{a \in \mathbb{R}}
			\int_{|u| < B_M}
			K\left( \frac{x - a}{\widetilde\lambda^{1/4} }\right)
			u\;
			d B_0(T(a,u))
		}_{\rho_M(x,t_n)}.
	\end{eqnarray}
	In \eqref{equ: proof -- bound u - u hat in part D meaning of FM}, $F_M(\cdot, \cdot) := F_M(\cdot, \cdot | t_n)$ is the empirical c.d.f of $(x,u)$ with a fixed $t_n$, and
	$
	Z_M(a,u) = \sqrt{M}(F_M(a,u) - F(a,u))
	$ is a two-dimensional empirical process \cite[see][]{tusnady1977remark, mack1982KernelRegression}.
	In \eqref{equ: proof -- bound u - u hat in part D meaning of B0}, $B_0(T(a, u))$ is a sample path of two-dimensional Brownian bride.
	And $T(a, u): \mathbb R^2 \to [0,1]^2$ is the transformation defined by \cite{rosenblatt1952remarks}, i.e.,
	$
	T(a,u)
	=
	(F_A(x), F_{U|A}(u|a)),
	$
	where $F_A$ is the marginal c.d.f of $A$ and $F_{U|A}$ is the conditional c.d.f of $U$ given $A$ \cite[see][Proposition 2]{mack1982KernelRegression}.
	
	Through the above decomposition of $\mathcal D$, we have
	$$
	P\left( \sup|D| > \frac{\epsilon}{4}\right)
	\leq
	P\left( \sup|e_M(x,t_n)| > \frac{\epsilon}{8}\right)
	+
	P\left( \sup \frac{1}{\sqrt{M}} |\rho_M(x,t_n)| > \frac{\epsilon}{8}\right).
	$$

	For $e_M(x,t_n)$, we have
	\begin{eqnarray*}
		& &
		P\left( \sup |e_M(x,t_n)| > \frac{\epsilon}{8} \right) \\
		& = &
		P
		\left(
		\sup
		\left|
		\frac{1}{ \sqrt{M} \widetilde\lambda^{1/4}}
		\int_{a \in \mathbb{R}}
		\int_{|u| < B_M}
		K \left( \frac{x - a}{\widetilde\lambda^{1/4} }\right)
		u\;
		d (Z_M(a,u) - B_0(T(a,u)))
		\right|
		> \frac{\epsilon}{8} \right) \\
		& \leq &
		P
		\left(
		\frac{2 B_M K_{\max}}{ \sqrt{M} \widetilde\lambda^{1/4}}
		\sup_{a,u}
		\left| Z_M(a,u) - B_0(T(a,u)) \right|
		>
		\frac{\epsilon}{8}
		\right). \\
	\end{eqnarray*}
	Proved by Theorem 1 in \cite{tusnady1977remark}, we know that, for any $\gamma$, we have
	$$
	P
	\left(
	\sup_{a,u}
	\left|
	Z_M(a,u) - B_0(T(a,u))
	\right|
	>
	\frac{ (C \log M + \gamma) \log M  }{\sqrt{M}}
	\right)
	\leq
	Q e^{-L \gamma},
	$$
	where  $C,Q,L$ are absolute positive constants which is independent of temporal resolution $N$ and spatial resolution $M$.
	Thus, when
	$
	\frac{\epsilon}{8}
	\geq
	\frac{2B_M K_{\max}}{ \sqrt{M} \widetilde\lambda^{1/4}}  \frac{ (C \log M + \gamma) \log M  }{\sqrt{M}},
	$
	we have
	$$
	P\left(\sup |e_M(x,t_n)| > \frac{\epsilon}{8} \right) < Q e^{-L\gamma}.
	$$

	For $\rho_M(x,t_n)$, by equation (7) in \cite{mack1982KernelRegression}, we have
	\begin{equation*}
		\begin{array}{ccc}
			\frac{\widetilde\lambda^{1/8}\sup |\rho_M(x,t_n)| }{ \sqrt{  \log(1/\widetilde\lambda^{1/4}) }   }
			&\leq&
			\underbrace{
				16 (\log V)^{1/2}
				S^{1/2}
				\left(
				\log \left(\frac{1}{\widetilde\lambda^{1/4}} \right)
				\right)^{-1/2}
				\int |\xi|^{1/2} |dK(\xi)|
			}_{\mathcal W_{1,M}} + \\
			&&
			\underbrace{
				16\sqrt{2}
				\widetilde\lambda^{-1/8}
				\left(
				\log \left(\frac{1}{\widetilde\lambda^{1/4}} \right)
				\right)^{-1/2}
				\int q(S \widetilde\lambda^{1/4} |\tau|) |d(K(\tau))|
			}_{\mathcal W_{2,M}}, \\
		\end{array}
	\end{equation*}
	where $V$ is a random variable satisfying
	$
	E(V) \leq 4\sqrt{2} \eta^4
	$
	for
	$
	\eta^2 = \max_{i=0,\ldots,M-1, n = 0, \ldots, N-1} E(U_i^n)^2,
	$
	$
	S = \sup_x \int u^2 f(x, u) du
	$
	with $f(\cdot, \cdot)$ as the distribution function of $(x_i, u_i^n)$, and
	$
	q(z) = \int_{0}^{z} \frac{1}{2} \sqrt{\frac{1}{y} \log\left( \frac{1}{y} \right)} dy.
	$
	So we have the following system of equations:
	\begin{eqnarray}
		&   & \nonumber
		P\left( \sup \frac{1}{\sqrt{M}} |\rho_M(x,t_n)| > \frac{\epsilon}{8}\right) \\
		& = &  \nonumber
		P
		\left(
		\frac{\widetilde\lambda^{1/8} \sup |\rho_M(x,t_n)| }{\sqrt{\log(1/\widetilde\lambda^{1/4})}}
		>
		\frac{\sqrt{M} \widetilde\lambda^{1/8} \epsilon}{8 \sqrt{\log(1/\widetilde\lambda^{1/4})}}
		\right) \\
		& \leq &  \nonumber
		P
		\left(
		\mathcal W_{1,M} + \mathcal W_{2,M}
		>
		\frac{\sqrt{M} \widetilde\lambda^{1/8} \epsilon}{8 \sqrt{\log(1/\widetilde\lambda^{1/4})}}
		\right) \\
		& \leq &  \label{equ: proof - bound rho}
		P
		\left(
		\mathcal W_{1,M}
		\geq
		\frac{\sqrt{M} \widetilde\lambda^{1/8} \epsilon}{16 \sqrt{\log(1/\widetilde\lambda^{1/4})}}
		\right)
		+
		P
		\left(
		\mathcal W_{2,M}
		\geq
		\frac{\sqrt{M} \widetilde\lambda^{1/8} \epsilon}{16 \sqrt{\log(1/\widetilde\lambda^{1/4})}}
		\right)
	\end{eqnarray}
	
	Now let us bound
	$
	P
	\left(
	\mathcal W_{1,M}
	\geq
	\frac{\sqrt{M} \widetilde\lambda^{1/8} \epsilon}{16 \sqrt{\log(1/\widetilde\lambda^{1/4})}}
	\right),
	P
	\left(
	\mathcal W_{2,M}
	\geq
	\frac{\sqrt{M} \widetilde\lambda^{1/8} \epsilon}{16 \sqrt{\log(1/\widetilde\lambda^{1/4})}}
	\right)
	$
	in \eqref{equ: proof - bound rho} separately.
	
	\begin{enumerate}
		\item
		For the first term in \eqref{equ: proof - bound rho}, we have
		\begin{eqnarray}
			&& \nonumber
			P
			\left(
			\mathcal W_{1,M}
			\geq
			\frac{\sqrt{M} \widetilde\lambda^{1/8} \epsilon}{16 \sqrt{\log(1/\widetilde\lambda^{1/4})}}
			\right) \\
			& = & \nonumber
			P
			\left(
			16 (\log V)^{1/2}
			S^{1/2}
			\left(
			\log \left(\frac{1}{\widetilde\lambda^{1/4}} \right)
			\right)^{-1/2}
			\int |\xi|^{1/2} |dK(\xi)|
			\geq
			\frac{\sqrt{M} \widetilde\lambda^{1/8} \epsilon}{16 \sqrt{\log(1/\widetilde\lambda^{1/4})}}
			\right)  \\
			& = & \nonumber
			P
			\left(
			(\log V)^{1/2}
			\geq
			\frac{\sqrt{M} \widetilde\lambda^{1/8} \epsilon}{16^2 S^{1/2} \int |\xi|^{1/2} |dK(\xi)| }
			\right)\\
			& = & \nonumber
			P
			\left(
			\log V
			\geq
			\left(
			\frac{\sqrt{M} \widetilde\lambda^{1/8} \epsilon}{16^2 S^{1/2} \int |\xi|^{1/2} |dK(\xi)| }
			\right)^2
			\right)\\
			& = & \nonumber
			P
			\left(
			V
			\geq
			\exp
			\left[
			\left(
			\frac{\sqrt{M} \widetilde\lambda^{1/8} \epsilon}{16^2 S^{1/2} \int |\xi|^{1/2} |dK(\xi)| }
			\right)^2
			\right]
			\right)\\
			& \leq & \label{equ: proof -- W1 -- markov inequality}
			\frac{E(V)}{
				\exp
				\left[
				\left(
				\frac{\sqrt{M} \widetilde\lambda^{1/8} \epsilon}{16^2 S^{1/2} \int |\xi|^{1/2} |dK(\xi)| }
				\right)^2
				\right]}\\
			& \leq & \label{equ: proof -- W1 -- EV less than}
			\frac{4\sqrt{2} \eta^4}{
				\exp
				\left[
				\left(
				\frac{\sqrt{M} \widetilde\lambda^{1/8} \epsilon}{16^2 S^{1/2} \int |\xi|^{1/2} |dK(\xi)| }
				\right)^2
				\right]}\\
			& = & \label{equ: proof -- W1 -- plug epsilon}
			4\sqrt{2} \eta^4
			\widetilde\lambda^{\omega/4}
		\end{eqnarray}
		Here inequality \eqref{equ: proof -- W1 -- markov inequality} is due to Markov's inequality, and inequality \eqref{equ: proof -- W1 -- EV less than} is due to the fact that $E(V) \leq 4\sqrt{2} \eta^4$.
		Equality \eqref{equ: proof -- W1 -- plug epsilon} is because we set
		$
		\frac{\sqrt{M} \widetilde\lambda^{1/8} \epsilon}{16 \sqrt{\log(1/\widetilde\lambda^{1/4})}}
		=
		\sqrt{\omega} \widetilde{C} (t_n, \sigma, \|u\|_{L^\infty}(\Omega)),
		$
		where
		$$
		\widetilde{C} (t_n, \sigma, \|u\|_{L^\infty}(\Omega))
		:=
		16 \sqrt{S} \int |\xi|^{1/2} |dK(\xi)|
		$$
		and $\omega >1$ is an arbitrary scaler.
		
		\item
		For the second term of \eqref{equ: proof - bound rho}, it converges to
		$
		\widetilde{C} (t_n, \sigma, \|u\|_{L^\infty}(\Omega))
		$
		by using arguments similar to \cite{silverman1978weak} (page. 180-181) under the condition in Lemma \ref{lemma: u_x and u_x hat difference bound} that
		$
		\int \sqrt{|x \log(|x|) |} |dK(x)| < + \infty.
		$
		Here we add
		$
		(t_n, \sigma, \|u\|_{L^\infty (\Omega)})
		$
		after $\bar{C}$ to emphasize that the constant
		$
		\bar{C}(t_n, \sigma, \|u\|_{L^\infty (\Omega)})
		$
		is dependent on $t_n, \sigma, \|u\|_{L^\infty (\Omega)}$.
		
		It should be noted that
		\begin{equation}
			\label{equ: proof -- bound C bar}
			\widetilde{C} (t_n, \sigma, \|u\|_{L^\infty(\Omega)}) < + \infty,
		\end{equation}
		given the reasons listed as follows.
		First, it can be easily verified that the term $\int |\xi|^{1/2} |dK(\xi)|$ in $\widetilde{C} (t_n, \sigma, \|u\|_{L^\infty}(\Omega))$ is bounded.
		Second, for
		$
		S = \sup_x \int u^2 f(x,u) du,
		$
		it is also bounded.
		The reasons are described as follows.
		For a general $\varrho > 0$, we have
		\begin{eqnarray*}
			&&
			\sup_{x \in [0,X_{\max}]}
			\int |u|^\varrho f(x, u) du \\
			& = &
			\sup_{x \in [0,X_{\max}]}
			\int |u|^\varrho \frac{1}{\sqrt{2 \pi \sigma^2}}
			\exp\left( -\frac{(u - u(x,t_n))^2}{2\sigma^2} \right)
			du\\
			& = &
			\sup_{x \in [0,X_{\max}]}
			\frac{1}{\sqrt{2}}
			\sigma^2 2^{\varrho/2}
			\Gamma\left( \frac{1 + \varrho}{2} \right)
			G\left(
			-\frac{\varrho}{2},
			\frac{1}{2},
			-\frac{1}{2} \left( \frac{u(x, t_n)}{\sigma}\right)^2
			\right),
		\end{eqnarray*}
		where $G(a,b,z)$ is Kummer's confluent hypergeometric function of $z \in \mathbb C$ with parameters $a,b \in \mathbb C$ \cite[see][]{winkelbauer2012moments}.
		Because $G\left( -\frac{\varrho}{2}, \frac{1}{2}, \cdot \right)$ is an entire function for fixed parameters, we have
		\begin{eqnarray*}
			&&
			\sup_{x \in [0,X_{\max}]}
			\int |u|^\varrho f(x, u) du\\
			& \leq &
			\sup_{x \in [0,X_{\max}]}
			\frac{1}{\sqrt{2}}
			\sigma^2 2^{\varrho/2}
			\Gamma\left( \frac{1+\varrho}{2} \right)
			\sup_{
				z
				\in
				\left[
				- \frac{\max_{t \in \Omega} u^2(x,t)}{2\sigma^2},
				- \frac{\min_{t \in \Omega} u^2(x,t)}{2\sigma^2}
				\right]}
			G\left( -\frac{\varrho}{2}, \frac{1}{2}, z \right)\\
			&<&
			+\infty.
		\end{eqnarray*}
		So we can bound
		$
		\sup_{x \in [0,X_{\max}]}
		\int |u|^\varrho f(x, u) du
		$
		by a constant.
		If we take $\varrho = 2$, we can obtain
		$
		S = \sup_x \int u^2 f(x,u) du
		$
		bounded by a constant.
		So we can declare the statement in \eqref{equ: proof -- bound C bar}.
		
		We would also like to declare that there exist a constant
		$
		\widetilde{C}(\sigma, \|u\|_{L^\infty(\Omega)}) > 0
		$
		such that for any $N \geq 1$, we have
		$$
		\max_{n=0, \ldots, N-1}
		\widetilde{C}(t_n, \sigma, \|u\|_{L^\infty (\Omega)})
		\leq
		\widetilde{C}(\sigma, \|u\|_{L^\infty (\Omega)}),
		$$
		where $\bar{C}(\sigma, \|u\|_{L^\infty (\Omega)})$ is independent of $t_n, x_i, M, N$, and only depends on the noisy data $\mathcal D$ itself.
		
		From the above discussion, we learn that $\mathcal{W}_{2,M}$ converges to $\widetilde{C}(t_n, \sigma, \|u\|_{L^\infty (\Omega)})$, which can be bounded by $\widetilde{C}(\sigma, \|u\|_{L^\infty (\Omega)})$.
		If we set
		$
		\frac{\sqrt{M} \widetilde\lambda^{1/8} \epsilon}{16 \sqrt{\log(1/\widetilde\lambda^{1/4})}}
		>
		\sqrt\omega \widetilde{C}(\sigma, \|u\|_{L^\infty (\Omega)})
		$
		with $\omega >1$, then there exists a positive integer $\dot M(\omega)$ such that as long as $M > \dot M(\omega)$, we have
		$
		P
		\left(
		\mathcal W_{2,M}
		\geq
		\frac{\sqrt{M} \widetilde\lambda^{1/8} \epsilon}{16 \sqrt{\log(1/\widetilde\lambda^{1/4})}}
		\right)
		=
		0.
		$
		
		For the value of $\omega$, we set it as $\omega = M^{2r}$ with $r > 0$.
		And we will discuss the value of $r$ later.
		
	\end{enumerate}
	By combining
	$
	P
	\left(
	\mathcal W_{1,M}
	\geq
	\frac{\sqrt{M} \widetilde\lambda^{1/8} \epsilon}{16 \sqrt{\log(1/\widetilde\lambda^{1/4})}}
	\right)
	,
	P
	\left(
	\mathcal W_{2,M}
	\geq
	\frac{\sqrt{M} \widetilde\lambda^{1/8} \epsilon}{16 \sqrt{\log(1/\widetilde\lambda^{1/4})}}
	\right)
	$
	together, we have when
	$
	\frac{\epsilon}{16}
	>
	\sqrt\omega
	\widetilde{C}(\sigma, \|u\|_{L^\infty(\Omega)})
	\sqrt{\frac{\log(1/\widetilde\lambda^{1/4})}{M \widetilde\lambda^{1/4}}}
	$
	and
	$
	M > \dot M(\omega),
	$
	we have
	$$
	P\left( \sup \left| \frac{1}{\sqrt{M}}\rho_M(x, t_n) \right| > \frac{\epsilon}{8} \right)
	<
	4\sqrt{2} \eta^4
	\widetilde{\lambda}^{\omega/4}.
	$$

	By combining the discussion on
	$
	P\left(\sup |\mathcal{A} | > \frac{\epsilon}{4} \right),
	P\left(\sup |\mathcal{B} | > \frac{\epsilon}{4} \right),
	P\left(\sup |\mathcal{C} | > \frac{\epsilon}{4} \right),
	$
	and
	$
	P\left(\sup |\mathcal{D} | > \frac{\epsilon}{4} \right),
	$
	we can conclude that when
	\begin{itemize}
		\item $\frac{\epsilon}{4}  > \frac{K_{\max}}{M \widetilde\lambda^{1/4}} B_M $
		\item $\frac{\epsilon}{4}  > AB_M^{1-s}  \;\; (s=2)$
		\item $\frac{\epsilon}{4} > \sqrt{2} \frac{d^3}{dx^3} f^*(0) \widetilde\lambda^{3/4}  $
		\item $\frac{\epsilon}{8} > \frac{2  B_M K_{\max} (C \log M +\gamma) \log M}{\widetilde\lambda^{1/4} M}$
		\item $\frac{\epsilon}{16} >  \sqrt\omega \widetilde{C}(\sigma, \|u\|_{L^\infty(\Omega)})  \sqrt{\frac{ \log(1/\widetilde\lambda^{1/4})}{M\widetilde\lambda^{1/4}}}$
	\end{itemize}
	we have
	\begin{equation}\label{equ: proof -- bound ABCD by Z1 Z2 Z3}
		P\left( \sup | \mathcal{A+B+C+D}| >  \epsilon \right)
		<
		\underbrace{
			2M e^{-\frac{C_M^2}{2\sigma^2}}
		}_{\mathcal Z_1}
		+
		\underbrace{
			Qe^{-L\gamma}
		}_{\mathcal Z_2}
		+
		\underbrace{
			4\sqrt{2} \eta^4 \widetilde{\lambda}^{\omega/4}
		}_{\mathcal Z_3}.
	\end{equation}
	
	Let
	\begin{equation*}
		\left\{
		\begin{array}{lll}
			E_1 &=& \frac{4 K_{\max} }{M \widetilde\lambda^{1/4}} B_M \\
			E_2 &=& 4 AB_M^{1-s} \\
			E_3 &=& 4 \sqrt{2} \frac{d^3}{dx^3} f^*(0) \widetilde\lambda^{3/4} \\
			E_4 &=& \frac{16  B_M K_{\max} (C \log M +\gamma) \log M}{\widetilde\lambda^{1/4} M} \\
			E_5 &=& 16\sqrt\omega \widetilde{C}(\sigma, \|u\|_{L^\infty(\Omega)})
			\sqrt{\frac{ \log(1/\widetilde\lambda^{1/4})}{M \widetilde\lambda^{1/4}}}
		\end{array}
		\right.,
	\end{equation*}
	by setting $\widetilde\lambda = M^{-a}, B_M = M^b$ with $a,b>0$,  we have
	\begin{equation*}
		\left\{
		\begin{array}{ccl}
			E_1
			& = &
			\frac{4 K_{\max} }{M \widetilde\lambda^{1/4}}B_M
			=
			\frac{4 K_{\max} }{M ^{1-a/4-b}} \\
			E_2
			& = &
			4  AB_M^{1-s}  = 4  A \frac{1}{M^{b(s-1)}}\\
			E_3
			& = &
			4 \sqrt{2} \frac{d}{dx} f^*(0) \widetilde\lambda^{3/4}
			=
			4 \sqrt{2} \frac{d}{dx} f^*(0) M^{-3a/4}  \\
			E_4
			& = &
			\frac{16  B_M K_{\max} (C \log M +\gamma) \log M}{\widetilde\lambda^{1/4} M}
			=
			\frac{16 K_{\max} (C \log M + \gamma) \log(M)}{ M^{1 - a/4 - b}} \\
			E_5
			& = &
			16  \sqrt\omega \widetilde{C}(\sigma, \|u\|_{L^\infty(\Omega)})
			\sqrt{\frac{ \log(1/\widetilde\lambda^{1/4})}{M \widetilde\lambda^{1/4}}}
			=
			8 \sqrt{a \omega}
			\widetilde{C}(\sigma, \|u\|_{L^\infty(\Omega)})
			\sqrt{\frac{ \log(M)}{M ^{1-a/4}}}.\\
		\end{array}
		\right.
	\end{equation*}
	To guarantee that $E_1,E_2,E_3,E_4, E_5 \to 0$ as $M \to +\infty$, we can set
	\begin{equation*}
		\left\{
		\begin{array}{l}
			1 - a/4 - b = 3a/4  \\
			b(s-1) > 0 \\
			\frac{1}{2}(1 - a/4) = 3a/4 \\
			a,b > 0\\
			s = 2
		\end{array}
		\right..
	\end{equation*}
	then we have
	\begin{equation*}
		\left\{
		\begin{array}{l}
			a = 4/7 \\
			b = 3/7 \\
			s = 2
		\end{array}
		\right..
	\end{equation*}
	Accordingly, we have
	\begin{equation*}
		\left\{
		\begin{array}{lll}
			E_1 = \frac{4 K_{\max} }{ M^{3/7} } \\
			E_2 = 4 A M^{-3/7} \\
			E_3 = 4 \sqrt{2} \frac{d^3}{dx^3} f^*(0) M^{-3/7} \\
			E_4 = \frac{16 K_{\max} (C \log M + \gamma) \log(M)}{ M^{3/7}} \\
			E_5 = 16 \sqrt{\frac{\omega}{7}} \widetilde{C}(\sigma, \|u\|_{L^\infty(\Omega)})
			\frac{\sqrt{\log(M)}}{M^{3/7}}
		\end{array}
		\right.,
	\end{equation*}
	where
	$$
	E_1 ,
	E_2 ,
	E_3 ,
	E_5 \lesssim E_4
	$$
	as $M \to +\infty$.
	Here, the operator $\lesssim$ means that when $M \to +\infty$, the order of the left side hand of $\lesssim$ will be much smaller than that on the right side hand.
	So we can declare that when $M$ is sufficiently large and
	\begin{equation*}
		\begin{array}{cccl}
			\epsilon
			& > &
			\max&
			\left\{
			\frac{4 K_{\max} }{ M^{3/7} },
			4 A M^{-3/7},
			4 \sqrt{2} \frac{d^3}{dx^3} f^*(0) M^{-3/7},
			\frac{16 K_{\max} (C \log M + \gamma) \log(M)}{ M^{3/7}} ,
			\right. \\
			&  & &
			\left.
			16 \sqrt{\frac{\omega}{7}} \widetilde{C}(\sigma, \|u\|_{L^\infty(\Omega)})
			\frac{\sqrt{\log(M)}}{M^{3/7}}
			\right\}, 
		\end{array}
	\end{equation*}
	we have
	\begin{eqnarray*}
		P\left( \sup | \mathcal{A+B+C+D}| >  \epsilon \right)
		& \leq &
		2M e^{-\frac{C_M^2}{2\sigma^2}}
		+
		Qe^{-L\gamma}
		+
		4\sqrt{2} \eta^4 \widetilde{\lambda}^{\omega/4}\\
		& = &
		2M e^{-\frac{( M^{3/7} - \| U \|_{L^{\infty}(\Omega)})^2 }{2\sigma^2}}
		+
		Q e^{-L \gamma}
		+
		4\sqrt{2} \eta^4 M^{-\omega/7}. 
	\end{eqnarray*}
\end{proof}

\subsubsection{Proof of Lemma 	C.2}
\label{proof: bound of tau}

\begin{proof}
	
	For the estimation error $\| \nabla_t \myu - \myX \mybeta^* \|_\infty$, we have
	\begin{eqnarray}
		\| \nabla_t \myu - \myX \mybeta^* \|_\infty
		& = &    \nonumber
		\| \nabla_t \myu - \nabla_t \myu^* + \nabla_t \myu^* - \myX \mybeta^* \|_\infty   \\
		& = &    \nonumber
		\| \nabla_t \myu - \nabla_t \myu^* + \myX^*\mybeta^* - \myX \mybeta^* \|_\infty   \\
		& \leq & \label{equ: proof -- bound of ut - X beta^* decomposition}
		\| \nabla_t \myu - \nabla_t \myu^* \|_\infty + \| (\myX^*- \myX) \mybeta^* \|_\infty.
	\end{eqnarray}
	So accordingly, we have
	\begin{eqnarray}
		P\left(
		\| \nabla_t \myu - \myX \mybeta^* \|_\infty
		>
		\epsilon
		\right)
		& \leq &    \nonumber
		P\left( \| \nabla_t \myu - \nabla_t \myu^* \|_\infty  > \frac{\epsilon}{2} \right)
		+
		P\left( \| (\myX^*- \myX) \mybeta^* \|_\infty \right).
	\end{eqnarray}
	
	In the remaining of the proof, we will discuss the bound of
	$
	P\left( \| \nabla_t \myu - \nabla_t \myu^* \|_\infty  > \frac{\epsilon}{2} \right)
	$
	and
	$
	P\left( \| (\myX^*- \myX) \mybeta^* \|_\infty \right)
	$
	separately.
	
	\begin{itemize}
		\item First let us discuss the bound of
		$
		P\left( \| \nabla_t \myu - \nabla_t \myu^* \|_\infty  > \frac{\epsilon}{2} \right).
		$
		Because
		\begin{eqnarray}
			\label{equ: proof -- support set recovary -- bound ut}
			P
			\left(
			\| \nabla_t \myu - \nabla_t \myu^* \|_\infty > \frac{\epsilon}{2}
			\right)
			& \leq &  \nonumber
			P
			\left(
			\max\limits_{i=0,\ldots,M-1}
			\sup\limits_{t \in [0,T_{\max}]}
			\left| \widehat{ \frac{\partial}{\partial t} u(x_i,t)} - \frac{\partial}{\partial t} u(x_i,t) \right|
			>
			\frac{\epsilon}{2}
			\right)  \\
			& \leq &  \nonumber
			\sum\limits_{i=0}^{M-1}
			P
			\left(
			\sup\limits_{t \in [0,T_{\max}]}
			\left| \widehat{ \frac{\partial}{\partial t} u(x_i,t)} - \frac{\partial}{\partial t} u(x_i,t) \right|
			>
			\frac{\epsilon}{2}
			\right) ,
		\end{eqnarray}
		if we set
		\begin{equation}
			\label{equ: proof -- epsilon/2 with response vector}
			\begin{array}{cccl}
				\frac{\epsilon}{2}
				& > &
				\mathscr C_{ (\sigma, \|u\|_{L^\infty(\Omega)})}
				\max&
				\left\{
				4 K_{\max} N^{-3/7},
				4 \bar A N^{-3/7},
				4 \sqrt{2} \frac{d^3}{dx^3} \bar f^*(0) N^{-3/7},
				\right. \\
				&  & &
				\frac{
					16 K_{\max}
					\left[
					C_{ (\sigma, \|u\|_{L^\infty(\Omega)})} \log (N)
					+
					\gamma_{ (N) }
					\right]
					\log(N)}{ N^{3/7}} ,\\
				& & &
				\left.
				16 \sqrt{\frac{\omega_{ (N) } }{7}}
				\widetilde{C}_{ (\sigma, \|u\|_{L^\infty(\Omega)})}
				\frac{\sqrt{\log(N)}}{N^{3/7}}
				\right\},
			\end{array}
		\end{equation}
		then we have
		\begin{eqnarray}
			&   & \nonumber
			P
			\left(
			\| \nabla_t \myu - \nabla_t \myu^* \|_\infty > \frac{\epsilon}{2}
			\right)\\
			& \leq & \label{equ: proof -- bound of ut - X beta^* bound1}
			M
			\left[
			2 N e^{-\frac{( N^{3/7} - \| U \|_{L^{\infty}(\Omega)})^2 }{2\sigma^2}}
			+
			Q_{ (\sigma, \|u\|_{L^\infty(\Omega)})}
			e^{-L \gamma_{ (N) } }
			+
			4\sqrt{2} \eta^4 N^{-\omega_{ (N) }/7}
			\right],
		\end{eqnarray}
		where inequity \eqref{equ: proof -- bound of ut - X beta^* bound1} is derived according to Corollary \ref{lemma: u_t and u_t hat difference bound}.

		\item Second, let us discuss the bound of
		$
		P\left( \| (\myX^*- \myX) \mybeta^* \|_\infty \right) .
		$
		Because
		\begin{eqnarray}
			&& \nonumber
			P
			\left(
			\| (\myX^*- \myX) \mybeta^* \|_\infty
			>
			\frac{\epsilon}{2}
			\right)\\
			& \leq & \nonumber
			P
			\left(
			\| \mybeta^*\|_\infty
			\max\limits_{n = 0,\ldots, N-1}
			\sup_{x \in [0,X_{\max}]}
			\sum_{k=1}^{K}
			\| (\myX_k^*(x,t_n)- \myX_k(x,t_n)) \|_\infty
			>
			\frac{\epsilon}{2}
			\right) \\
			& = &  \nonumber
			P
			\left(
			\max\limits_{n = 0,\ldots, N-1}
			\sup_{x \in [0,X_{\max}]} \sum_{k=1}^{K}
			\| (\myX_k^*(x,t_n)- \myX_k(x,t_n)) \|_\infty
			>
			\frac{\epsilon}{2 \| \mybeta^*\|_\infty}
			\right) \\
			& \leq &  \nonumber
			\sum_{n=0}^{N-1}
			\sum_{k=1}^{K}
			P
			\left(
			\sup_{x \in [0,X_{\max}]}
			\| (\myX_k^*(x,t_n)- \myX_k(x,t_n)) \|_\infty > \frac{\epsilon}{2 K \| \mybeta^*\|_\infty}
			\right),
		\end{eqnarray}
		if we set
		\begin{equation}
			\label{equ: proof -- epsilon/2 with design matrix}
			\begin{array}{cccl}
				\frac{\epsilon}{2K \| \mybeta^* \|_{\infty}}
				& > &
				\mathscr C_{ (\sigma, \|u\|_{L^\infty(\Omega)}) }
				\max&
				\left\{
				4 K_{\max} M^{-3/7} ,
				4 A M^{-3/7},
				4 \sqrt{2} \frac{d^3}{dx^3} f^*(0) M^{-3/7},
				\right. \\
				& & &
				\frac{
					16
					\left[ C_{ (\sigma, \|u\|_{L^\infty(\Omega)}) } \log M + \gamma_{(M)}\right]
					\log(M)
				}{ M^{3/7}} ,\\
				&  & &
				\left.
				16 \sqrt{\frac{\omega_{(M)}}{7}}
				\widetilde{C}(\sigma, \|u\|_{L^\infty(\Omega_{(M)})})
				\frac{\sqrt{\log(M)}}{M^{3/7}}
				\right\},
			\end{array}
		\end{equation}
		then we have
		\begin{eqnarray}
			&& \nonumber
			P
			\left(
			\| (\myX^*- \myX) \mybeta^* \|_\infty
			>
			\frac{\epsilon}{2}
			\right)\\
			& \leq & \label{equ: proof -- bound of ut - X beta^* bound2}
			NK
			\left[
			2M e^{-\frac{( M^{3/7} - \| U \|_{L^{\infty}(\Omega)})^2 }{2\sigma^2}}
			+
			Q_{ (\sigma, \|u\|_{L^\infty(\Omega)}) } e^{-L \gamma_{(M)}}
			+
			4\sqrt{2} \eta^4 M^{-\omega_{(M)}/7}
			\right].
		\end{eqnarray}
		Inequality \eqref{equ: proof -- bound of ut - X beta^* bound2} is derived by Lemma \ref{lemma: u_x and u_x hat difference bound}.
	\end{itemize}
	
	By combining the results in \eqref{equ: proof -- epsilon/2 with response vector}, \eqref{equ: proof -- bound of ut - X beta^* bound1}, \eqref{equ: proof -- epsilon/2 with design matrix}, \eqref{equ: proof -- bound of ut - X beta^* bound2}, we have that when
	\begin{equation*}
		\begin{array}{ccll}
			\frac{\epsilon}{2}
			& > &
			\mathscr C_{ (\sigma, \|u\|_{L^\infty(\Omega)}) }
			\;\max&
			\left\{
			4 K_{\max} M^{-3/7},
			4 K K_{\max} \| \mybeta^* \|_{\infty} N^{-3/7},
			\right.\\
			& & &
			\left.
			4 A M^{-3/7},
			4 K \| \mybeta^* \|_{\infty}
			\bar A N^{-3/7},
			\right. \\
			& & &
			\left.
			4 \sqrt{2} \frac{d^3}{dx^3} f^*(0) M^{-3/7},
			4 \sqrt{2} K \| \mybeta^* \|_{\infty} \frac{d^3}{dx^3} \bar f^*(0) N^{-3/7},
			\right. \\
			& & &
			\frac{
				16 K K_{\max}
				\| \mybeta^* \|_{\infty}
				\left[ C_{ (\sigma, \|u\|_{L^\infty(\Omega)}) } \log(M) + \gamma_{ (M) } \right]
				\log(M)
			}{ M^{3/7}},\\
			& & &
			\frac{
				16 K_{\max}
				\left[ C_{ (\sigma, \|u\|_{L^\infty(\Omega)}) } \log(N) + \gamma_{(N)} \right]
				\log(N)
			}{ N^{3/7}}, \\
			& & &
			16 \sqrt{\frac{\omega_{(M)}}{7}}
			\widetilde{C}_{ (\sigma, \|u\|_{L^\infty(\Omega)}) }
			\frac{\sqrt{\log(M)}}{M^{3/7}}, \\
			& & &
			\left.
			16 K \| \mybeta^* \|_{\infty}
			\sqrt{\frac{\omega_{(N)}}{7}}
			\widetilde{C}_{ (\sigma, \|u\|_{L^\infty(\Omega)}) }
			\frac{\sqrt{\log(N)}}{N^{3/7}}
			\right\},
		\end{array}
	\end{equation*}
	we have
	\begin{eqnarray*}
		&  &
		P\left(
		\| \nabla_t \myu - \myX \mybeta^* \|_\infty
		>
		\epsilon
		\right)\\
		& \leq &
		M
		\left[
		2 N e^{-\frac{( N^{3/7} - \| U \|_{L^{\infty}(\Omega)})^2 }{2\sigma^2}}
		+
		Q_{ (\sigma, \|u\|_{L^\infty(\Omega)})}
		e^{-L \gamma_{ (N) } }
		+
		4\sqrt{2} \eta^4 N^{-\omega_{ (N) }/7}
		\right] + \\
		&&
		NK
		\left[
		2M e^{-\frac{( M^{3/7} - \| U \|_{L^{\infty}(\Omega)})^2 }{2\sigma^2}}
		+
		Q_{ (\sigma, \|u\|_{L^\infty(\Omega)}) } e^{-L \gamma_{(M)}}
		+
		4\sqrt{2} \eta^4 M^{-\omega_{(M)}/7}
		\right]
	\end{eqnarray*}
	
	Now, let us do some simplification of the above results.
	Let $M = N^{\kappa}, \gamma_{(M)} = \gamma_{(N)} = \frac{1}{L} N^r, \omega_{(M)} = \omega_{(N)} = N^{2r}$, and
	\begin{equation*}
		\left\{
		\begin{array}{ccl}
			\mathcal J_1
			& = & 4
			K K_{\max} \| \mybeta^* \|_{\infty} N^{-3\kappa/7}  \\
			\mathcal J'_1
			& = &
			4 K_{\max} N^{ -3/7}  \\
			\mathcal J_2
			& = &
			4 A K \| \mybeta^* \|_{\infty} N^{-3 \kappa/7} \\
			\mathcal J'_2
			& = &
			4 \bar A N^{-3/7}\\
			\mathcal J_3
			& = &
			4 \sqrt{2} K \| \mybeta^* \|_{\infty} \frac{d^3}{dx^3} f^*(0) N^{-3 \kappa / 7} \\
			\mathcal J'_3
			& = &
			4 \sqrt{2} \frac{d^3}{dx^3} \bar f^*(0) N^{-3/7} \\
			\mathcal J_4
			& = &
			\frac{
				16 K K_{\max}
				\| \mybeta^* \|_{\infty}
				\left[
				C_{(\sigma, \|u\|_{L^\infty(\Omega)})} (\log(\kappa)
				+
				\log(N))
				+
				N^{r}/L
				\right]
				(\log(\kappa) + \log(N) )}{ N^{3 \kappa / 7}} \\
			\mathcal J'_4
			& = &
			\frac{
				16 K_{\max}
				\left[
				C_{(\sigma, \|u\|_{L^\infty(\Omega)})} \log(N)
				+
				N^{r}
				\right]
				\log(N)
			}{ N^{3/7}}  \\
			\mathcal J_5
			& = &
			16 K \| \mybeta^* \|_{\infty}
			\sqrt{\frac{N^{2r}}{7}} \widetilde{C}_{ (\sigma, \|u\|_{L^\infty(\Omega)})}
			\frac{\sqrt{\log(\kappa) + \log(N)}}{ N^{3\kappa/7}} \\
			\mathcal J'_5 & = & 16 \sqrt{\frac{N^{2r}}{7}} \widetilde{C}_{(\sigma, \|u\|_{L^\infty(\Omega)})}
			\frac{\sqrt{\log(N)}}{N^{3/7}}
		\end{array}
		\right..
	\end{equation*}
	To guarantee that
	$
	\mathcal J_1, \mathcal J'_1,
	\mathcal J_2, \mathcal J'_2,
	\mathcal J_3, \mathcal J'_3,
	\mathcal J_4, \mathcal J'_4,
	\mathcal J_5, \mathcal J'_5
	\to 0,
	$
	as $N \to +\infty$, we need
	\begin{equation*}
		\left\{
		\begin{array}{l}
			3\kappa/7 - r > 0 \\
			3/7 - r > 0  \\
		\end{array}
		\right.,
	\end{equation*}
	where the optimal $\kappa$ is $\kappa = 1$.
	Accordingly, we have
	$$
	\mathcal J_1, \mathcal J'_1,
	\mathcal J_2, \mathcal J'_2,
	\mathcal J_3, \mathcal J'_3,
	\mathcal J_5, \mathcal J'_5
	\lesssim
	\mathcal J_4, \mathcal J'_4.
	$$
	
	Based on the above discussion, we can declare that when $N$ is sufficiently large, with
	$$
	\epsilon
	>
	\mathscr C_{ (\sigma, \|u\|_{L^\infty(\Omega)}) }
	\frac{\log(N)}{N^{3/7 - r}}
	$$
	for any $r \in \left(0, \frac{3}{7} \right)$ and $M = O(N)$, we have
	\begin{eqnarray*}
		&&
		P\left \| \nabla_t \myu - \myX \mybeta^* \|_\infty > \epsilon \right) \\
		& \leq &
		M
		\left[
		2 N e^{-\frac{( N^{3/7} - \| U \|_{L^{\infty}(\Omega)})^2 }{2\sigma^2}}
		+
		Q_{ (\sigma, \|u\|_{L^\infty(\Omega)})}
		e^{-L \gamma_{ (N) } }
		+
		4\sqrt{2} \eta^4 N^{-\omega_{ (N) }/7}
		\right] + \\
		&&
		NK
		\left[
		2M e^{-\frac{( M^{3/7} - \| U \|_{L^{\infty}(\Omega)})^2 }{2\sigma^2}}
		+
		Q_{ (\sigma, \|u\|_{L^\infty(\Omega)}) } e^{-L \gamma_{(M)}}
		+
		4\sqrt{2} \eta^4 M^{-\omega_{(M)}/7}
		\right]\\
		& = &
		M
		\left[
		2 N e^{-\frac{( N^{3/7} - \| U \|_{L^{\infty}(\Omega)})^2 }{2\sigma^2}}
		+
		Q_{ (\sigma, \|u\|_{L^\infty(\Omega)})}
		e^{- N^r }
		+
		4\sqrt{2} \eta^4 N^{- N^{2r}/7}
		\right] + \\
		&&
		NK
		\left[
		2M e^{-\frac{( M^{3/7} - \| U \|_{L^{\infty}(\Omega)})^2 }{2\sigma^2}}
		+
		Q_{ (\sigma, \|u\|_{L^\infty(\Omega)}) } e^{-N^r}
		+
		4\sqrt{2} \eta^4 M^{- N^{2r}/7}
		\right]\\
		& = &
		O(N e^{-N^r}). 
	\end{eqnarray*}
	Thus, we finish the proof of the theorem.
\end{proof}

\subsubsection{Proof of Theorem \ref{theo: support set recovary}}
\label{proof: Support Set Recovery}
\begin{proof}
	By KKT-condition, any minimizer $\mybeta$ of \eqref{equ: LASSO model - matrix algebra} must satisfies:
	$$
	-\frac{1}{MN} \myX^\top(\nabla_t \myu - \myX \mybeta) + \lambda \myz = 0 \;\; \text{for} \;\; \myz \in \partial\|\mybeta\|_1,
	$$
	where $\partial\|\mybeta\|_1$ is the sub-differential of $\|\mybeta\|_1$.
	The above equation can be equivalently transformed into
	\begin{equation}
		\label{equ: proof -- first order condition of LASSO}
		\myX^\top \myX(\mybeta - \mybeta^*)
		+
		\myX^\top \left[ (\myX - \myX^*)\mybeta^* - (\nabla_t \myu - \nabla_t \myu^*) \right]
		+
		\lambda MN \myz
		=
		0.
	\end{equation}
	Here matrix $\myX \in \mathbb R^{MN \times K}$ is defined in \eqref{equ: estimated design matrix}, and matrix $\myX^* \in \mathbb R^{MN \times K}$ is defined as
	\begin{equation*}
		\begin{array}{ccccccccc}
			\myX^*
			& = &
			\left( \right.
			\myx_0^0 &
			\myx_1^0 &
			\ldots &
			\myx_{M-1}^0 &
			\myx_1^0 &
			\ldots &
			\myx_{M-1}^{N-1}
			\left. \right)^\top,
		\end{array}
	\end{equation*}
	with
	\begin{equation*}
		\begin{array}{ccc}
			\myx_i^n
			& = &
			\left(
			\begin{array}{cccccccccc}
				1, &
				u(x_i,t_n)  ,&
				\frac{\partial u(x_i, t_n)}{\partial x} ,&
				\frac{\partial^2 u(x_i, t_n)}{\partial x^2}, &
				\left( \widehat{u(x_i,t_n)} \right)^2 ,&
				\ldots,
				\left( \frac{\partial^2 u(x_i, t_n)}{\partial x^2} \right)^{p_{\max}}
			\end{array}
			\right)^\top
			\in \mathbb R^K.
		\end{array}
	\end{equation*}
	And vector $\mybeta^* = (\beta_1, \ldots, \beta_K) \in \mathbb R^{K}$ is the ground truth coefficients. Besides, vector $\nabla_t \myu \in \mathbb R^{MN}$ is defined in \eqref{equ: estimated response vector}, and vector $\nabla_t \myu^* \in \mathbb R^{K}$ is the ground truth, i.e.,
	\begin{equation*}
		\nabla_t \myu^*
		=
		\left(
		\begin{array}{cccccccccccc}
			\frac{\partial u(x_0, t_0)}{\partial t}, &
			\frac{\partial u(x_1, t_0)}{\partial t}, &
			\ldots, &
			\frac{\partial u(x_{M-1}, t_0)}{\partial t}, &
			\frac{\partial u(x_0, t_1)}{\partial t}, &
			\ldots,&
			\frac{\partial u(x_{M-1}, t_{N-1})}{\partial t}
		\end{array}
		\right)^\top.
	\end{equation*}
	
	Let us denote $\mS = \{i: \mybeta^*_i \neq 0 \;\forall\; i = 0, 1, \ldots, K\}$, then we can decompose $\myX$ into $\myX_{\mS}$ and $\myX_{\mS^c}$, where $\myX_{\mS}$ is the columns of $\myX$ whose indices are in $\mS$ and $\myX_{\mS^c}$ is the complement of $\myX_{\mS}$.
	And we can also decompose $\mybeta$ into $\mybeta_{\mS}$ and $\mybeta_{\mS^c}$, where $\mybeta_{\mS}$ is the subvector of $\mybeta$ only contains elements whose indices are in $\mS$ and $\mybeta_{\mS^c}$ is the complement of $\mybeta_{\mS}$.
	
	By using the decomposition, we can rewrite \eqref{equ: proof -- first order condition of LASSO} as
	\begin{equation}
		\label{equ: proof -- first order condition of LASSO after decomposition}
		\begin{array}{ccl}
			\begin{pmatrix}
				\mathbf{0} \\
				\mathbf{0}
			\end{pmatrix}
			& = &
			\begin{pmatrix}
				\myX_\mS^\top \myX_\mS & \myX_\mS^\top \myX_{\mS^c}\\
				\myX_{\mS^c}^\top \myX_{\mS} & \myX_{\mS^c}^\top \myX_{\mS^c}
			\end{pmatrix}
			\begin{pmatrix}
				\mybeta_{\mS} -\mybeta^*_{\mS} \\
				\mybeta_{\mS^c}
			\end{pmatrix}
			+
			\begin{pmatrix}
				\myX^\top_\mS \\
				\myX^\top_{\mS^c}
			\end{pmatrix}
			\left[ (\myX - \myX^*)_\mS \mybeta^*_\mS - (\nabla_t \myu - \nabla_t \myu^*) \right]
			+\\
			&  &
			\lambda MN
			\begin{pmatrix}
				\myz_\mS \\
				\myz_{\mS^c}
			\end{pmatrix}
		\end{array}
	\end{equation}
	Suppose the primal-dual witness (PDW) construction gives us an solution $(\check{\mybeta}, \check{\myz}) \in \mathbb R^K \times \mathbb R^K$, where $\check{\mybeta}_{\mS^c} = 0$ and $\check{\myz} \in \partial \|\check{\mybeta}\|_1$.
	By plugging $(\check{\mybeta}, \check{\myz})$ into the above equation, we have
	\begin{eqnarray}
		\check{\myz}_{\mS^c}
		& = & \nonumber
		\myX_{\mS^c}^\top \myX_{\mS}
		(\myX_{\mS}^\top \myX_{\mS} )^{-1} \myz_{\mS}
		-
		\myX_{\mS^c}^\top
		\underbrace{(\myI - \myX_{\mS}(\myX_{\mS}^\top \myX_{\mS} )^{-1} \myX_{\mS}^\top ) }_{\myH_{X_s}}
		\frac{\left[ (\myX - \myX^*)_\mS \mybeta^*_\mS - (\nabla_t \myu - \nabla_t \myu^*) \right]}{\lambda MN} \\
		& = & \label{equ: proof -- PDW -- z Sc}
		\myX_{\mS^c}^\top \myX_{\mS}
		(\myX_{\mS}^\top \myX_{\mS} )^{-1} \myz_{\mS}
		-
		\frac{1}{\lambda MN}
		\myX_{\mS^c}^\top
		\myH_{X_s}
		\underbrace{  \left( \myX _\mS \mybeta^*_\mS - \nabla_t \myu  \right)}_{\mytau}. 
	\end{eqnarray}
	
	From \eqref{equ: proof -- PDW -- z Sc}, we have
	\begin{eqnarray*}
		P(\|\check{\myz}_{\mS^c}\|_\infty \geq 1)
		& = &
		P\left(
		\left\|
		\myX_{\mS^c}^\top \myX_{\mS}
		(\myX_{\mS}^\top \myX_{\mS} )^{-1} \myz_{\mS}
		-
		\frac{1}{\lambda MN}
		\myX_{\mS^c}^\top
		\myH_{X_s}
		\mytau
		\right\|_\infty>1
		\right)  \\
		& \leq &
		P\left(
		\left\|
		\myX_{\mS^c}^\top \myX_{\mS}
		(\myX_{\mS}^\top \myX_{\mS} )^{-1} \myz_{\mS}
		\right\|_\infty > 1 -\mu
		\right)+ \\
		&   &
		P\left(
		\left\|
		\frac{1}{\lambda MN}
		\myX_{\mS^c}^\top
		\myH_{X_s}
		\mytau
		\right\|_\infty > \mu
		\right)  .
	\end{eqnarray*}
	If we denote
	$
	\widetilde{Z_j}
	=
	\frac{1}{\lambda MN}
	(\myX_{\mS^c})_j^\top
	\myH_{X_s}
	\mytau ,
	$
	where
	$
	(\myX_{\mS^c})_j$ is the $j$-th column of $\myX_{\mS^c}
	$,
	then we have
	\begin{equation}
		\label{equ: proof -- PDW -- z Sc two compoents}
		P(\|\check{\myz}_{\mS^c}\|_\infty \geq 1)
		\leq
		P\left(
		\left\|
		\myX_{\mS^c}^\top \myX_{\mS}
		(\myX_{\mS}^\top \myX_{\mS} )^{-1}
		\right\|_\infty > 1 -\mu
		\right)
		+
		P\left(
		\max_{j \in \mS^c} |\widetilde{Z_j}| > \mu
		\right)  .
	\end{equation}
	Now let us discuss the upper bound of the second term, i.e.,
	$
	P
	\left(
	\max_{j \in \mS^c} |\widetilde{Z_j}| > \mu
	\right) .
	$
	Because
	\begin{eqnarray}
		P\left(
		\max_{j \in \mS^c} |\widetilde{Z_j}|
		>
		\mu
		\right)
		& = & \nonumber
		P\left(
		\left\|
		\frac{1}{\lambda MN}
		\myX_{\mS^c}^\top
		\myH_{X_s}
		\mytau
		\right\|_{\infty}
		>
		\mu
		\right) \\
		& \leq & \nonumber
		P\left(
		\left\|
		\frac{1}{\lambda MN}
		\myX_{\mS^c}^\top
		\myH_{X_s}
		\mytau
		\right\|_2
		>
		\mu
		\right) \\
		& \leq & \nonumber
		P\left(
		\left\|
		\frac{1}{\lambda MN}
		\myX^\top
		\myH_{X_s}
		\mytau
		\right\|_2
		>
		\mu
		\right) \\
		& \leq & \nonumber
		P\left(
		\frac{1}{\lambda MN}
		\left\|
		\myX
		\right\|_2
		\left\|
		\mytau
		\right\|_2
		>
		\mu
		\right) \\
		& \leq & \nonumber
		P\left(
		\left\|
		\mytau
		\right\|_2
		>
		\lambda \mu \sqrt{\frac{MN}{K}}
		\right) \\
		& \leq & \label{equ: bound Zj}
		P\left(
		\left\|
		\mytau
		\right\|_\infty
		>
		\lambda \mu \frac{1}{\sqrt{K}}
		\right). 
	\end{eqnarray}
	
	By Lemma \ref{lemma: bound y-Xb*}, we know when
	$$
	\lambda \mu \frac{1}{\sqrt{K}}
	>
	\mathscr C(\sigma, \|u\|_{L^\infty(\Omega)}) \frac{\log(N)}{N^{3/7 - r}},
	$$
	we have
	\begin{equation*}
		P
		\left(
		\| \nabla_t \myu - \myX \mybeta^* \|_\infty
		> \epsilon
		\right)
		<
		N e^{-N^r}.
	\end{equation*}
	So we know that
	\begin{eqnarray}
		P
		\left(
		\| \mytau \|_\infty
		> \lambda \mu \frac{1}{\sqrt{K}}
		\right)
		& = & \nonumber
		P
		\left(
		\| \nabla_t \myu - \myX_{\mathcal S} \mybeta^*_{\mathcal S} \|_\infty
		> \lambda \mu \frac{1}{\sqrt{K}}
		\right)\\
		& \leq & \nonumber
		P
		\left(
		\| \nabla_t \myu - \myX \mybeta^* \|_\infty
		> \lambda \mu \frac{1}{\sqrt{K}}
		\right) \\
		& < & \label{equ: proof -- PDW -- bound tau}
		N e^{-N^r}. 
	\end{eqnarray}
	
	By plugging the results in \eqref{equ: bound Zj} and \eqref{equ: proof -- PDW -- bound tau} into \eqref{equ: proof -- PDW -- z Sc two compoents}, we have
	\begin{eqnarray*}
		P(\|\check{\myz}_{\mS^c}\|_\infty \geq 1)
		& \leq &
		P
		\left(
		\left\|
		\myX_{\mS^c}^\top \myX_{\mS}
		(\myX_{\mS}^\top \myX_{\mS} )^{-1}
		\right\|_\infty > 1 -\mu
		\right)
		+
		P
		\left(
		\max_{j \in \mS^c} |\widetilde{Z_j}| > \mu
		\right)  \\
		& \leq &
		P
		\left(
		\left\|
		\myX_{\mS^c}^\top \myX_{\mS}
		(\myX_{\mS}^\top \myX_{\mS} )^{-1}
		\right\|_\infty > 1 -\mu
		\right)
		+
		P\left(
		\left\|
		\mytau
		\right\|_\infty
		>
		\lambda \mu \frac{1}{\sqrt{K}}
		\right)\\
		& \leq &
		P
		\left(
		\left\|
		\myX_{\mS^c}^\top \myX_{\mS}
		(\myX_{\mS}^\top \myX_{\mS} )^{-1}
		\right\|_\infty > 1 -\mu
		\right)
		+
		N e^{-N^r}. 
	\end{eqnarray*}
	
	The probability for proper support set recovery is
	\begin{eqnarray*}
		P(\|\check \myz_{\mS^c}\|_\infty < 1)
		&=&
		1 - P(\|\check \myz_{\mS^c}\|_\infty \geq 1) \\
		&\geq&
		1-
		\left[
		P\left(
		\left\|
		\myX_{\mS^c}^\top \myX_{\mS}
		(\myX_{\mS}^\top \myX_{\mS} )^{-1}
		\right\|_\infty > 1 -\mu
		\right)
		+
		N e^{-N^r}
		\right] \\
		& = &
		P\left(
		\left\|
		\myX_{\mS^c}^\top \myX_{\mS}
		(\myX_{\mS}^\top \myX_{\mS} )^{-1}
		\right\|_\infty \leq 1 -\mu
		\right)
		- N e^{-N^r} \\
		& \leq &
		P_{\mu}
		-
		N e^{-N^r}.
	\end{eqnarray*}
	Thus, we finish the proof.
\end{proof}

\subsubsection{Proof of Theorem \ref{theo: beta estimation error bound}}
\label{proof: Estimation Error Bound}
\begin{proof}
	By equation \eqref{equ: proof -- first order condition of LASSO after decomposition},
	we can solve $\mybeta_{\mS} - \mybeta_{\mS}^*$ as
	$$
	\mybeta_{\mS} - \mybeta_{\mS}^*
	=
	(\myX_{\mS}^\top \myX_{\mS})^{-1}
	\left[
	-
	\myX_{\mS}^\top
	\left(
	\myX_{\mS} - \myX_{\mS}^*
	\right)
	\mybeta^*_{\mS}
	+
	\myX_{\mS}^\top
	\left( \nabla_t \myu - \nabla_t \myu^*   \right)
	-
	\lambda MN \myz_{\mS}
	\right].
	$$
	Thus, we have the following series of equations:
	\begin{eqnarray}
		& & \nonumber
		\max_{k \in \mS} |\mybeta_k - \mybeta_k^*| \\
		& \leq & \nonumber
		\left\|(\myX_\mS^\top \myX_\mS)^{-1} \right\|_\infty
		\left\|
		\myX_\mS^\top
		\left[
		\nabla_t \myu - \nabla_t \myu^* -(\myX_\mS-\myX_\mS^*)\mybeta^*_\mS
		\right]
		-
		\lambda MN \myz_{\mS}
		\right\|_\infty  \\
		& \leq &  \nonumber
		\left\|(\myX_\mS^\top \myX_\mS)^{-1} \right\|_\infty
		\left[
		\left\|
		\myX_\mS^\top
		\left[
		\nabla_t \myu - \nabla_t \myu^* -(\myX_\mS-\myX_\mS^*)\mybeta^*_\mS
		\right]
		\right\|_\infty
		+
		\lambda MN
		\left\| \myz_{\mS} \right\|_\infty
		\right] \\
		& = &  \label{equ: proof -- estimation error 1}
		\left\|(\myX_\mS^\top \myX_\mS)^{-1} \right\|_\infty
		\left[
		\left\|
		\myX_\mS^\top \left( \nabla_t \myu -  \myX_\mS \mybeta^*_\mS  \right)
		\right\|_\infty
		+
		\lambda MN \left\| \myz_{\mS} \right\|_\infty
		\right] \\
		& \leq &  \label{equ: proof -- estimation error 2}
		\left\|\left( \frac{\myX_\mS^\top \myX_\mS }{MN}\right)^{-1} \right\|_\infty
		\left(
		\frac{
			\left\|
			\myX_\mS^\top \left( \nabla_t \myu -  \myX_\mS \mybeta^*  \right)
			\right\|_\infty}{MN} + \lambda
		\right)  \\
		& \leq &  \label{equ: proof -- estimation error 3}
		\sqrt{K}C_{\min}
		\left(
		\frac{
			\left\|
			\myX_\mS^\top \left( \nabla_t \myu -  \myX_\mS \mybeta^*  \right)
			\right\|_\infty}{MN} + \lambda
		\right)  \\
		& \leq & \label{equ: proof -- estimation error 4}
		\sqrt{K}C_{\min}
		\left(
		\frac{
			\| \myX_\mS \|_{\infty, \infty}
			\left\|
			\nabla_t \myu -  \myX_\mS \mybeta^*
			\right\|_\infty}{MN} + \lambda
		\right)  \\
		& \leq & \nonumber 
		\sqrt{K}C_{\min}
		\left(
		\frac{
			\| \myX_\mS \|_{F}
			\left\|
			\nabla_t \myu -  \myX_\mS \mybeta^*
			\right\|_\infty}{\sqrt{MN}} + \lambda
		\right)  \\
		& \leq & \label{equ: proof -- estimation error 6}
		\sqrt{K}C_{\min}
		\left(
		\frac{
			\sqrt{MNK}
			\left\|
			\nabla_t \myu -  \myX_\mS \mybeta^*
			\right\|_\infty}{\sqrt{MN}} + \lambda
		\right)  \\
		& = & \nonumber 
		\sqrt{K} C_{\min}
		\left(
		\sqrt{K}
		\left\|
		\nabla_t \myu -  \myX_\mS \mybeta^*
		\right\|_\infty + \lambda
		\right)  \\
		& \leq & \label{equ: proof -- estimation error 8}
		\sqrt{K}C_{\min}
		\left(
		\sqrt{K}
		\mathscr C_{ (\sigma, \|u\|_{L^\infty(\Omega)}) }
		\frac{\log(N)}{N^{3/7 - r}}
		+
		\lambda
		\right)
	\end{eqnarray}
	Equation \eqref{equ: proof -- estimation error 1} is because $\nabla_t \myu^* = \myX_{\mS} \mybeta_{\mS}$.
	Inequality \eqref{equ: proof -- estimation error 2} is because $\left\| \myz_{\mS} \right\|_\infty = 1$.
	Inequality \eqref{equ: proof -- estimation error 3} is because of Condition \ref{assumption -- PDW -- minimal eigenvalue condition}.
	Inequality \eqref{equ: proof -- estimation error 4} is because for a matrix $\myA$ and a vector $\myx$, we have $\|\myA \myx\|_{q} \leq \|\myA \|_{p,q} \|\myx\|_{p}$.
	Here the matrix norm for matrix $\myA \in \mathbb R^{m \times n}$ in $\| \myA \|_{\infty, \infty}  = \|\text{vector}(\myA)\|_{\infty}$.
	In inequality \eqref{equ: proof -- estimation error 6}, the  norm of matrix $\myA \in \mathbb R^{m \times n}$ is that
	$
	\left\| \myA \right\|_F
	=
	\sqrt{ \sum_{i =1}^{m} \sum_{j =1}^{n} |A_{ij}|^2} ,
	$
	and the norm of vector $\mya \in \mathbb R^d$ is
	$
	\|\mya \|_{\infty} = \max_{1\leq i \leq d} |a_i|.
	$
	Inequality \eqref{equ: proof -- estimation error 6} is because we normalized columns of $\myX$.
	Inequality \eqref{equ: proof -- estimation error 8} is due to Lemma Lemma \ref{lemma: bound y-Xb*} under probability
	$
	1 - O(N e^{-N^r}) \to 1.
	$
\end{proof}

\subsection{The full model used in Section 4}
\label{appendix: simulation full model}
The full model used in Section \ref{sec: PDE paper -- simulation} is 
$$
  \begin{array}{ccl}
  \frac{\partial}{\partial t} u(x,t) & = &
  \beta_1^* + 
  \beta_2^* u(x, t) + 
  \beta_3^* \frac{\partial}{\partial x} u(x,t) +
  \beta_4^* \frac{\partial^2}{\partial x^2} u(x,t) + \\
  & &
  \beta_5^* \left[u(x, t)\right]^2 +
  \beta_6^* \left[\frac{\partial}{\partial x} u(x,t)\right]^2 +
  \beta_7^* \left[\frac{\partial}{\partial x} u(x,t) \right]^2 +\\
  & & 
  \beta_8^* u(x, t) \frac{\partial}{\partial x} u(x,t) +
  \beta_9^* u(x, t) \frac{\partial^2}{\partial x^2} u(x,t) + \\
  & &
  \beta_{10}^* \frac{\partial}{\partial x} u(x,t) \frac{\partial^2}{\partial x^2} u(x,t)
  \end{array}.
$$

\subsection{Checking Conditions of Example 1,2,3}
\label{sec: model diagnosis}
In this section, we check Condition \ref{assumption -- PDW -- mutual incoherence condition} - Condition \ref{assumption -- spline -- decreasing penalty parameter} of the above three examples: (1) example 1 (the transport equation), (2) example 2 (the inviscid Burgers' equation) and (3) example 3 (the viscous Burgers' equation).

\subsubsection{Verification of Condition\ref{assumption -- PDW -- mutual incoherence condition}, \ref{assumption -- PDW -- minimal eigenvalue condition}}

In this section, we check the Condition \ref{assumption -- PDW -- mutual incoherence condition} - Condition \ref{assumption -- PDW -- minimal eigenvalue condition} under example 1,2,3, though the applicability of the results is by no means restricted to these.

\begin{figure}[htbp]
	\centering
	\begin{tabular}{ccc}
		\includegraphics[width = 0.3\textwidth]{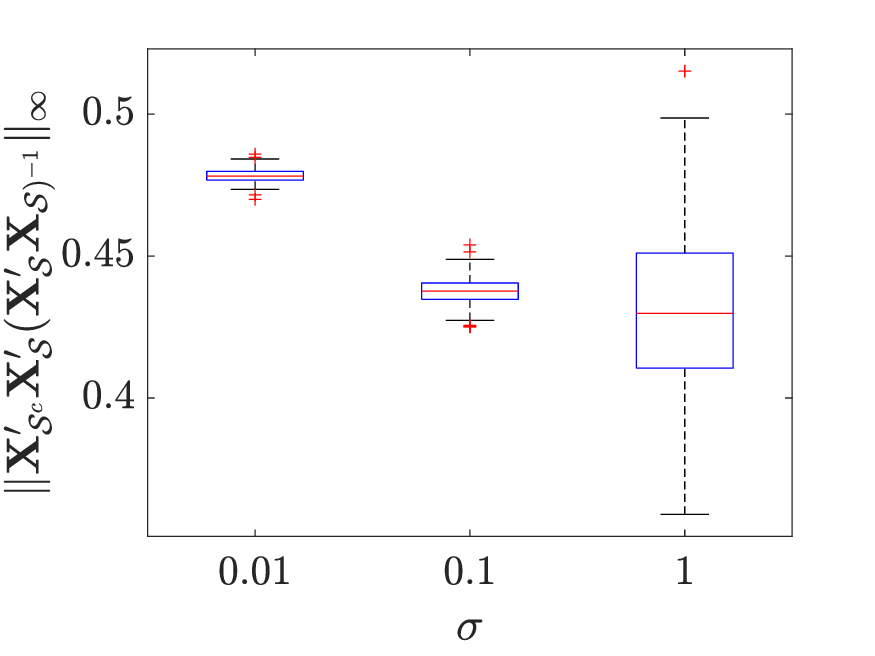} &
		\includegraphics[width = 0.3\textwidth]{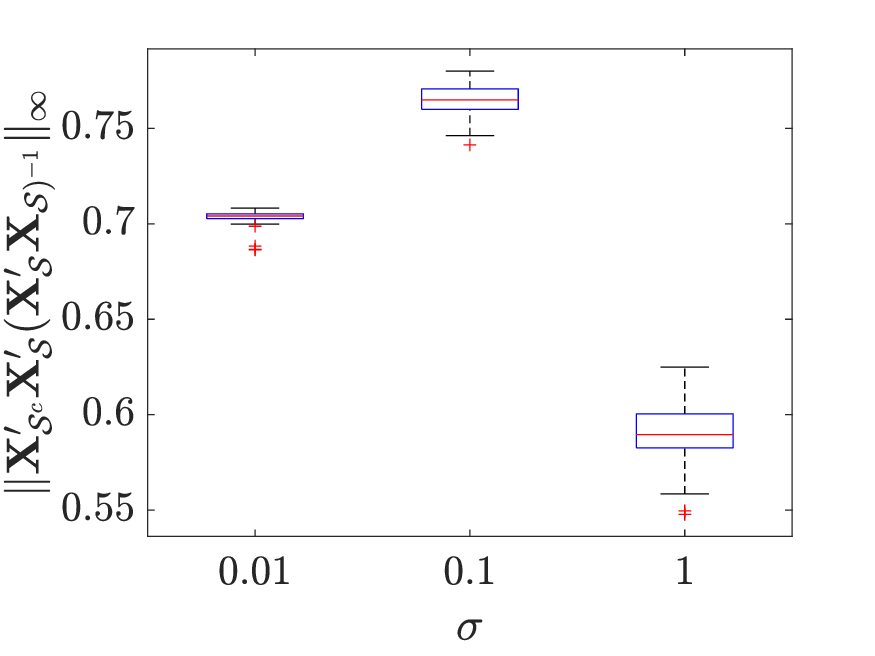} &
		\includegraphics[width = 0.3\textwidth]{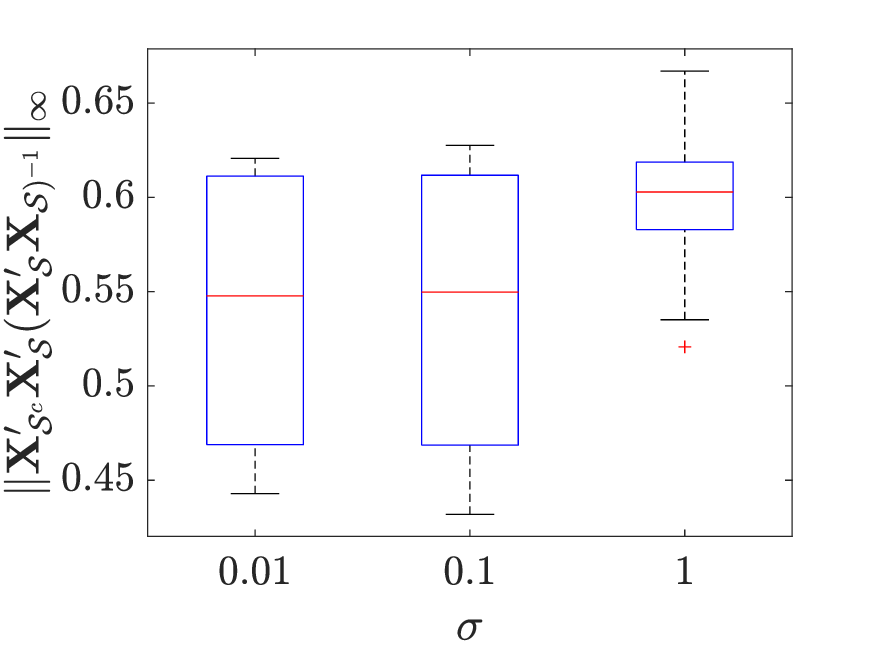} \\
		(a) example 1: &
		(b) example 2: &
		(c) example 3: \\
	\end{tabular}
	\caption{Box plots of
		$
		\left\|
		\myX_{\mathcal S^c}^\top \myX_\mathcal S( \myX_{\mathcal S}^\top \myX_\mathcal S)^{-1}
		\right\|_\infty
		$
		under $\sigma = 0.01, 0.1, 1$ when $M=N=100$.
		\label{fig: verify PDW conditions - mutual_incoherence} }
\end{figure}

\begin{figure}[htbp]
	\centering
	\begin{tabular}{ccc}
		\includegraphics[width = 0.3\textwidth]{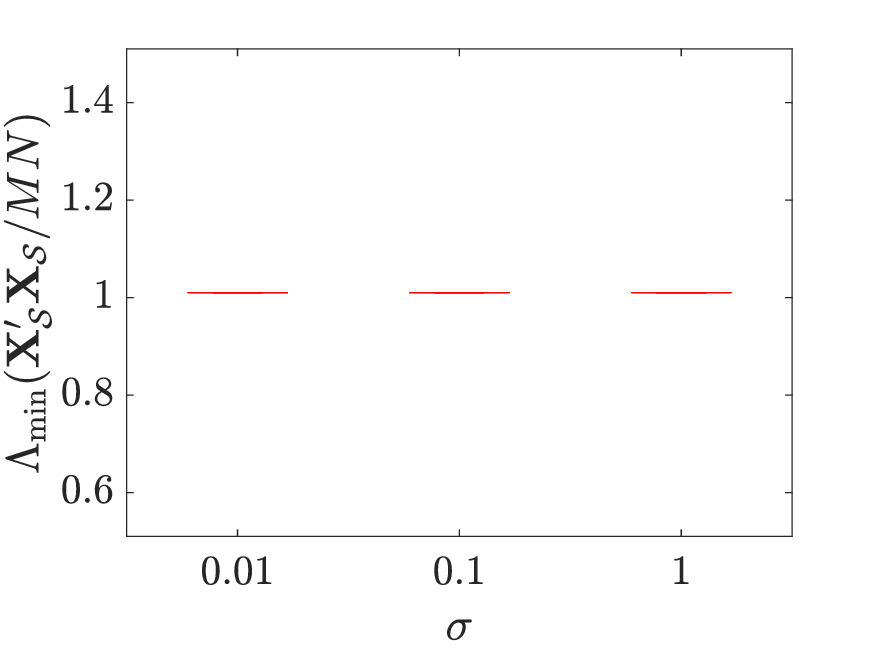} &
		\includegraphics[width = 0.3\textwidth]{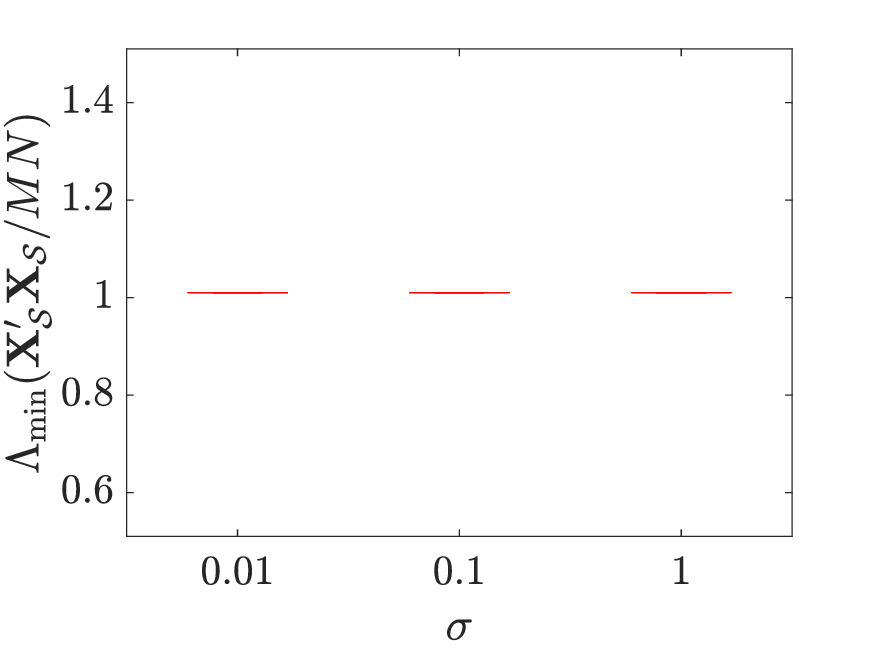} &
		\includegraphics[width = 0.3\textwidth]{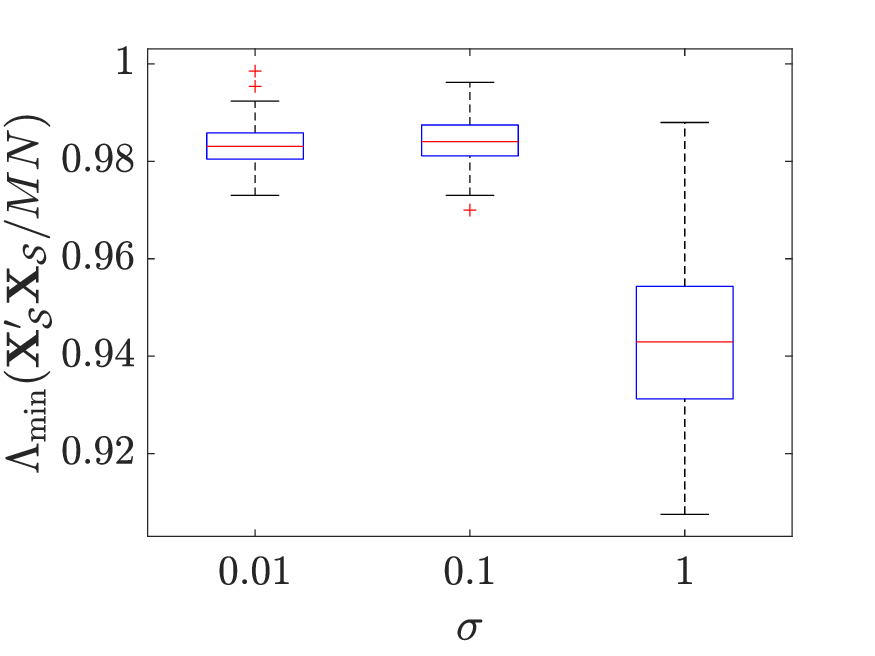} \\
		(a) example 1: &
		(b) example 2: &
		(c) example 3: \\
	\end{tabular}
	\caption{Box plots of the minimal eigenvalue of matrix
		$
		\frac{1}{NM} \myX_\mathcal S^\top \myX_\mathcal S
		$
		under $\sigma = 0.01, 0.1, 1$ when $M=N=100$.
		\label{fig: verify PDW conditions - min eigenvalue} }
\end{figure}

The verification results can be found in Fig. \ref{fig: verify PDW conditions - mutual_incoherence} and Fig. \ref{fig: verify PDW conditions - min eigenvalue}, where (a),(b),(c) are the box plot of
$
\left\|
\myX_{\mathcal S^c}^\top \myX_\mathcal S( \myX_{\mathcal S}^\top \myX_\mathcal S)^{-1}
\right\|_\infty
$
and the minimal eigenvalue of matrix $\frac{1}{NM} \myX_\mathcal S^\top \myX_\mathcal S$ of these three examples under $\sigma = 0.01, 0.1, 1$, respectively.
From Fig. \ref{fig: verify PDW conditions - mutual_incoherence}, we find the value of
$
\left\|
\myX_{\mathcal S^c}^\top \myX_\mathcal S( \myX_{\mathcal S}^\top \myX_\mathcal S)^{-1}
\right\|_\infty
$
is smaller than 1, so there exist a $\mu \in (0, 1]$ such that Condition \ref{assumption -- PDW -- mutual incoherence condition} is met.
From Fig. \ref{fig: verify PDW conditions - min eigenvalue}, we find the minimal eigenvalue of matrix $\frac{1}{MN} \myX_{\mS}^\top \myX$ are all strictly larger than 0, so we declare Condition \ref{assumption -- PDW -- minimal eigenvalue condition} is satisfied.

\subsubsection{Verification of Condition \ref{assumption -- spline -- convergence cdf} and Condition \ref{assumption -- spline -- bounded pdf}}
In example 1,2,3, the design points $x_0, x_1, \ldots, x_{M-1}$ and $t_0, t_1, \ldots, t_{N-1}$ are equally spaced, i.e., $x_0 = 1/M, x_1 = 2/M, \ldots, x_{M-1} = 1$ and $t_0 = 0.1/N, t_1 = 0.2/N, \ldots, t_{N-1} = 0.1$.
Under this scenario, there exist an absolutely continuous distribution $F(x) = x$ for $x \in[1/M, 1]$ and $G(t) = 0.1 t$ for $t \in [0.1/N, 0.1]$, where the empirical c.d.f. of the design points $x_0, x_1, \ldots, x_{M-1}$ and $t_0, t_1, \ldots, t_{N-1}$ will converge to $F(x), G(t)$, respectively, as $M,N \to +\infty$.
For the $F(x), G(t)$, we know their first derivatives is bounded for $x \in[1/M, 1]$ and $t \in [0.1/N, 0.1]$, respectively.
In the simulation of this paper, we take the equally spaced design points as an illustration example, and its applicability is by no means restricted to this case.

\subsubsection{Verification of Condition \ref{assumption -- spline -- decreasing penalty parameter}.}
The Condition \ref{assumption -- spline -- decreasing penalty parameter} ensures that the smoothing parameter does not tend to zero too rapidly.
\cite{silverman1984spline} shows that for the equally spaced design points, this condition meets.
For other types of design points, for instance, randomly and independently distributed design points, it can also be verified that Condition \ref{assumption -- spline -- decreasing penalty parameter} is satisfied \cite[see][Section 2]{silverman1984spline}.

\subsection{Details of the Case Study}
The header of the CALIPSP dataset and its visualization can be found in Table \ref{table: header of CALIPSP} and Fig. \ref{fig: case study -- curve and solution path}(a), respectively.
Fig. \ref{fig: case study -- curve and solution path}(a) shows presents the curves of the dynamic in the CALIPSP dataset, where the x-axis is the longitude and the y-axis is the value of the observed temperature.
Here the black curve plots the observed temperature in January 2017, and the lighter color presents the later month.
As seen from Fig. \ref{fig: case study -- curve and solution path}, we find overall there is an increasing trend of the temperature in the first half-year and then the temperature decreases.

\begin{table}[htbp]
	\caption{The header of the CALIPSP dataset}
	\label{table: header of CALIPSP}
	\centering
	\begin{adjustbox}{max width=0.95\textwidth}
		\centering
		\begin{threeparttable}
			\begin{tabular}{c|cccccccccccc}
				\hline
				&\multicolumn{7}{c}{longitude}\\
				& -177.5 & -172.5 & -167.5 & $\ldots$ & 167.5 & 172.5& 177.5 \\
				\hline
				Jan 2017 & -46.5103  & -48.4720  & -44.6581  & $\ldots$ & -44.2778    & -47.3354  & -44.0146\\
				Feb 2017 & -46.2618  & -43.1994  & -47.6370  & $\ldots$ &  -46.8409   &  -46.2727 &  -40.5556\\
				$\vdots$ & $\vdots$  & $\vdots$& $\vdots$& $\vdots$ & $\vdots$  & $\vdots$& $\vdots$ \\
				Dec 2017 & -47.3145    & -47.6505  &  -50.8332 & $\ldots$ &  -43.9705   & -43.0475 & -46.0618\\
				\hline
			\end{tabular}
			\begin{tablenotes}
				\footnotesize
				\item[1] The data is downloaded from \url{https://asdc.larc.nasa.gov/data/CALIPSO/LID_L3_Tropospheric_APro_CloudFree-Standard-V4-20/} (registration is required).
				\item[2] The negative and positive longitude refer to the west and east longitude, respectively.
			\end{tablenotes}
		\end{threeparttable}
	\end{adjustbox}
\end{table}

\subsection{More details of Fig. \ref{fig: case study -- surface}}
The three-dimension surface plot of the observed temperature in 2017 can be found in Fig. \ref{fig: case study -- surface}(a.1), whose fitted value can be found in Fig. \ref{fig: case study -- surface}(a.2).
The three-dimension surface plot of the residual between the observed temperature and the fitted temperature can be found in Fig. \ref{fig: case study -- surface}(a.3).
Seeing from Fig. \ref{fig: case study -- surface}(a.1)-(a.3), we find the fitted temperature captures the dynamic trend of the raw data well.
Although the magnitude of the residual is not small, it is still satisfying given the following reasons.
The fitted value by using the explicit Euler method only serves as the baseline method, which is not accurate enough for fitting.
In this paper, we focus on PDE identification, so we use the most simple method -- the explicit Euler method -- to check if the identified PDE model in \eqref{equ: case study -- identified model} can capture the features of the underlying PDE model.
The more advanced method --  Runge-Kutta fourth order method (RK4) --  is not implementable in our content, and the reasons are explained in online supplementary material.

The value to plot Fig. \ref{fig: case study -- surface}(a.2) is calculated as follows.
First, we use the identified PDE model in \eqref{equ: case study -- identified model} to predict the value of $\frac{\partial}{\partial t}u(x,t)$ in January 2017.
Then, we use the explicit Euler method \cite[][]{butcher2008numerical} to predict the future value from February 2017 to December 2017, i.e., we have 
$
u(x, t+\Delta t) = u(x,t) + \frac{\partial}{\partial t}u(x,t) \Delta t.
$

\end{document}